\documentclass[a4paper,11pt]{article}

\usepackage{jheppub} 

\DeclareMathOperator{\sh}{\mathrm{sh}}
\DeclareMathOperator{\ch}{\mathrm{ch}}
\DeclareMathOperator{\tnh}{\mathrm{th}}

\newcommand{\tht}{\vartheta}
\newcommand{\ph}{\varphi}

\newcommand{\tens}[1]{{#1}}
\newcommand{\grad}{{\tens{d}}}

\newcommand{\cv}[1]{{\tens{\partial}}_{#1}}

\newcommand{\pcovd}{{\tens{\triangledown}}}
\newcommand{\Lob}{{\scriptscriptstyle\mathrm{Lob}}}
\newcommand{\AdS}{{\scriptscriptstyle\mathrm{AdS}}}
\newcommand{\crit}{{\scriptstyle\mathrm{cr}}}
\newcommand{\mx}{{\scriptstyle\mathrm{max}}}
\newcommand{\plane}{{\scriptstyle\mathrm{hp}}}
\newcommand{\ren}{{\scriptstyle\mathrm{ren}}}
\newcommand{\tot}{{\scriptstyle\mathrm{tot}}}
\newcommand{\comp}{{\scriptstyle\mathrm{comp}}}
\newcommand{\scri}{{\mathcal{I}}}

\newcommand{\realn}{{\mathbb{R}}}

\newcommand{\tP}{{\bar{t}}}
\newcommand{\xP}{{\bar{x}}}
\newcommand{\yP}{{\bar{y}}}
\newcommand{\zP}{{\bar{z}}}
\newcommand{\rP}{{\bar{r}}}

\renewcommand{\tens}[1]{{\boldsymbol{#1}}}

\newcommand{\imgdir}{}   

\title{Minimal surfaces and entanglement entropy in anti-de~Sitter space}

\author[a]{Pavel Krtou\v{s}}
\author[b]{Andrei Zelnikov}

\affiliation[a]{Institute of Theoretical Physics,
Faculty of Mathematics and Physics, Charles University in Prague,
V~Hole\v{s}ovi\v{c}k\'ach 2, Prague, Czech Republic}

\affiliation[b]{Theoretical Physics Institute, Department of Physics,
University of Alberta,\\
Edmonton, Alberta, T6G 2E1 Canada }

\emailAdd{Pavel.Krtous@utf.mff.cuni.cz}
\emailAdd{zelnikov@ualberta.ca}

\abstract{%
According to Ryu and Takayanagi, the entanglement entropy in conformal
field theory (CFT) is related through the AdS/CFT correspondence to
the area of a minimal surface in the bulk. We study this holographic
geometrical method of calculating the entanglement entropy in the
vacuum case of a CFT which is holographically dual to empty
anti-de~Sitter (AdS) spacetime. Namely, we investigate the minimal
surfaces spanned on boundaries of spherical domains at infinity of
hyperbolic space, which represents a time-slice of AdS spacetime. We
consider a generic position of two spherical domains: two disjoint
domains, overlapping domains, and touching domains. In all these cases
we find the explicit expressions for the minimal surfaces and the
renormalized expression for the area. We study also the embedding of
the minimal surfaces into full AdS spacetime and we find that for a
proper choice of the static Killing vector we can model a dynamical
situation of ``tearing'' of the minimal surface when the domains on
which it is spanned are moved away from each other. }


\begin{document}
\maketitle
\flushbottom

\newpage

\section{Introduction}
\label{sc:intro}

\subsection{Black hole entropy and entanglement entropy}
\label{ssc:bhentgent}

The connection between different areas of physics is of most importance in
fundamental physics. One of the most remarkable achievements of general
relativity is the discovery that an entropy is an inherent property of the
black hole horizons. In fact, the famous Bekenstein-Hawking formula
\cite{Bekenstein:1972tm,Bekenstein:1973ur,Bekenstein:1974ax,Hawking:1971vc,
Hawking:1974rv,Hawking:1974sw}
\begin{equation}\label{Bekenstein-Hawking}
S_{BH}= \frac{k_B c^3}{\hbar}\frac{{A}}{4 G}
\end{equation}
is applicable to any Killing horizon and
provides the relation of the entropy of the gravitational system and the area
of the horizon ${A}$. This remarkable formula connects thermodynamics,
gravity
and relativistic quantum field theory. This relation is valid not only in
four dimensions but in higher dimensions too. In higher dimensions the
gravitational constant $G$ is the ${D}$-dimensional one and ${A}$ is
the volume of $(D{-}2)$-dimensional surface of the horizon.

An entanglement entropy is known to have a
very similar dependence on the area of a surface separating two subsystems of
a quantum mechanical system
\cite{Bombelli:1986rw,Frolov:1993ym,Srednicki:1993im}. This
resemblance of the entanglement entropy with the horizon entropy
has deep roots and is related to the problem of statistical-mechanical
explanation of black hole entropy \cite{Frolov:1993ym}. Let us consider
quantum fields described by a wave function
in a stationary black hole spacetime. The black hole horizon is
the surface which separates the interior of the black hole from its exterior.
Then one can show \cite{Barvinsky:1994jca} that the corresponding entanglement entropy
reduces to the Bekenstein-Hawking entropy. For that one has to take into
account that quantum fields on a curved background lead to the
renormalization of the effective gravitational constant and, at the same time,
these
quantum fields also contribute to the entanglement entropy of the horizon.
It's amazing that the renormalized entropy per unit area of a horizon is
governed by the same formulas as
the quantum corrections to the gravitational coupling \cite{Susskind:1994sm}.
As the result \eqref{Bekenstein-Hawking} remains valid after taking account of
quantum corrections, one just has to substitute $G$ with
$G_{\mathrm{ren}}$.

The interpretation of the Bekenstein-Hawking formula
\eqref{Bekenstein-Hawking} as an entanglement entropy becomes even more
convincing
in the framework of induced or emergent gravity models
\cite{Sakharov:1967pk,Jacobson:1994iw,Frolov:1996aj,Frolov:1996qh,Frolov:2003ed}.
In these models the Einstein--Hilbert action is the leading term  to the
low-energy effective gravitational action, where the gravitational coupling
and a cosmological constant, are completely generated by quantum fluctuations
of matter fields living on a curved background spacetime. The gravitational
constant $G$ entering the \eqref{Bekenstein-Hawking} is then the induced
Newton constant $G_{\mathrm{ind}}$.

In the case of static black
holes the event horizon coincides with the Killing horizon and is the minimal
area surface defined on the Einstein-Rosen bridge.
In the paper \cite{Barvinsky:1994jca} it was proposed that the
the minimal area surface on the $t=\mbox{const}$ slice of the spacetime may
play an important role in defining the entanglement entropy of black holes
in a more general setup of the problem. Note that the minimal area
surface is a more general notion than just a horizon of a static black
hole. The trace of the extrinsic curvature vanishes both for the minimal
area surface and the horizon, but in the case of the horizon all components of
the extrinsic curvature vanish.


\subsection{Entanglement entropy and minimal surfaces}
\label{ssc:eentrms}

Recently holographic computation of the entanglement entropy in conformal
field theories (CFT) got a lot of attention and developments, especially in
the frameworks of the AdS/CFT correspondence. Ryu and Takayanagi
\cite{Ryu:2006bv,Ryu:2006ef,Nishioka:2009un} proposed that in a static
configuration the entanglement entropy of a subsystem
localized in a domain $\Omega$ is given by the elegant formula\footnote{From
now on we
use
$k_{\scriptscriptstyle B}=c=\hbar=1$ system of units.}
\begin{equation}\label{Ryu-Takayanagi}
S_{\Omega}=\frac{{A}_{\Sigma_\Omega}}{4 \,G} .
\end{equation}
Given a static time slice
(the \mbox{${(D{-}1)}$-dimensional} bulk space), the
\mbox{$({D{-}2})$-dimensional}
domain ${\Omega}$ belongs to infinite boundary ${\mathcal{I}}$ of the bulk and
${A_{\Sigma_\Omega}}$ in Eq.~\eqref{Ryu-Takayanagi} is
the area of a ${(D{-}2)}$-dimensional minimal surface $\Sigma_{\Omega}$
in the bulk spanned on the boundary ${\partial \Omega}$ of the subsystem
(i.e., ${\partial\Sigma_{\Omega}=\partial \Omega}$).
One may consider the bulk surface
$\Sigma_\Omega$ to be homologous to the boundary
region $\Omega$ \cite{Fursaev:2006ih,Headrick:2010zt}.

In the case of the Einstein gravity in the bulk and static backgrounds this
conjecture was recently proved \cite{Lewkowycz:2013nqa}. In a more general
case, when the gravitational action contains higher curvature corrections, a
formula similar to the Wald entropy was proposed \cite{Myers:2013lva}.

The holographic derivation of the Eq.~\eqref{Ryu-Takayanagi} for the
entanglement entropy was proposed in \cite{Fursaev:2006ih} using the replica
trick. This approach works well in application to the von Neumann entropy. A
more general notion of the Renyi entanglement entropy appears naturally in the
replica method. But the derivation of the relation of the Renyi
entanglement entropy with the area of a minimal surface needs different
approach \cite{Hartman:2013mia,Faulkner:2013yia}.

There is another interesting question: Is there formula similar to
Eq.~\eqref{Ryu-Takayanagi},
when the domain $\Omega$ consists of a set of disjoint domains? In
this case the minimal surfaces in the bulk may not be unique. The existence of a
set of different solutions for minimal surfaces with the same boundaries
${\partial\Sigma_{\Omega}=\partial \Omega}$ may lead to a new physics in the
context of AdS/CFT correspondence. A natural generalization is to consider the
set of surfaces with the absolute minimum of their total area taken as a measure
of the entanglement of disconnected regions. This choice satisfies the strong
sub-additivity property \cite{Headrick:2007km}, that any physically acceptable
entropy function has to satisfy.

Recently there have been
discussions of different generalizations of the
Eq.~\eqref{Ryu-Takayanagi} in application to the entanglement entropy for
disconnected regions \cite{Hubeny:2007re,Tonni:2010pv} that still respect  the
strong sub-additivity condition. A closely related notion  of `differential
entropy' has been proposed in
\cite{Balasubramanian:2013lsa} in application to a set of
closed curves in the bulk of $\mbox{AdS}_3$. It describes uncertainty about the
quantum state of two-dimensional CFT  left by the collection of local,
finite-time observables.
In \cite{Myers:2014jia} the notion of `differential entropy' has been extended
to higher dimensions.

Nontrivial physics appears already in the case of only two disjoint domains.
Entanglement entropy for a quantum subsystem  localized in two domains can be
used as a probe of confinement \cite{Klebanov:2007ws,Lewkowycz:2012mw}. In
general, minimal surfaces in the bulk are not uniquely defined by the
condition ${\partial\Sigma_{\Omega}=\partial \Omega}$ at the AdS infinity, if
$\partial \Omega$ is the boundary of the disjoint regions. In addition to the
solution describing two disconnected minimal surfaces in the bulk, there can be
a tubelike minimal area surface, connecting the boundaries of both domains.
The existence of such solutions depends on the distance between the domains an
on their size. There is a maximum distance between components beyond which the
tubelike minimal surface cease to exist
\cite{Klebanov:2007ws,Hirata:2006jx}.

\subsection{Plan of the work}
\label{ssc:plan}

In this paper we study minimal surfaces in the pure AdS spacetime. We found
exact solutions for all types of minimal surfaces spanned on one or two
spherical boundaries at conformal infinity. The relative positions and the
sizes of these spherical boundaries are considered to be arbitrary.
We show that even in the pure AdS background there is a critical
behavior of the entanglement entropy that was
demonstrated \cite{Klebanov:2007ws} for the asymptotically AdS spacetimes with
a black hole in the bulk.
Some of these results have been already announced in a short overview
\cite{Krtous:2013vha},
here we present detail derivation and more thorough discussion.

In the following section we consider minimal surfaces in a warped space with an additional
symmetry. The next section is the overview of various facts from the hyperbolic geometry
which appears as a geometry of the time slice of the AdS spacetime. The section~\ref{sc:msLobsp}
contains the main results: we find the minimal surfaces spanned on the boundaries of two
spherical domains at infinity. Three qualitatively different cases of mutual positions
of the spherical domains are considered: (i) two disjoint domains,
(ii) overlapping domains, and (iii) touching domains. In the first case we find that
for close spherical domains there exists a tube-like minimal surface joining the boundaries of
these domains. In the section~\ref{sc:msAdS} we discuss embeddings of the minimal
surfaces into AdS spacetime. We show that the embedding of the tube-like minimal surface
using the Killing vector associated with observers with the acceleration larger
than the cosmological one can model ``tearing'' of the minimal surface into
two pieces when the the domains are moved far away from each other.
The paper is concluded by the summary.

\section{Minimal surfaces in warped spaces}
\label{sc:warpspc}

\subsection{Warped space}

In mathematics a problem of finding a minimal surface with a given boundary is
known as Plateau's problem. In general, the variational principle leads to the
local
condition of vanishing trace of extrinsic curvature
\begin{equation}\label{zerotrk}
    k = 0\;.
\end{equation}

However, it is difficult to find an explicit solution for general boundary
conditions in an arbitrary curved space. Therefore, we will discuss only highly
symmetric spaces and surfaces aligned to their symmetry. Namely, we start with
the warped space with the metric
\begin{equation}\label{warpedmtrc}
    \tens{g} = p_{ij}(x^k)\,\grad x^i\grad x^j
    + R^2(x^k)\, q_{\mu\nu}(y^\kappa)\,\grad y^\mu\grad x^\nu\,
\end{equation}
Here, the ${D}$-dimensional space is covered by coordinates ${\{x^i,y^\mu\}}$, with ${i=1,2}$ and $\mu=1,\dots,D{-}2$. We speak about 2-dimensional \mbox{${x}$-plane} with the metric ${\tens{p}}$ and \mbox{${(D{-}2)}$-dimensional} `symmetry' \mbox{${y}$-space} with the inner metric ${\tens{q}}$. Mixing between \mbox{${x}$-plane} and \mbox{${y}$-space} is encoded only in the `radial' function ${R(x^k)}$.

\begin{figure}[b]
\begin{center}
\includegraphics[width=3in]{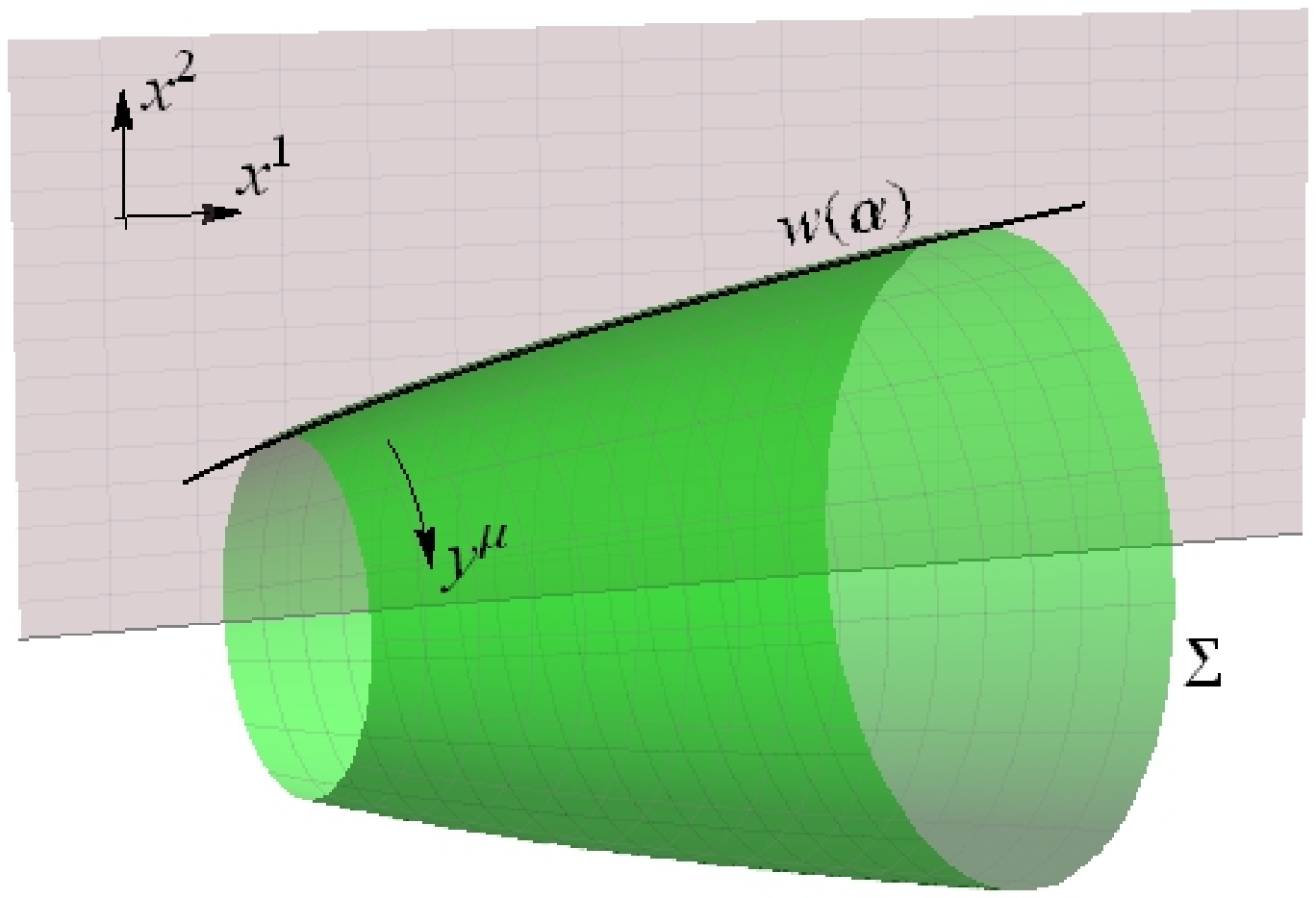}%
\includegraphics[width=3in]{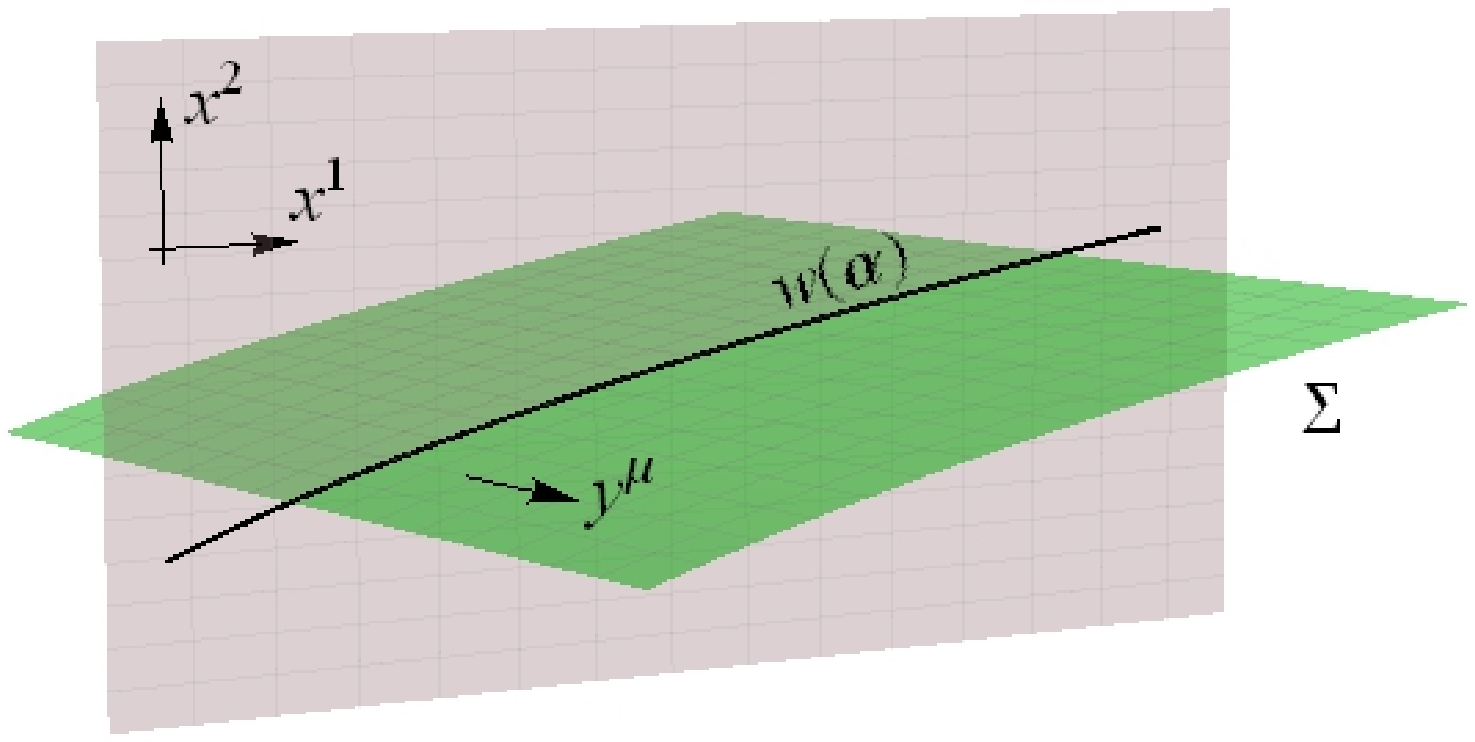}%
\\[-6ex]
{\small(a)\hspace{2.6in}(b)}%
\end{center}
\caption{\label{fig:warpspc}\small
{\bf Warped space.}\quad Warped space is a generalization of idea that rotationally symmetric space can be obtained by a rotation of a ${x}$-plane around the axis. The orbits of the rotation form so called ${y}$-space, cf.~(a). Alternatively, the ${y}$-directions can have a character of a translation (b). Of course, in higher dimensions one can have more general situations. The surfaces aligned to the warp symmetry are given by a profile function ${w(\alpha)}$ in the ${x}$-plane propagated freely in ${y}$-directions.}
\end{figure}

In 3 dimensions, there is only one $y$-coordinate and it is aligned to a Killing symmetry of the spacetime; for example, ${y^1=\ph}$ and ${\tens{q}=\grad\ph^2}$ for the rotational symmetry, see Fig.~\ref{fig:warpspc}.

In the warped space we can look for a minimal surface ${\Sigma}$ aligned to its symmetry. By the alignment of ${\Sigma}$ to the symmetry we mean that a surface ${\Sigma}$ is given by a profile curve ${w(\alpha)}$ in the ${x}$-plane, with coordinates ${y^\mu}$ being unrestricted.

In the 3-dimensional example above, the rotation-symmetric surface is given by the rotation of the profile curve ${w(\alpha)}$ around the axis, cf.~Fig.~\ref{fig:warpspc}a.

Substituting the ansatz \eqref{warpedmtrc} into a definition of the extrinsic curvature of the surface, a straightforward  derivation gives an expression in terms of quantities living on the ${x}$-plane:
\begin{equation}\label{reducedk}
    k = \bigl( - \dot{w}^j \pcovd_{\!j} ( s^{-2} \dot{w}_i ) + \partial_i R^{D-2}\bigr)\, n^i\;.
\end{equation}
Here ${\dot{\tens{w}}}$ is a vector tangent to the profile curve ${w(\alpha)}$, ${s^2=\dot{w}^i\dot{w}^j p_{ij}}$, and ${\tens{n}}$ is a unit normal of the profile curve in the $x$-plane (${n_i \dot{w}^i = 0}$, ${n^i n^j p_{ij}=1}$). The covariant derivative ${\pcovd}$ (a smaller nabla) is the metric derivative associated with the 2-dimensional metric ${\tens{p}}$, living just in the 2-dimensional \mbox{${x}$-plane}.

The task of finding a minimal surface thus leads to the second order equation ${k=0}$ for the curve ${w(\alpha)}$ in the 2-dimensional ${x}$-plane.

\subsection{Additional symmetry}

Now we restrict the geometry even more. We assume an additional Killing symmetry in the ${x}$-plane and a diagonal form of the metric ${\tens{p}}$,
\begin{equation}\label{addsym}
\tens{p} = h_{(1)}^2(x^2) (\grad x^1)^2 + h_{(2)}^2(x^2) (\grad x^2)^2\;,\qquad R=R(x^2)\;.
\end{equation}
${x^1}$ is thus the Killing coordinate.

We may fix the parametrization of the profile curve, namely we use ${x^2}$ coordinate as the parameter
\begin{equation}\label{gaugefix}
    w^2(\alpha) = \alpha\;.
\end{equation}
We are thus looking just for the coordinate ${w^1(\alpha)}$.

Substituting these assumptions into \eqref{reducedk}, the condition for a minimal surface \eqref{zerotrk} becomes
\begin{equation}\label{EoMS}
    \ddot{w}^1 +
    (\dot{w}^1)^3 \frac{h^2_{(1)}}{h_{(2)}^2} \frac{(R^{D{-}2}h_{(1)})\dot{}}{R^{D{-}2}h_{(1)}} +
    \dot{w}^1 \frac{(R^{D{-}2} h_{(1)}^2 h_{(2)}^{-1})\dot{}}{R^{D{-}2} h_{(1)}^2 h_{(2)}^{-1}}
    = 0\;.
\end{equation}
Here, ${h_{(j)}(\alpha)}$ and ${R(\alpha)}$ depend only on the parameter ${\alpha}$ (the coordinate ${w^2}$), and the dot denotes the derivative with respect of ${\alpha}$.

Thanks to the additional Killing symmetry this equation does not contain ${w^1}$, just its derivatives. It is thus the first order differential equation for ${\dot{w}^1}$ which, actually, can be integrated:
\begin{equation}\label{firstintw}
    \dot{w}^1 = \frac{h_{(2)}}{h_{(1)}} \frac{c}{\sqrt{R^{D{-}2} h_{(1)}^2-c^2}}\;.
\end{equation}
Here, ${c}$ is an integration constant.

Integrating this expression once more, we get the profile curve ${w(\alpha)}$ for the minimal surface. Before doing it in explicit examples, we derive a general expression for the area of the minimal surface.

The metric ${\tens{h}}$ induced on the aligned surface ${\Sigma}$ is
\begin{equation}\label{inducedmtr}
    \tens{h} = s^2 \grad\alpha\grad\alpha+R^2 q_{\mu\nu}\grad y^\mu \grad y^\nu\;.
\end{equation}
The corresponding volume element ${\mathfrak{h}^{1/2}}$ is
\begin{equation}\label{inducedvol}
    \mathfrak{h}^{1/2} = s R^{D{-}2} d\alpha\,\mathfrak{q}^{1/2}\;,
\end{equation}
where ${\mathfrak{q}^{1/2}}$ is the volume element on ${y}$-space given by the metric ${\tens{q}}$. Taking into account \eqref{firstintw}, the area of the surface ${\Sigma}$ becomes
\begin{equation}\label{areams}
    A =\int \mathfrak{h}^{1/2} = \mathcal{A} \int s R^{D{-}2} d\alpha\\
      = \mathcal{A} \int \frac{h_{(1)}h_{(2)}R^{2(D{-}2)}}{\sqrt{R^{2(D{-}2)} h_{(1)}^2-c^2}}d\alpha\;.
\end{equation}
Here, ${\mathcal{A}=\int{\mathfrak{q}^{1/2}}}$ is the volume of the ${y}$-space (for example, ${\mathcal{A}=2\pi}$ in the 3-dimensional example discussed above), and the integral in \eqref{areams} must be taken in appropriate limits.


\section{Lobachevsky space -- spatial section of the anti-de~Sitter spacetime}
\label{sc:Lobsp}

\subsection{Static Killing vectors in AdS}

Our aim is to study minimal surfaces in static regions of the 4-dimensional AdS spacetime. AdS is maximally symmetric space with a constant curvature which defines a length scale~${\ell}$. Since we are interested in global view of the AdS spacetime, we specify the metric in global cosmological coordinates ${\tau,\,r,\,\tht,\,\ph}$:
\begin{equation}\label{AdSmtrc}
    \tens{g}_{\AdS} = \ell^2\,\bigl(-\ch^2\!r\,\grad\tau^2+\grad r^2+
    \sh^2\!r\,(\grad\tht^2+\sin^2\!\tht\,\grad\ph^2)\bigr)\;.
\end{equation}

It is useful to visualize the AdS spacetime as a tube ${\realn\times B^3}$. The
vertical direction ${\realn}$ corresponds to time and the horizontal ball ${B^3}$
represents a spatial section with its hyperbolic geometry compactified to unit ball
\cite{Hawking:1973uf}. More details on its geometry and various coordinates can be
found Appendix~\ref{apx:AdS}.

At this moment  it is sufficient to  mention that AdS possesses three qualitatively different Killing vectors which have a static region. Orbits of such Killing vectors are worldlines of uniformly accelerated observers.

\begin{figure}
\begin{center}
\hspace*{0.1in}%
\includegraphics[width=1.25in]{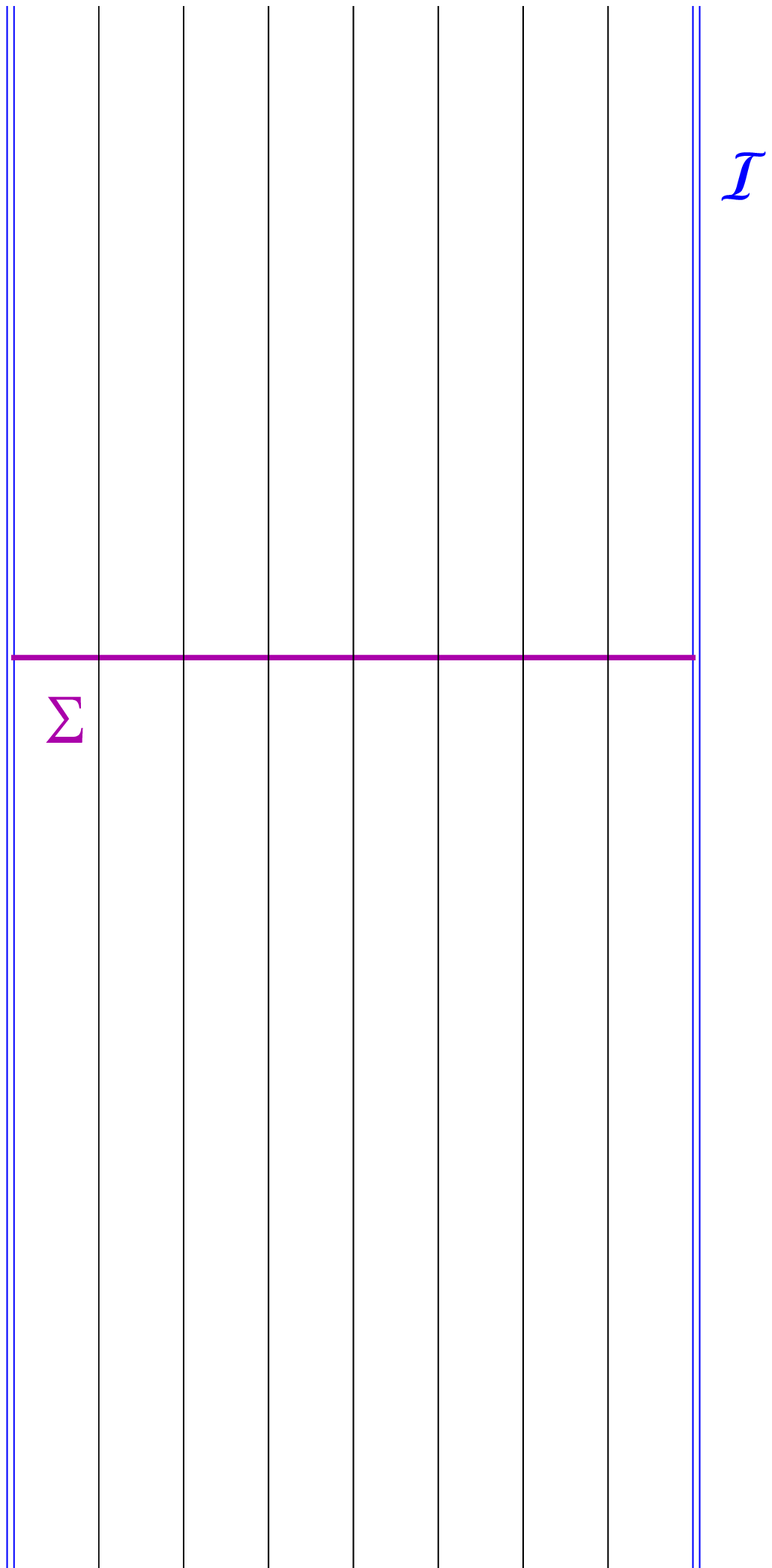}\hspace*{0.7in}%
\includegraphics[width=1.25in]{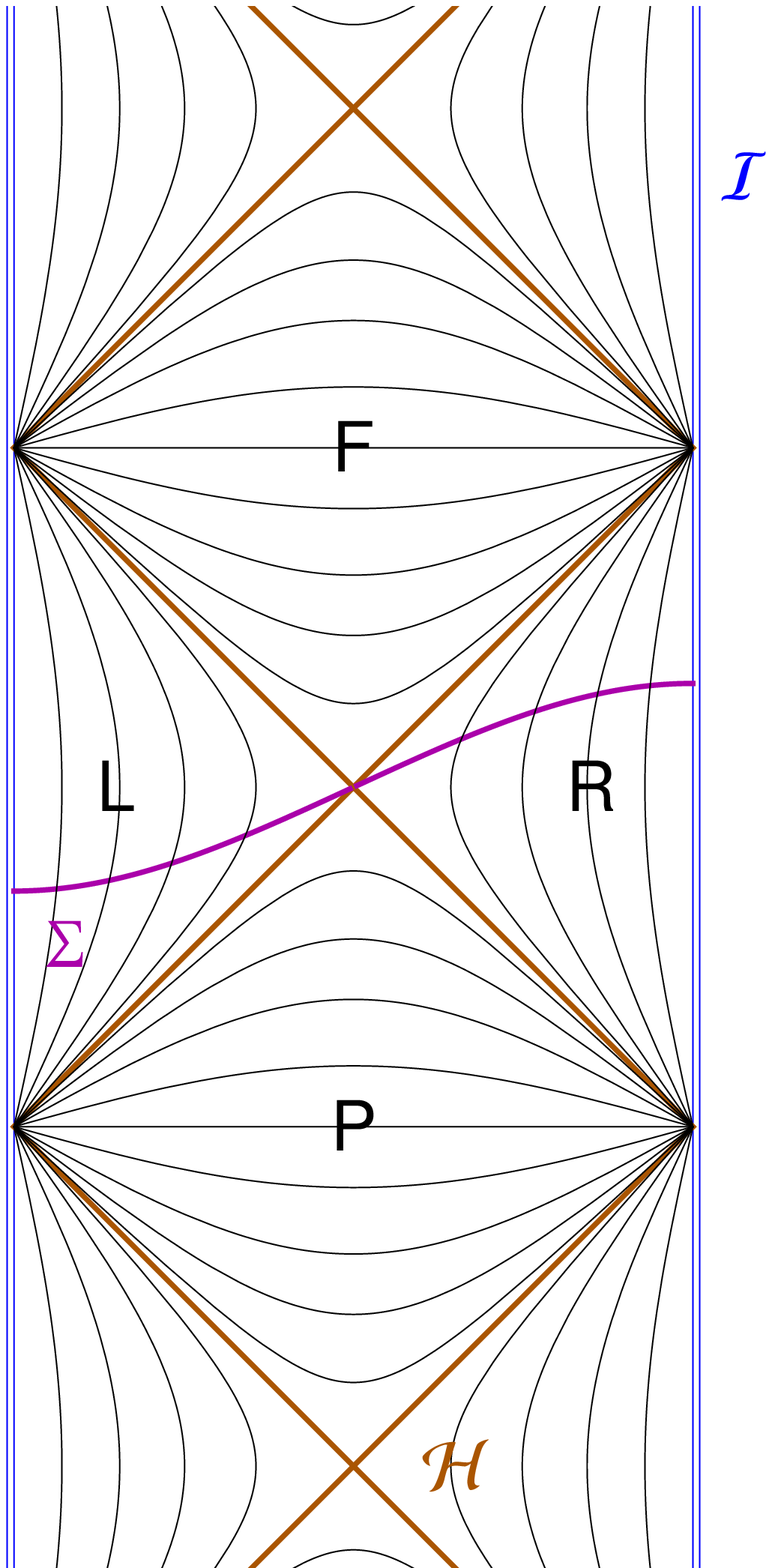}\hspace*{0.7in}%
\includegraphics[width=1.25in]{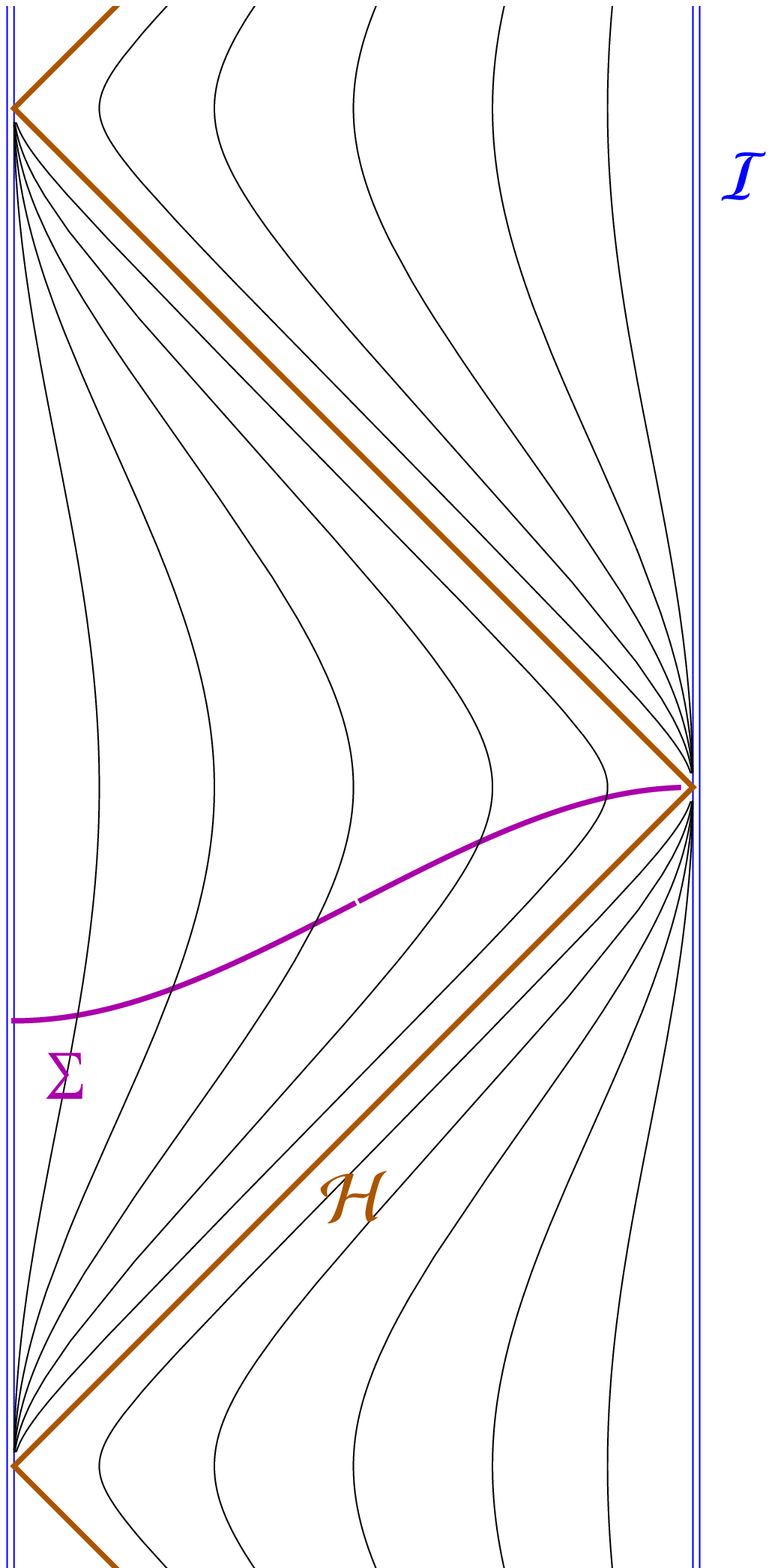}%
\\[1ex]
{\small(a)\hspace{128pt}(b)\hspace{128pt}(c)}%
\end{center}
\caption{\label{fig:KVAdS}\small
{\bf Killing vectors in AdS.}\quad
AdS spacetime can be visualized as a tube ${\realn\times B^3}$ with time in vertical direction. Spatial directions are appropriately compactified and the surface ${\realn\times S^3}$ of the tube corresponds to infinity ${\scri}$ of AdS. 2-dimensional diagrams here represent just sections ${\tht=0}$, ${\ph=0,\pi}$.
The diagrams show orbits of three qualitatively different Killing vectors which are at least somewhere timelike. Surfaces $\Sigma$ orthogonal to the orbits of the Killing vectors, representing one instant of static times, are also shown.\quad
(a)~Static Killing vector of type I. It is a globally smooth Killing vector. Its orbits have a uniform acceleration smaller than ${1/\ell}$. \quad
(b)~Static Killing vector of type II. It has a bifurcation structure repeating in temporal direction. The spacetime is divided into various domains separated by Killing horizons $\mathcal{H}$. The Killing vector is timelike only in regions \textsf{R} and \textsf{L}, it is spacelike in the domains \textsf{P} and \textsf{F}. The orbits of this vector have a uniform acceleration bigger than ${1/\ell}$. \quad
(c)~Poincar\'{e} static Killing vector. It is timelike everywhere except the
horizons where it is null. The acceleration of the orbits is exactly ${1/\ell}$.
The horizons $\mathcal{H}$ divide spacetime into separate patches covered by
Poincar\'{e} coordinates.
}
\end{figure}

Let us denote the Killing vector with the orbit acceleration smaller than ${1/\ell}$ the static Killing vector of type I. It is globally smooth vector field which is timelike in the whole AdS (see Fig.~\ref{fig:KVAdS}a). The standard prototype of such Killing vector is the time coordinate vector ${\cv{\tau}}$ in the cosmological coordinates introduced above.

The Killing vector with the orbit acceleration larger than ${1/\ell}$ will be called the static Killing vector of type II. It is not globally smooth, it has a bifurcation character and it resembles (in the bifurcation area) the boost Killing vector of the Minkowski spacetime (cf.~Fig.~\ref{fig:KVAdS}b). In the aligned static coordinates \eqref{tauchiTRrel}, it is given by ${\cv{T}}$.

Finally, the Killing vector with the orbit acceleration exactly ${1/\ell}$ will
be called the Poincar\'{e} Killing vector since the associated coordinate system
is formed by the well-known Poincar\'{e} coordinates \eqref{tauchitrrel}. In
these coordinates it is given by ${\cv{\tP}}$. Its orbits are shown in
Fig.~\ref{fig:KVAdS}c.

In all these three cases spatial sections orthogonal to the Killing vectors (namely ${\tau=\text{const}}$, ${T=\text{const}}$ or ${\tP=\text{const}}$) have the geometry of a maximally symmetric 3-dimen\-sional space of a constant negative curvature, i.e., of  the Lobachevsky space (also, the hyperbolic space). Now, we will give a short description of its geometry; for further details see Appendix~\ref{apx:Lobsp}.

\subsection{Lobachevsky space and its symmetries}

The geometry of the hyperbolic space can be given by the metric in spherical coordinates:
\begin{equation}\label{Lobmtrcsph}
    \frac1{\ell^2}\,
    \tens{g}_{\Lob} = \grad r^2 + \sh^2\!r\, \bigl( \grad\tht^2+\sin^2\!\tht\,\grad\ph^2\bigr)\;.
\end{equation}
${r}$ is the radial distance from the origin. We can introduce also a rescaled coordinate ${\chi}$ given by ${\sh r = \tan\chi}$. Using this coordinate the metric takes a form conformal to the metric on hemisphere, cf. \eqref{LobMconfsph}.

The symmetry group of the 3-dimensional Lobachevsky space is SO(3,1). All isometries can be generated by three rotations and three translations. Orbits of the rotations are circles around the axis of rotation, the orbits of the translations are exocycles -- curves equidistant from the axis of the translation.

The translation and the rotation with a common axis commute. Therefore, it is possible to find coordinates adjusted to both these symmetries, which we naturally call the cylindrical coordinates. The metric in the cylindrical coordinates reads:
\begin{equation}\label{Lobmtrccyl}
    \frac1{\ell^2}\,
    \tens{g}_{\Lob} = \grad \rho^2 + \ch^2\!\rho\,\grad\zeta^2+\sh^2\!\rho\,\grad\ph^2\;.
\end{equation}
${\rho}$ is a distance from the axis, ${\zeta}$ is a coordinate running in the direction of the translation, and ${\ph}$ in the direction of the rotation. We will use also an axial coordinate ${P=\sh\rho\in(0,\infty)}$ measured by the circumference of a circle around the axis, the metric is then given by
\begin{equation}\label{LobmtrccylP}
    \frac1{\ell^2}\,
    \tens{g}_{\Lob} = \frac1{1+P^2}\,\grad P^2 + (1+P^2)\,\grad\zeta^2+P^2\,\grad\ph^2\;.
\end{equation}
Yet another axial coordinate ${Z=\ch\rho\in(1,\infty)}$ leads to the metric
\begin{equation}\label{LobmtrccylZ}
    \frac1{\ell^2}\,
    \tens{g}_{\Lob} = \frac1{Z^2-1}\,\grad Z^2 + Z^2\,\grad\zeta^2+(Z^2-1)\,\grad\ph^2\;.
\end{equation}
For relation between the spherical and cylindrical coordinates see Appendix~\ref{apx:Lobsp}, eqs.~\eqref{chithtbarrel}, \eqref{barchizetarel}, \eqref{axialcoorrel}.

Beside translations and rotations, there exists also a special isometry type: the horocyclic symmetry.\footnote{%
Although this symmetry is not either a rotation or a translation, its generator can be obtained as a linear combination of generators of a rotation and a translation.}
Its orbits are horocycles with a common center at infinity (vaguely said, the
horocycles are circles with the center moved just to infinity; all orbits have
exactly one common improper point at infinity coinciding with their common
center). All horocyclic symmetries with the same center at infinity commute with
each other. One can thus find coordinates adjusted to two horocyclic symmetries:
Poincar\'{e} coordinates ${\xP,\,\yP,\,\zP}$, cf.\ eq.~\eqref{zetachirbar},
\eqref{xyzrPrel}. The metric in terms of these coordinates reads
\begin{equation}\label{LobmtrcPoinc}
    \tens{g}_{\Lob} = \frac{\ell^2}{\zP^2}\,\bigl(\grad \xP^2 +\grad \yP^2+\grad\zP^2\bigr)\;.
\end{equation}
${\zP}$ labels various horospheres, coordinates ${\xP}$ and ${\yP}$ define two commuting horocyclic symmetries.

All three types of isometries can be used to understand Lobachevsky space as the warped geometry. Moreover, the complementary commuting symmetry can be understood as the additional  symmetry in a sense of \eqref{addsym} and it allows us to use the results \eqref{firstintw} and \eqref{areams} derived above.

\subsection{Various representations of Lobachevsky space}

\begin{figure}[t]
\begin{center}
\includegraphics[width=2in]{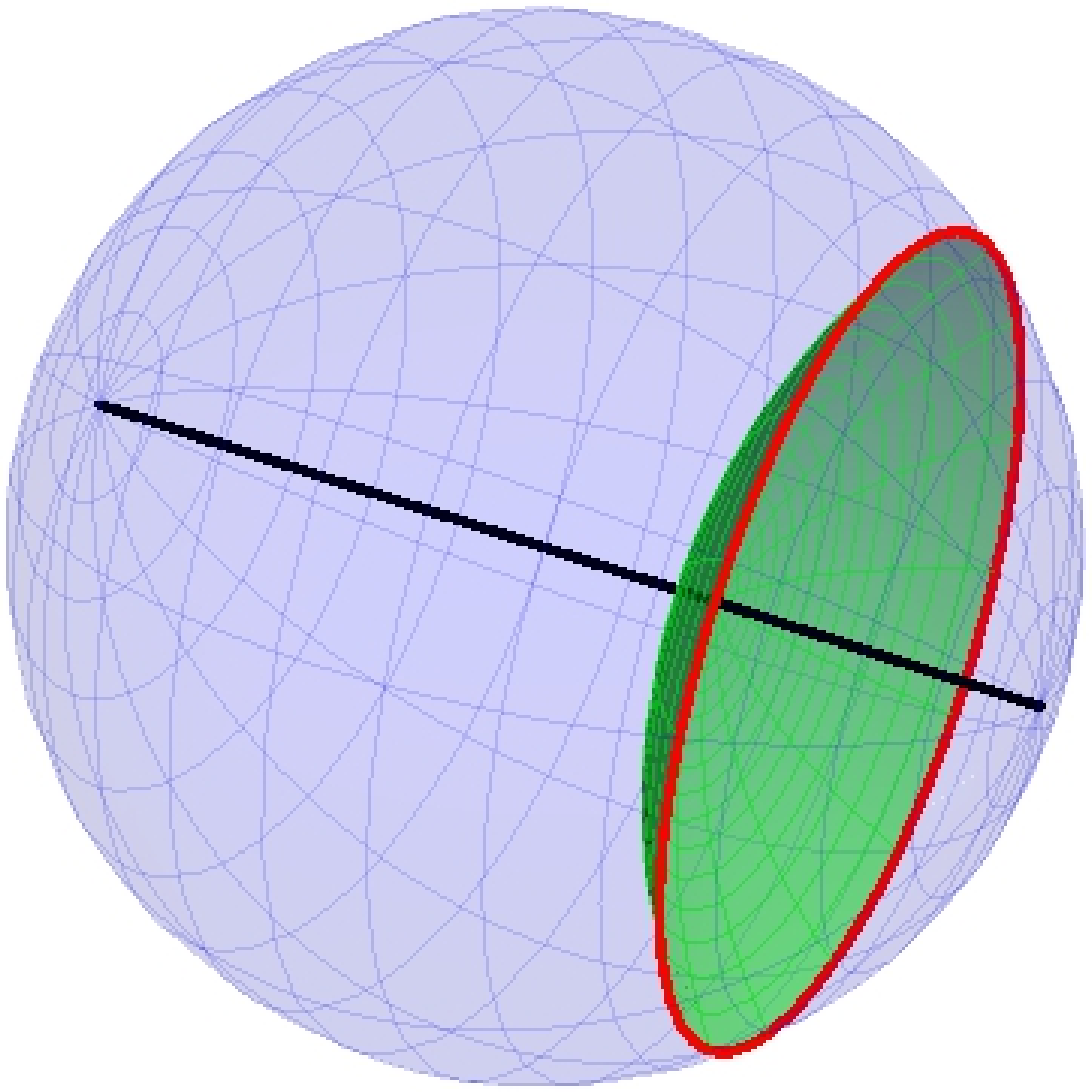}\hspace*{-2pt}%
\includegraphics[width=2in]{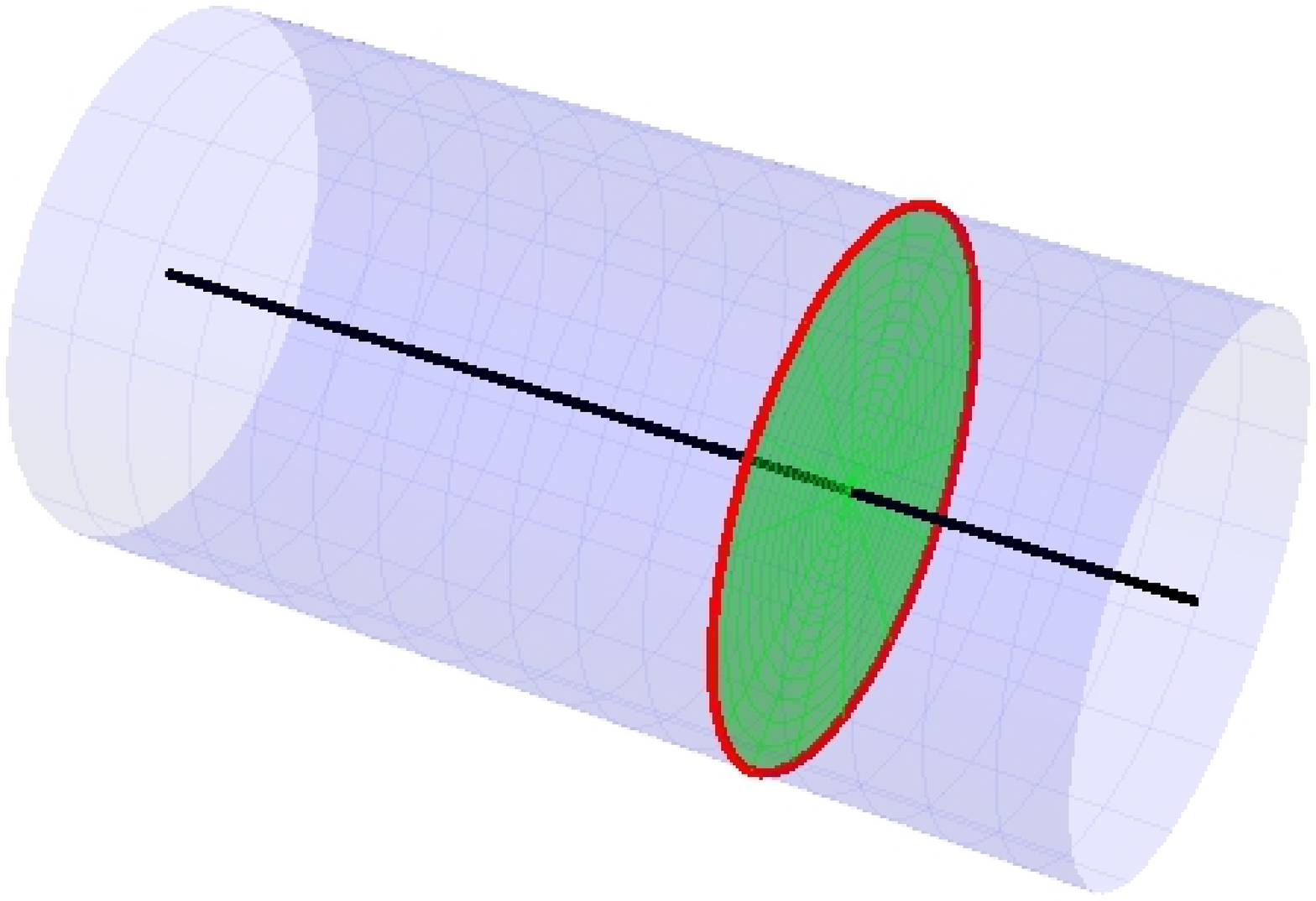}\hspace*{-2pt}%
\includegraphics[width=2in]{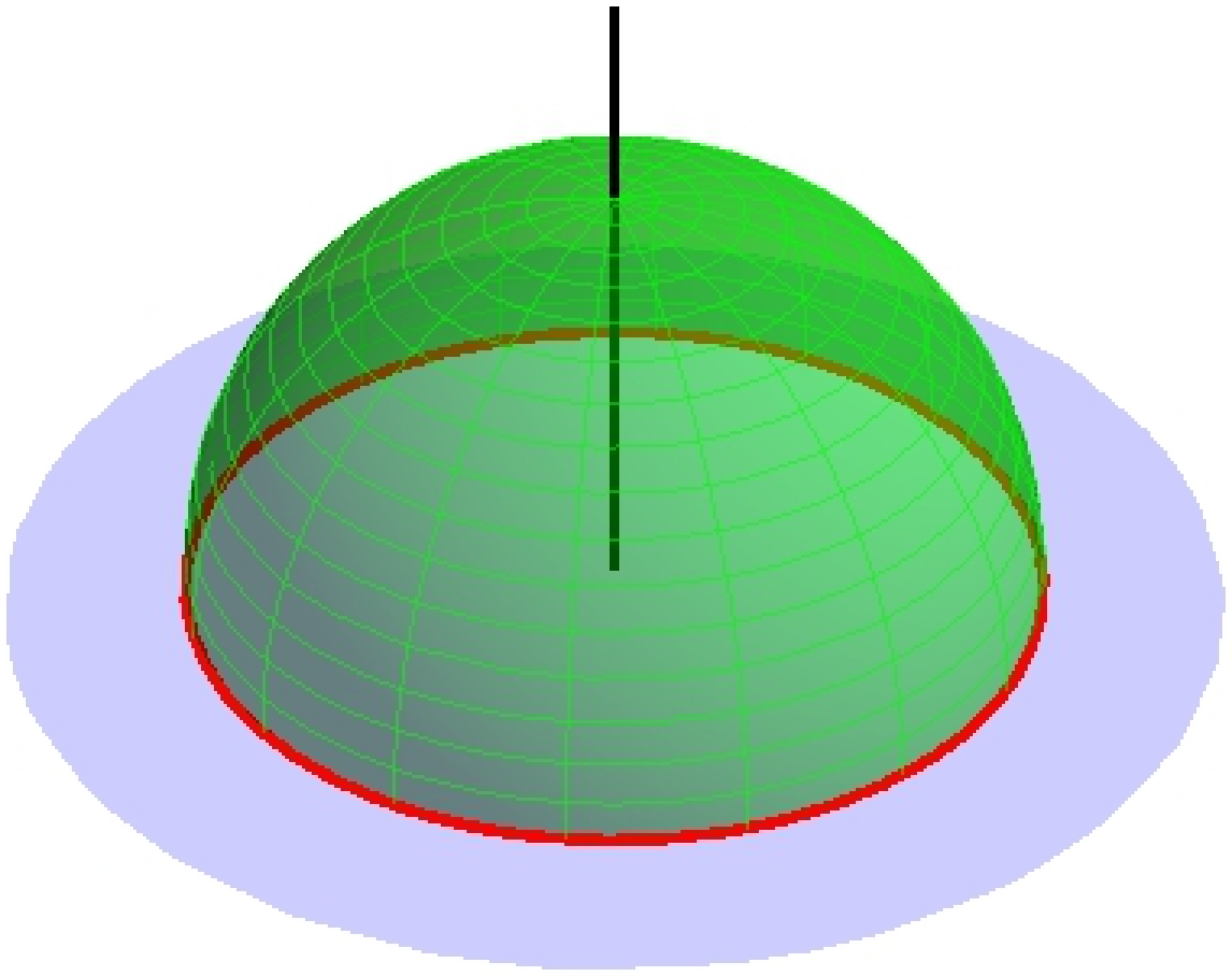}%
\\[1ex]
{\small(a)\hspace{130pt}(b)\hspace{130pt}(c)}%
\end{center}
\caption{\label{fig:Lobrepr}\small
{\bf Visualization of Lobachevsky space.}\quad
Lobachevsky space can be represented in Euclidean space in various ways,
emphasizing different symmetries of the hyperbolic geometry.\quad Diagram (a)
shows so called Poincar\'{e} spherical model in which the spherical symmetry is
emphasized. Whole Lobachevsky space is compactified into a unit ball with its
spherical boundary corresponding to infinity of the hyperbolic space. Geodesic
are represented as arcs orthogonal to infinity and hyperbolic planes as
spherical surfaces orthogonal to infinity. Planes reach infinity in circular
boundaries.\quad Diagram (b) emphasizes cylindrical symmetry of the hyperbolic
geometry. Whole Lobachevsky space is squeezed into cylinder. The infinity
corresponds to the surface of the cylinder and two improper points in both
directions along the axis. The lines parallel to the axis represent exocylcles
-- curves equidistant from the axis. Hyperbolic planes orthogonal to the axis
(and exocycles) are represented by flat discs. They reach infinity in boundaries
represented by a circle around the cylinder.\quad Diagram (c) is half-space
Poincar\'{e} model in which Lobachevsky space is mapped onto the half ${\zP>0}$
of
Euclidean space. The plane ${\zP=0}$  (together with one more improper point)
depicts infinity of Lobachevsky space. Shifts parallel to this plane (in
Euclidean sense) represent horocyclic symmetries of the hyperbolic geometry.
Geodesics are semicircles and hyperbolic planes hemispheres, both orthogonal to
infinity. Hyperbolic planes reach infinity again in circular boundaries.
}
\end{figure}

Before we proceed in looking for minimal surfaces, we describe how we will visualize hyperbolic space.

Hyperbolic space is spherically symmetric. It is demonstrated explicitly in
terms of spherical coordinates ${r,\,\tht,\,\ph}$. The spherical symmetry
suggests that we can  (non-isometrically) map whole Lobachevsky space into a
unit ball in Euclidean space by just identifying ${\tht,\,\ph}$ with the
standard Euclidean spherical angles and choosing a suitable compactifying
function for the radial coordinate. We will use so called Poincar\'{e}
spherical
model which is given by the compactifying function ${\tnh\frac{r}{2}}$, see
Fig.~\ref{fig:Lobrepr}a. The surface of the unit ball corresponds to infinity of
the hyperbolic space.

Another natural representation emphasizes the cylindrical symmetry. We can map
whole Lobachevsky space into interior of the cylinder identifying coordinates
${\zeta,\,\ph}$ with the standard Euclidean cylindrical coordinates and
employing suitable compactifying function of the coordinate ${\rho}$, namely
${\tnh\frac\rho2}$, in the direction from the axis, cf.~Fig.~\ref{fig:Lobrepr}b.
The surface of the cylinder again corresponds to infinity of the hyperbolic
space.

Finally, the fact that the metric in Poincar\'{e} coordinates
\eqref{LobmtrcPoinc}
has a conformally flat form suggests another representation, so called
Poincar\'{e}
half-space model. Identifying ${\xP,\, \yP,\,\zP}$ with the standard Cartesian
coordinates, it maps the Lobachevsky space onto half ${\zP>0}$ of the Euclidean
space, see Fig.~\ref{fig:Lobrepr}c. Infinity of the hyperbolic space corresponds
to the plane ${z=0}$.

\section{Minimal surfaces in Lobachevsky space}
\label{sc:msLobsp}

\subsection{Spherical/circular boundary at infinity}
\label{ssc:boundaries}

The entanglement entropy can be defined for an arbitrary domain at infinity of the hyperbolic space. However, we concentrate on special domains restricted just by simple spherical boundaries. (For 2-dimensional infinity of the ${D=3}$ bulk space these would be circular boundaries). By the spherical/circular boundary we mean a surface at infinity, which is obtained by projecting a hyperbolic plane in the bulk into infinity.

The infinity of the hyperbolic space has a structure of the sphere with a conformal geometry induced by the bulk geometry. For ${D=3}$, the conformal geometry of two-dimensional sphere is equivalent to the complex structure of the Riemann sphere. The holomorphic M\"{o}bius transformations preserve the notion of the circle, as can be also seen from their correspondence to isometries of the bulk.

The representation of the spherical/circular boundaries using hyperplanes in the bulk allows us to define the distance between two disjoint spherical/circular boundaries: it is the distance of the corresponding hyperplanes. Indeed, for not-crossing boundaries the hyperplanes are so called ultraparallel and there exists a common perpendicular line along which we measure the distance of both planes.

For two spherical/circular boundaries which intersect themselves, we can analogously define the angle between them as the angle of corresponding intersecting hyperplanes.

The last possibility is that the spherical/circular boundaries touch themselves
in one point. The corresponding hyperplanes are then asymptotic to each other.
In this case one cannot associate with these two hyperplanes any measure which
would estimate their relation. The reason is simple: all pairs of asymptotic
hyperplanes are isometric to each other. It means that any two touching
spherical/circular boundaries are equivalent and there is no scale which could
distinguish them.

The definition of spherical/circular boundary gives immediately also a solution of the minimal surface problem. The trivial minimal surface spanned on one spherical/circular boundary is just the hyperplane which defines the boundary.

Of course, we will be mainly interested in more complicated surfaces. Namely, in
surfaces spanned on two spherical/circular boundaries. However, the trivial planar
solution will be important for renormalization of the area of the minimal surface.
The area of the hyperplanes regularized in various ways will be given below. It can
be shown that in all cases it is proportional to regularized size of the boundary
\cite{Krtous:2013vha}.

\subsection{Surfaces with rotational symmetry}

As we mentioned in Sec.~\ref{sc:msLobsp}, the 3-dimensional Lobachevsky space can be viewed as a warped space in various ways. We start with the choice in which the symmetry ${y}$-space has the rotational $\ph$-symmetry and the additional symmetry of the ${x}$-plane is the translation $\zeta$-symmetry. For that, it is natural to employ the cylindrical coordinates with parametrization~\eqref{LobmtrccylP}.

\begin{figure}[t]
\begin{center}
\includegraphics[width=2in]{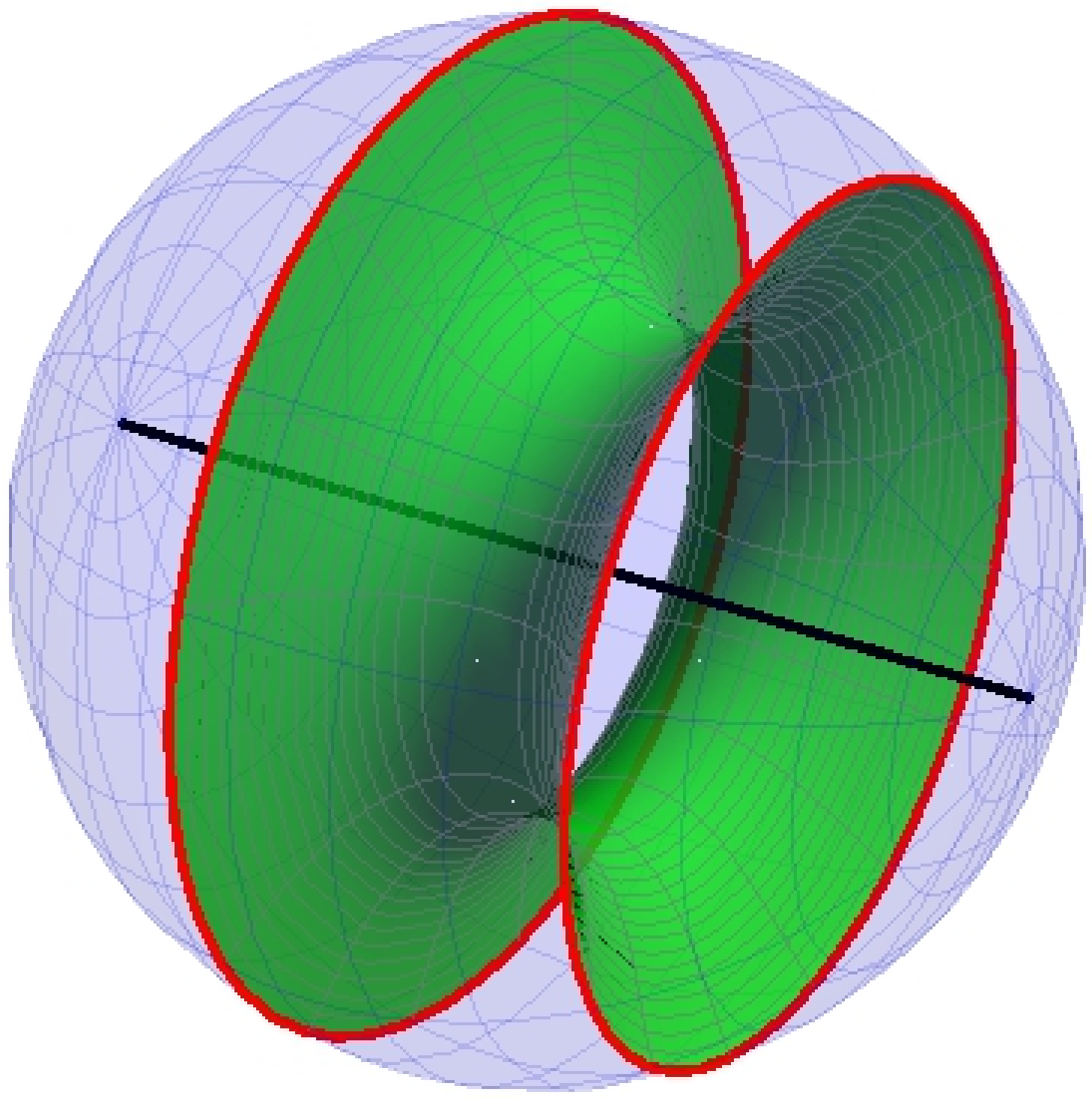}\hspace*{-2pt}%
\includegraphics[width=2in]{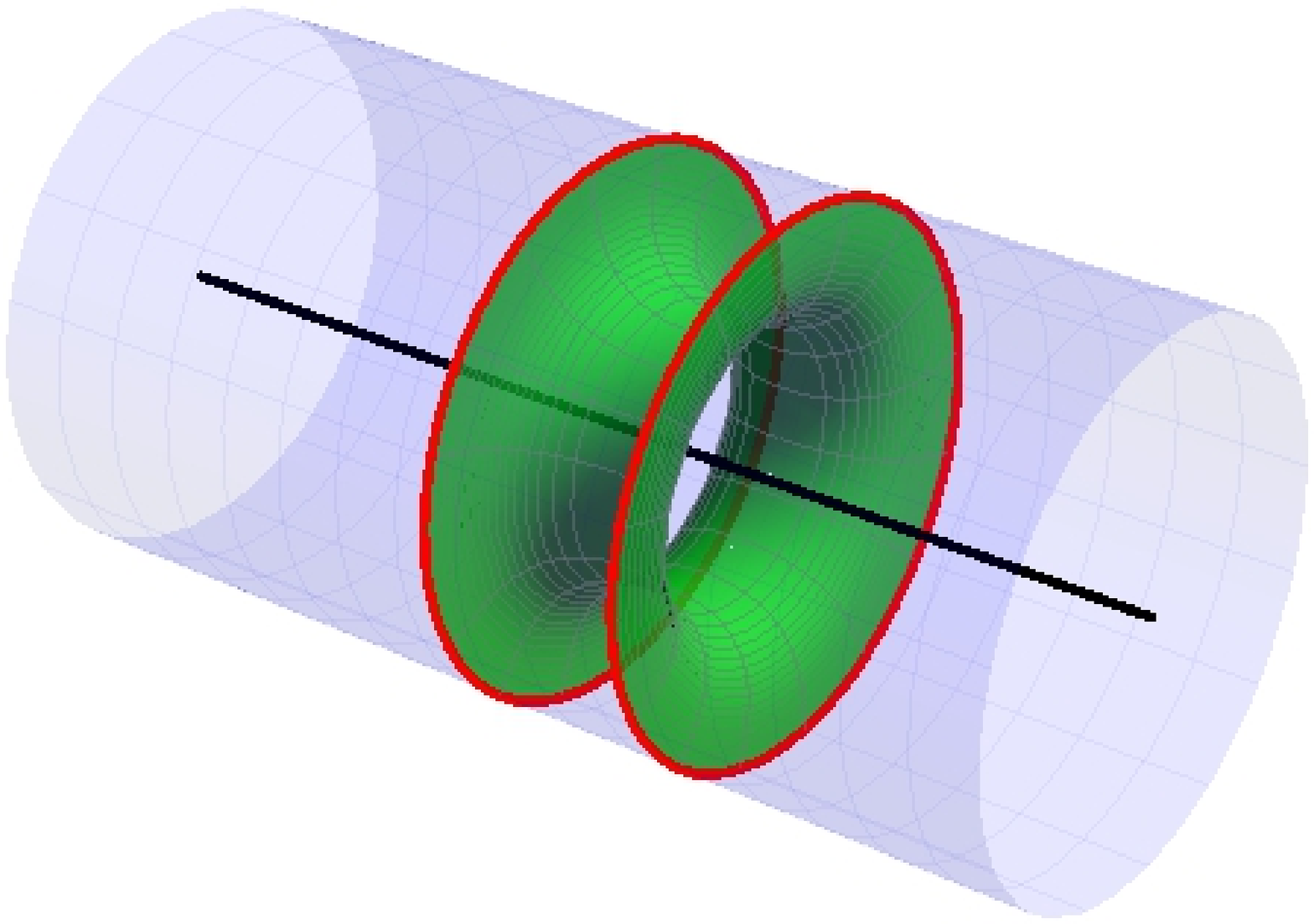}\hspace*{-2pt}%
\includegraphics[width=2in]{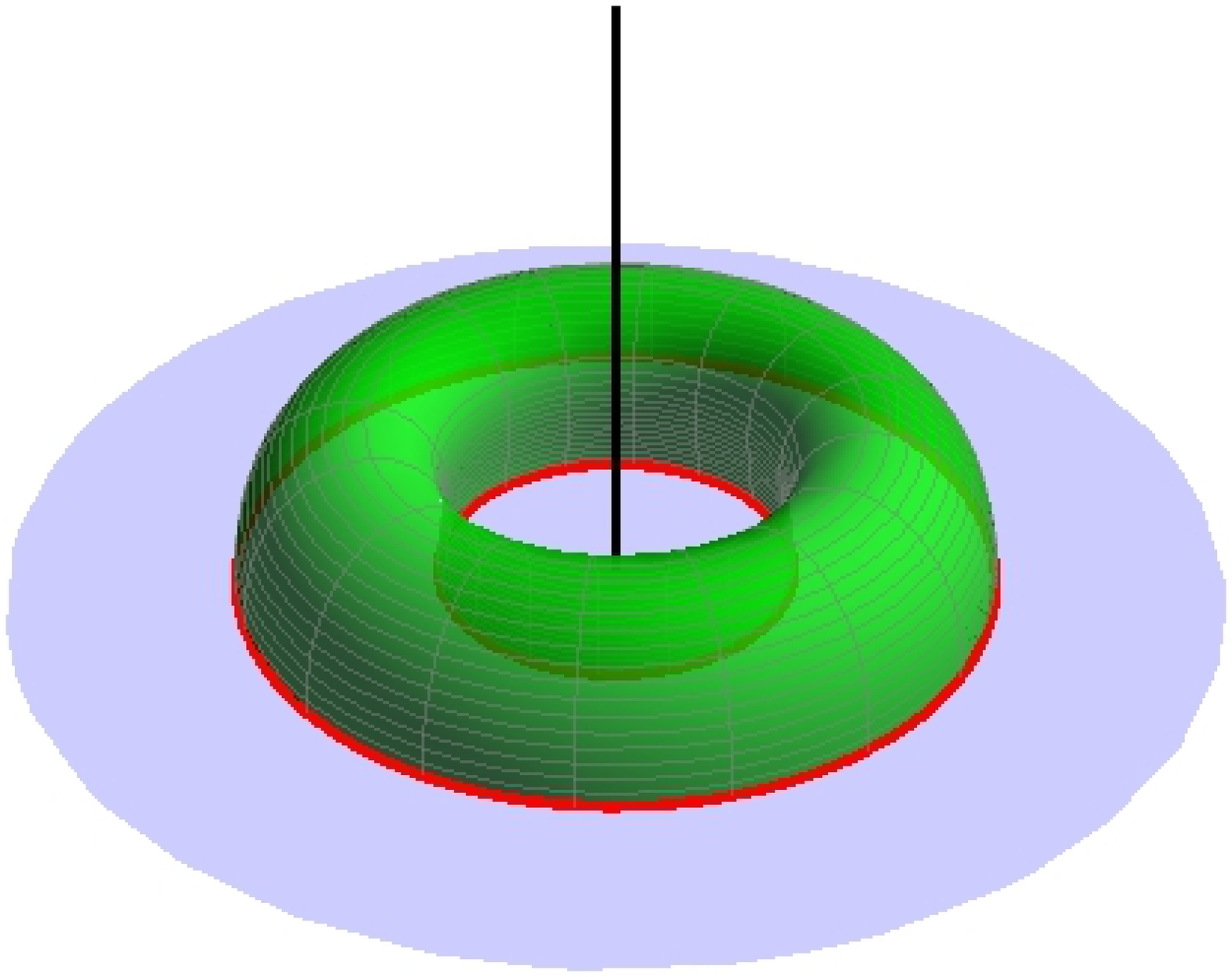}%
\\[1ex]
{\small(a)\hspace{130pt}(b)\hspace{130pt}(c)}%
\end{center}
\caption{\label{fig:msrot}\small
\textbf{Rotation-symmetric minimal surface spanned on two boundaries.}\quad
The surface is depicted using (a) spherical, (b) cylindric and (c) half-space visualization of Lobachevsky space (cf.~Fig.~\ref{fig:Lobrepr}). The cylindric visualization corresponds closely to the 2-dimensional diagram in Fig.~\ref{fig:prcrrot}.
}
\end{figure}

\begin{figure}[t]
\begin{center}
\includegraphics[width=3in]{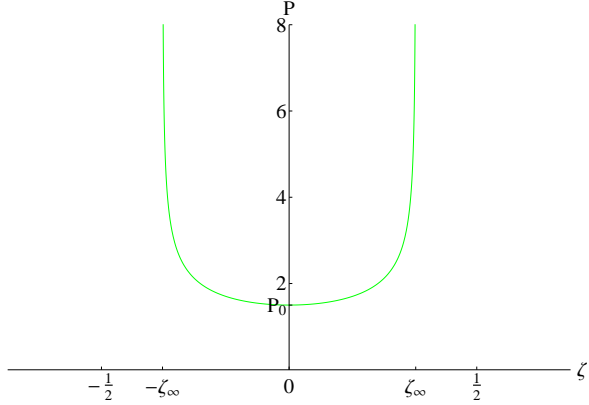}%
\end{center}
\caption{\label{fig:prcrrot}\small
\textbf{Profile curve for rotation-symmetric minimal surface.}\quad
The curve is drawn in the ${x}$-plane covered by coordinates ${\zeta,\, P}$. It is given by solution \eqref{rotsym-sol}. The corresponding minimal surface is shown in Fig.~\ref{fig:msrot}.
}
\end{figure}

To find a minimal surface, we substitute
\begin{equation}\label{rotsym-mtr}
\begin{gathered}
    x^1=\zeta\;,\qquad x^2=P\;,\qquad y^1=\ph\;\,\\
    h_{(1)} = \ell\sqrt{1+P^2}\;,\quad
    h_{(2)} = \frac\ell{\sqrt{1+P^2}}\;,\quad
    R=\ell\,P
\end{gathered}
\end{equation}
into equation \eqref{EoMS} for the profile curve. One obtains
\begin{equation}\label{rotsym-fi}
    \zeta'(P) = \pm\frac{P_0\sqrt{1+P_0^2}}{(1+P^2)\sqrt{P^4+P^2-P_0^4-P_0^2}}\;,
\end{equation}
where we conveniently redefined the integration constant. This equation can be
integrated in terms of elliptic integrals (cf.~3.157.5 of
\cite{GradshteinRyzhik:book}):
\begin{equation}\label{rotsym-sol}
\begin{split}
    \zeta(P) &= \zeta_0 \pm\frac{P_0}{\sqrt{1+P_0^2}\sqrt{1+2P_0^2}}\\
    &\mspace{20mu}
    \times\Bigl[
    (1+P_0^2)\,\mathsf{F}\Bigl({\textstyle\arccos\frac{P_0}{P},\sqrt{\frac{1+P_0^2}{1+2P_0^2}}}\Bigr)
    -P_0^2\, \mathsf{\Pi}\Bigl({\textstyle\arccos\frac{P_0}{P},\frac1{1+P_0^2},\sqrt{\frac{1+P_0^2}{1+2P_0^2}}}\Bigr)
    \Bigr]\;.
\end{split}
\end{equation}
The profile curve is thus parametrized by ${\zeta_0}$ and ${P_0}$. ${P}$ takes values in ${(P_0,\infty)}$. Two possible signs correspond to two symmetric parts of the same curve with a turning point at ${P=P_0}$, ${\zeta=\zeta_0}$.
Embedding of the corresponding rotation-symmetric minimal surface into the 3-dimensional Lobachevsky space is shown in Fig.~\ref{fig:msrot}. The graph of the profile curve itself is depicted in Fig.~\ref{fig:prcrrot}.

\begin{figure}[b]
\begin{center}

\includegraphics[width=3in]{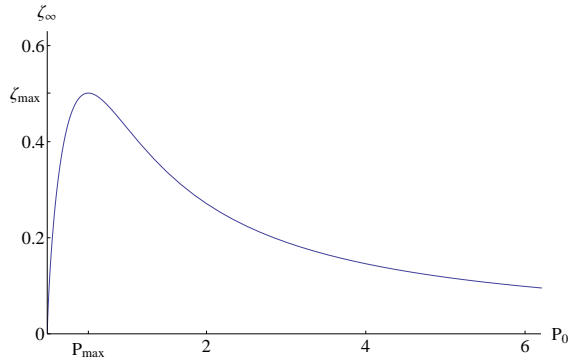}%
\end{center}
\caption{\label{fig:bnddist}\small
\textbf{Distance of the circular boundaries joined by the minimal surface}\quad
The minimal surface given by \eqref{rotsym-sol} reaches infinity in two circular boundaries which has distance ${s=2\ell\zeta_\infty}$. Diagram shows the dependence of ${\zeta_\infty}$ on the parameter ${P_0}$. For ${\zeta_\infty<\zeta_\mx}$, one has two values of ${P_0}$, i.e., two possible minimal surfaces joining such boundaries (see Fig.~\ref{fig:twoms}). For ${\zeta_\infty>\zeta_\mx}$, there is no minimal surfaces joining the boundaries.
}
\end{figure}

We see that the profile curve reaches infinity for two values of ${\zeta}$. It thus describes the minimal surface spanned on two circular boundaries.  Boundaries of the surface correspond to the hyperplanes given by ${\zeta=\zeta_0\pm\zeta_\infty}$, where
\begin{equation}\label{rotsym-bound}
    \zeta_\infty = \frac{P_0}{\sqrt{1+P_0^2}\sqrt{1+2P_0^2}}
    \Bigl[
    (1+P_0^2)\,\mathsf{K}\Bigl({\textstyle\sqrt{\frac{1+P_0^2}{1+2P_0^2}}}\Bigr)
    -P_0^2\, \mathsf{\Pi}\Bigl({\textstyle\frac1{1+P_0^2},\sqrt{\frac{1+P_0^2}{1+2P_0^2}}}\Bigr)
    \Bigr]\;.
\end{equation}

The graph  of the dependence of ${\zeta_\infty}$ on ${P_0}$ in Fig.~\ref{fig:bnddist} shows that there exists a maximal value of ${\zeta_\infty}$. It means that there exists a maximal distance of the circular boundaries for which these can be joined by a minimal surface. Numerically, this critical distance is ${s_\mx=2\ell\zeta_\infty\approx1.00229\,\ell}$, it is achieved for ${P_\mx\approx0.516334}$.

\begin{figure}
\begin{center}
\includegraphics[width=2.5in]{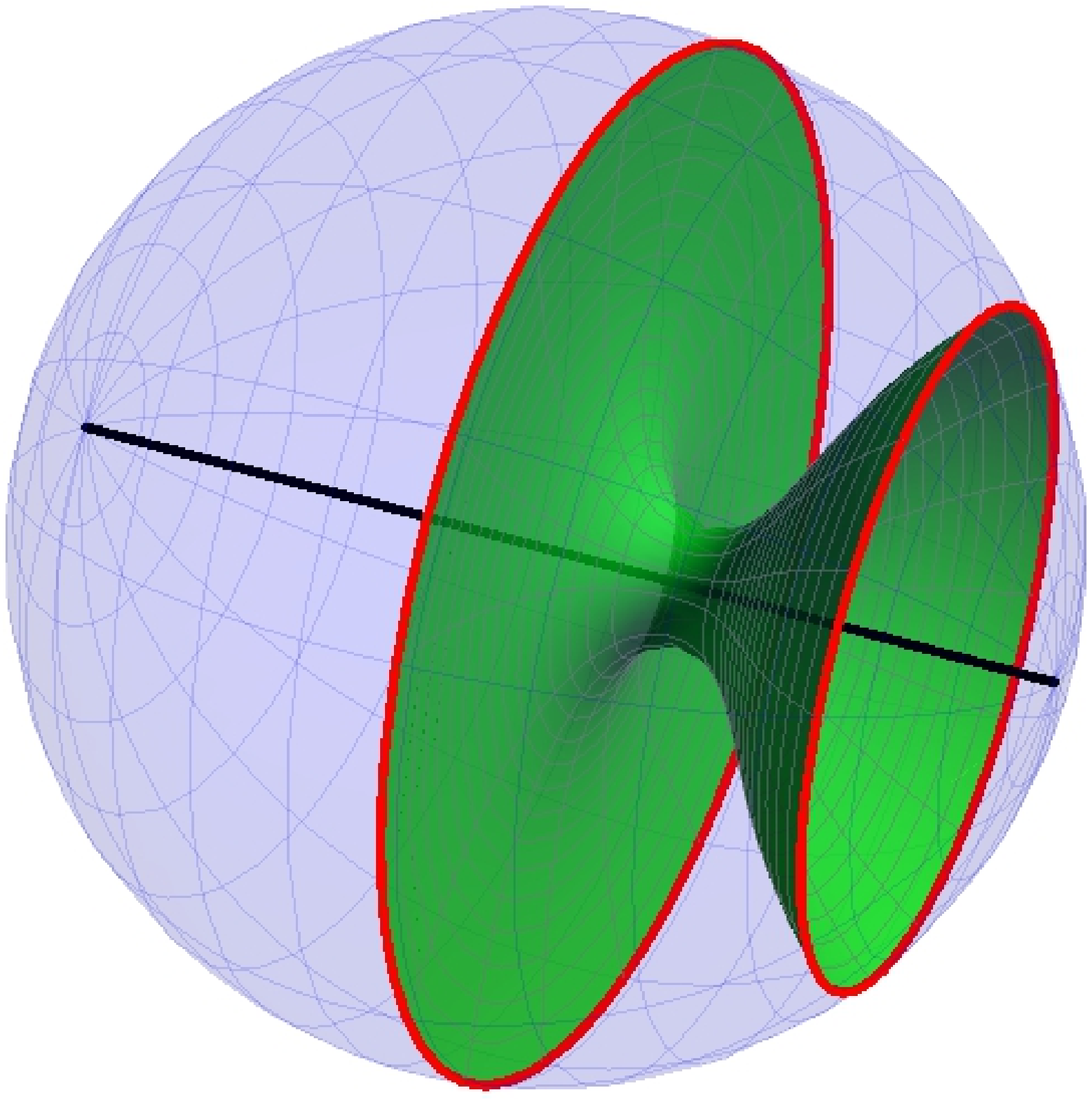}\qquad
\includegraphics[width=2.5in]{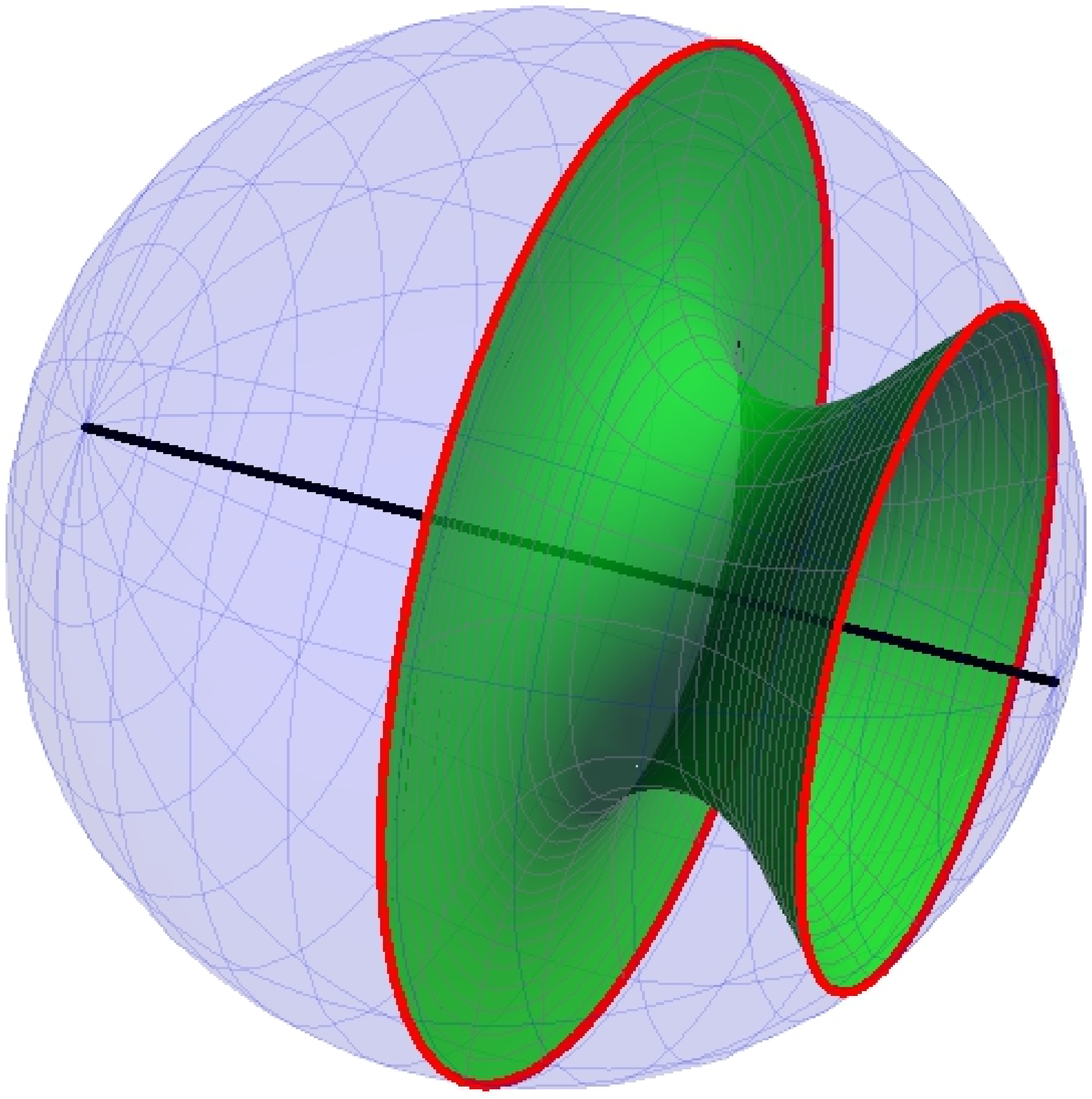}%
\end{center}
\caption{\label{fig:twoms}\small
\textbf{Two minimal surfaces spanned on the same boundaries.}\quad
Two circular boundaries with the mutual distance smaller than the distance ${s_{\mx}}$ can be joined by two minimal surfaces. One remains far from the axis, other reaches closer to the axis.
}
\end{figure}

The graph in Fig.~\ref{fig:bnddist} also reveals that for a given distance of two circular boundaries smaller than ${s_{\mx}}$ there exist two minimal surfaces spanned between them. One (that with larger value of ${P_0}$) is shallow, remaining further from the axis,  and other (with smaller ${P_0}$) is reaching closer to the axis, see Fig.~\ref{fig:twoms}. It indicates that the corresponding system at the AdS infinity can exist in two different non-trivial phases, both of them distinct from the trivial phase given by two hyperplanes.

Now we can proceed to evaluate the area of the minimal surface. Substituting
\eqref{rotsym-mtr} into \eqref{areams} (and taking into account both halves of the
surface given by \eqref{rotsym-sol}) we find that area up to radius ${P}$ is given by
\begin{equation}\label{rotsym-area}
  A(P)=4\pi\ell^2\int_{P_0}^P\frac{P^2}{\sqrt{P^4+P^2-P_0^4-P_0^2}}dP
      =\frac{4\pi\ell^2 P_0^2}{\sqrt{1+2P_0^2}}\,
      \mathsf{\Pi}\Bigl({\textstyle\arccos\frac{P_0}{P},1,\sqrt{\frac{1+P_0^2}{1+2P_0^2}}}\Bigr)
\end{equation}
(cf.~3.153.4 of \cite{GradshteinRyzhik:book} with 111.06 of \cite{ByrdFriedman:book}).

The area of the whole minimal surface ${A|_{P=\infty}}$ is diverging: the surface is reaching up to infinity. However, we can renormalize it by subtracting the area of the trivial solution spanned of the same boundaries, i.e., subtracting the area of two hyperplanes. The regularized area of one hyperplane (i.e., evaluated up to radius ${P}$) is
\begin{equation}\label{rotsym-planearea}
  A_{\plane}(P)=2\pi\ell^2\!\int_0^P\frac{P\,dP}{\sqrt{1+P^2}}=2\pi\ell^2(\sqrt{1+P^2}-1)\;.
\end{equation}
The expansion for large ${P}$ shows that the renormalized area of the surface \eqref{rotsym-sol} is finite:
\begin{equation}\label{rotsym-regarea}
    A_\ren = (A-2A_{\plane})|_{P\to\infty}
    =4\pi\ell^2\!\Bigl[ 1
    +{\textstyle\frac{P_0^2}{\sqrt{1{+}2P_0^2}}}\,
    \mathsf{K}\Bigl({\textstyle\sqrt{\frac{1{+}P_0^2}{1{+}2P_0^2}}}\Bigr)
    -\sqrt{1{+}2P_0^2}\,\mathsf{E}\Bigl({\textstyle\sqrt{\frac{1{+}P_0^2}{1{+}2P_0^2}}}\Bigr)
    \Bigr]\;.
\end{equation}

\begin{figure}
\begin{center}
\includegraphics[width=1.9in]{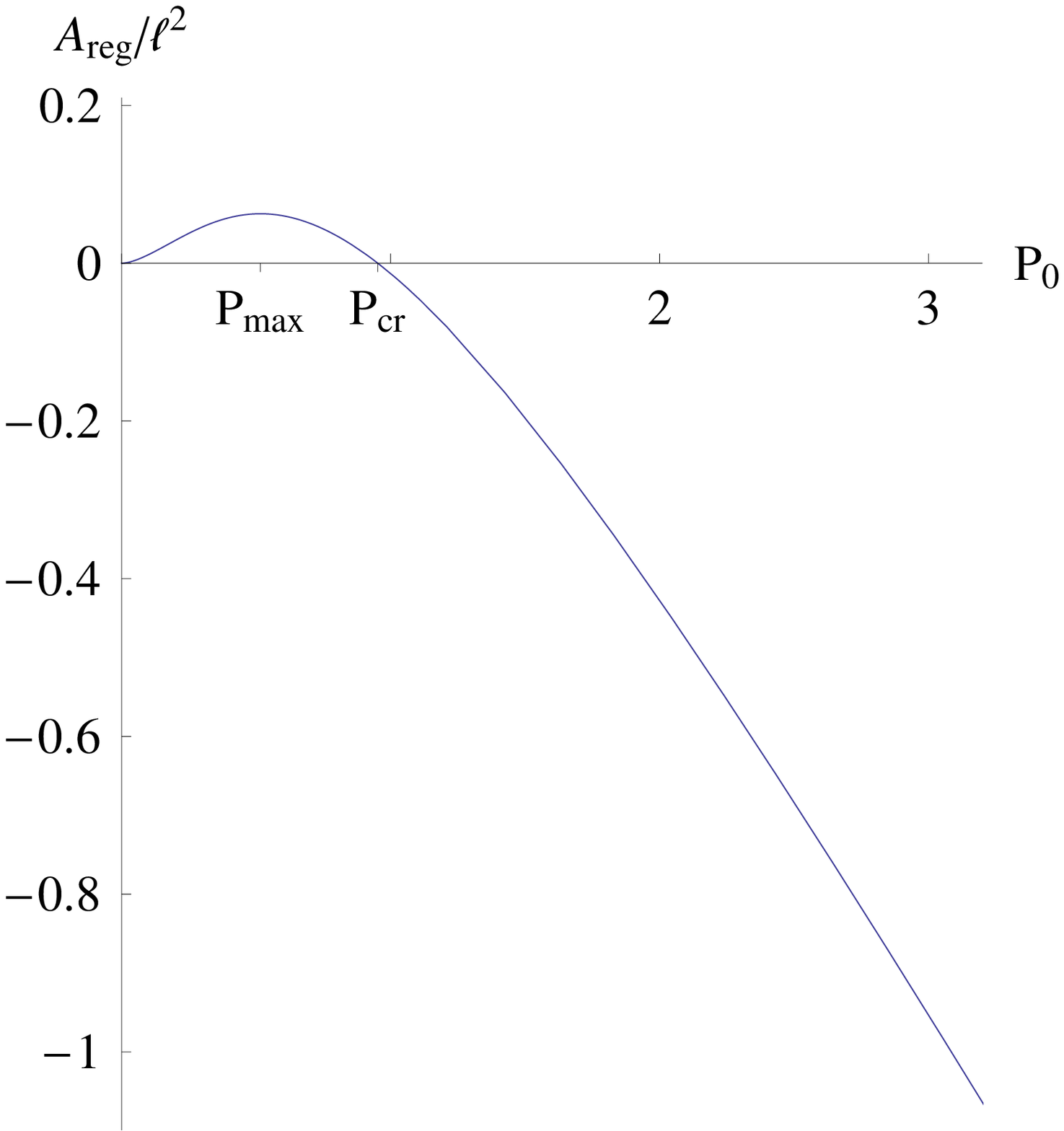}\hfill%
\includegraphics[width=1.9in]{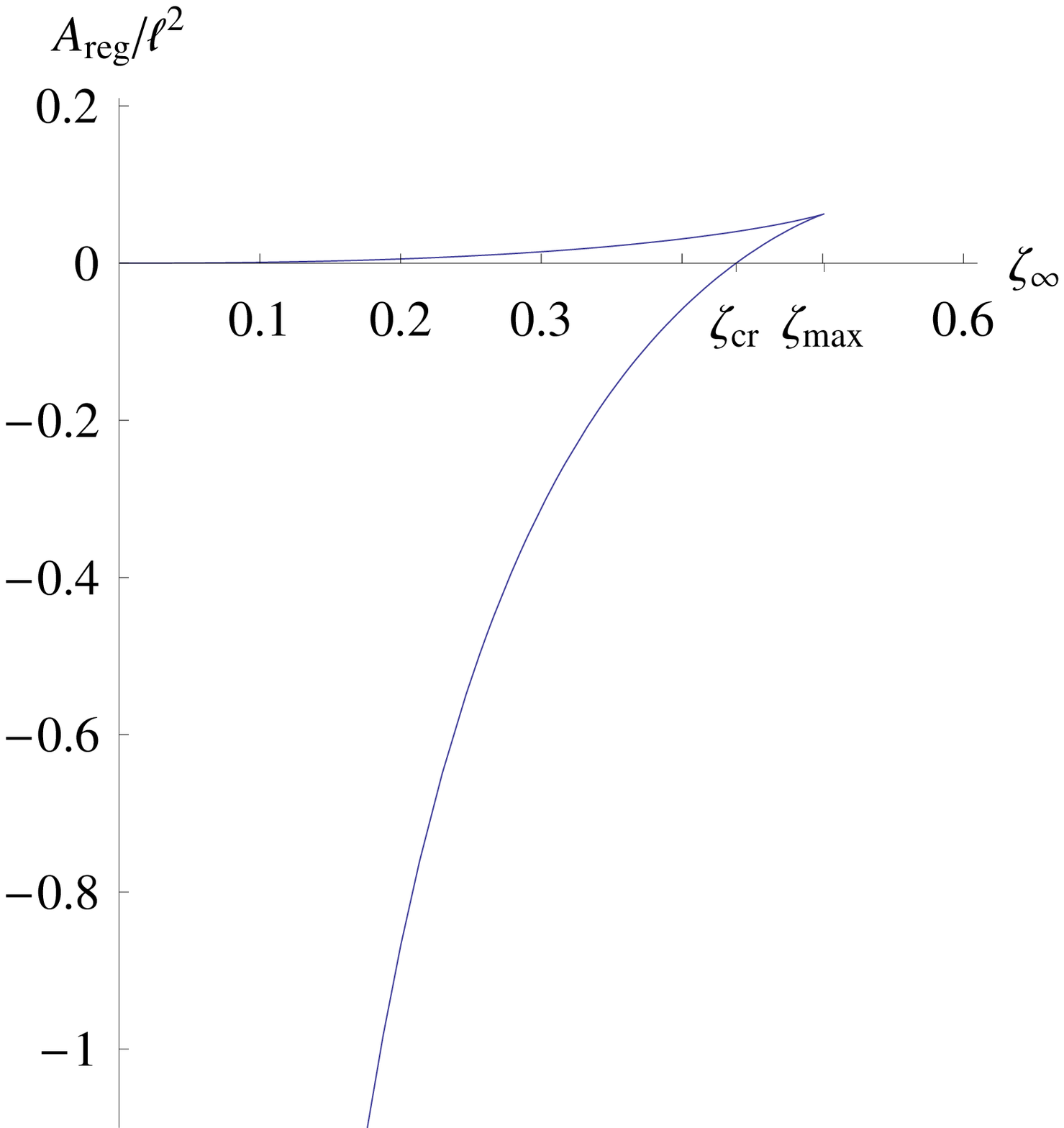}\hfill%
\includegraphics[width=1.9in]{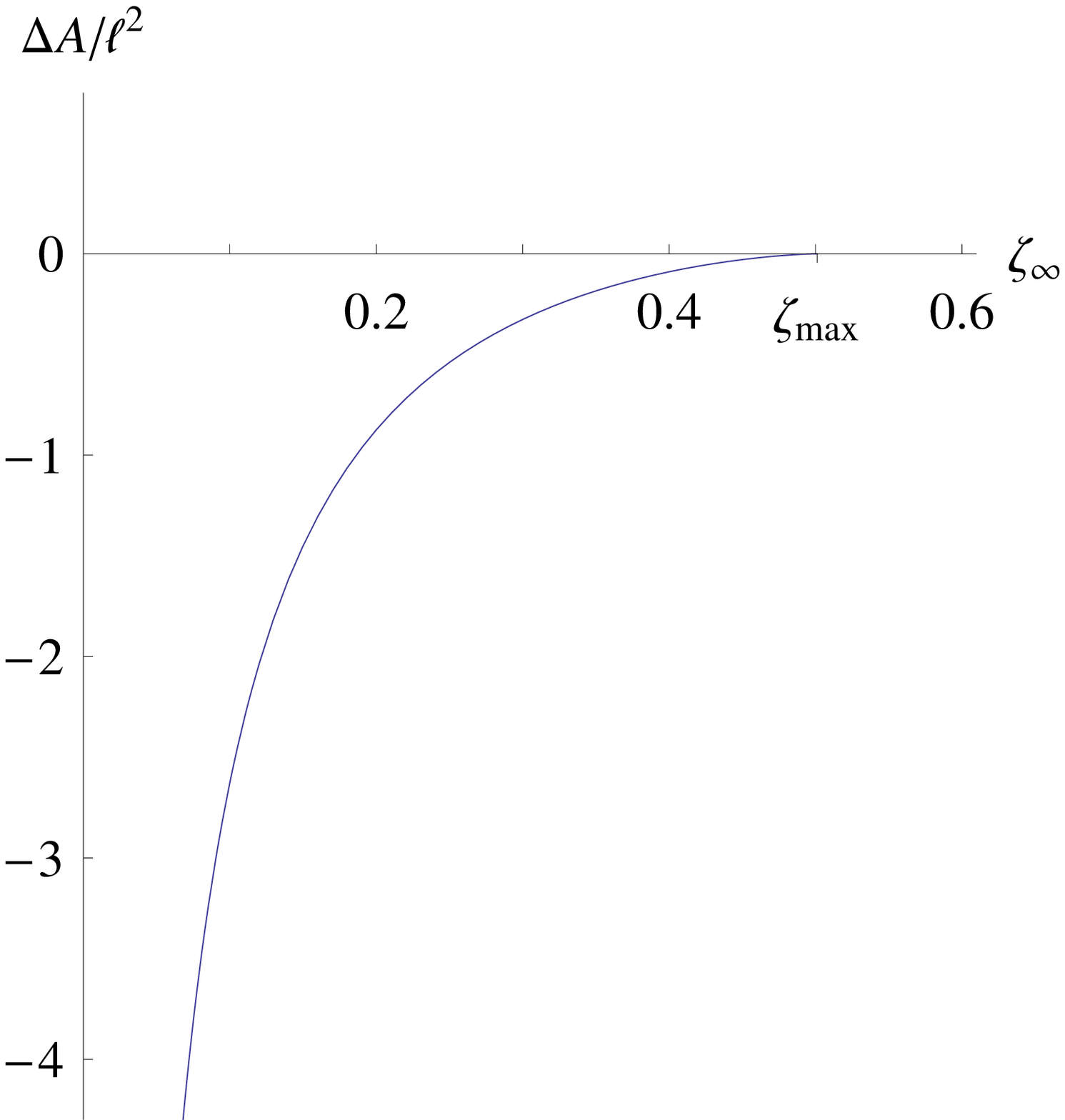}%
\\[1ex]
{\small(a)\hspace{130pt}(b)\hspace{130pt}(c)}%
\end{center}
\caption{\label{fig:arearot}\small
\textbf{Regularized area of the minimal surface spanned on two disjoint circular boundaries.}\quad
(a) The regularized area ${A_\ren}$ as a function of the parameter ${P_0}$ of the minimal surface.\quad
(b) The regularized area as a function of the distance~${s=2\ell\zeta_\infty}$ between the circular boundaries.\quad
(c) Difference ${\Delta A}$ between areas of two minimal surfaces spanned on the same boundaries.
}
\end{figure}

The renormalized area as a function of the parameter ${P_0}$ and of the distance ${s}$ between the boundaries is shown in Fig.~\ref{fig:arearot}. The first diagram shows that for ${P_0<P_\crit}$ the renormalized area is positive. In other words, the area of the minimal surface is larger than the area of two hyperplanes spanned on the same boundaries. For small values of ${P_0}$, the nontrivial phase has thus larger entanglement entropy than than the trivial one. The second diagram reveals that for the distance of the boundaries ${s\in(s_\crit,s_\mx)}$ there exist two nontrivial phases with entanglement entropy larger than the the trivial phase. For ${s>s_\mx}$ there exists only the trivial phase. A numerical estimate gives ${P_\crit\approx0.95264}$ and ${s_\crit\approx0.876895\,\ell}$.

Finally, for close boundaries, ${s<s_\mx}$, we can compute the difference ${\Delta A}$ between areas of two possible minimal surfaces. This difference is finite and independent of a renormalization of the areas. The graph of ${\Delta A}$ is shown in Fig.~\ref{fig:arearot}c.

\subsection{Surfaces with translation symmetry}

Lobachevsky space can be also viewed as a warped space with the symmetric ${y}$-space given by the translation $\zeta$-symmetry. The additional symmetry of the ${x}$-plane is then the rotational $\ph$-symmetry. Again, it is useful to work in cylindrical coordinates, however, an integration is simpler in coordinates \eqref{LobmtrccylZ}.

Substituting
\begin{equation}\label{trsym-mtr}
\begin{gathered}
    x^1=\ph\;,\qquad x^2=Z\;,\qquad y^1=\zeta\;,\\
    h_{(1)} = \ell\sqrt{Z^2-1}\;,\quad
    h_{(2)} = \frac\ell{\sqrt{Z^2-1}}\;,\quad
    R=\ell\,Z
\end{gathered}
\end{equation}
into equation \eqref{EoMS} for the profile curve we get
\begin{equation}\label{trsym-fi}
    \ph'(Z) = \pm\frac{Z_0\sqrt{Z_0^2-1}}{(Z^2-1)\sqrt{Z^4-Z^2-Z_0^4+Z_0^2}}\;.
\end{equation}
Integrating (cf.~3.157.5 of \cite{GradshteinRyzhik:book}), we obtain
\begin{equation}\label{trsym-sol}
\begin{split}
    \ph(Z) &= \ph_0 \pm\frac{Z_0}{\sqrt{Z^2-1}\sqrt{2Z_0^2-1}}\\
    &\mspace{20mu}\times\Bigl[
    {Z_0^2}\, \mathsf{\Pi}\Bigl({\textstyle\arccos\frac{Z_0}{Z},\frac1{1-Z_0^2},\sqrt{\frac{Z_0^2-1}{2Z_0^2-1}}}\Bigr)
    -(Z_0^2-1)\,\mathsf{F}\Bigl({\textstyle\arccos\frac{Z_0}{Z},\sqrt{\frac{Z_0^2-1}{2Z_0^2-1}}}\Bigr)
    \Bigr]\;.
\end{split}
\end{equation}

\begin{figure}[t]
\begin{center}
\includegraphics[width=2in]{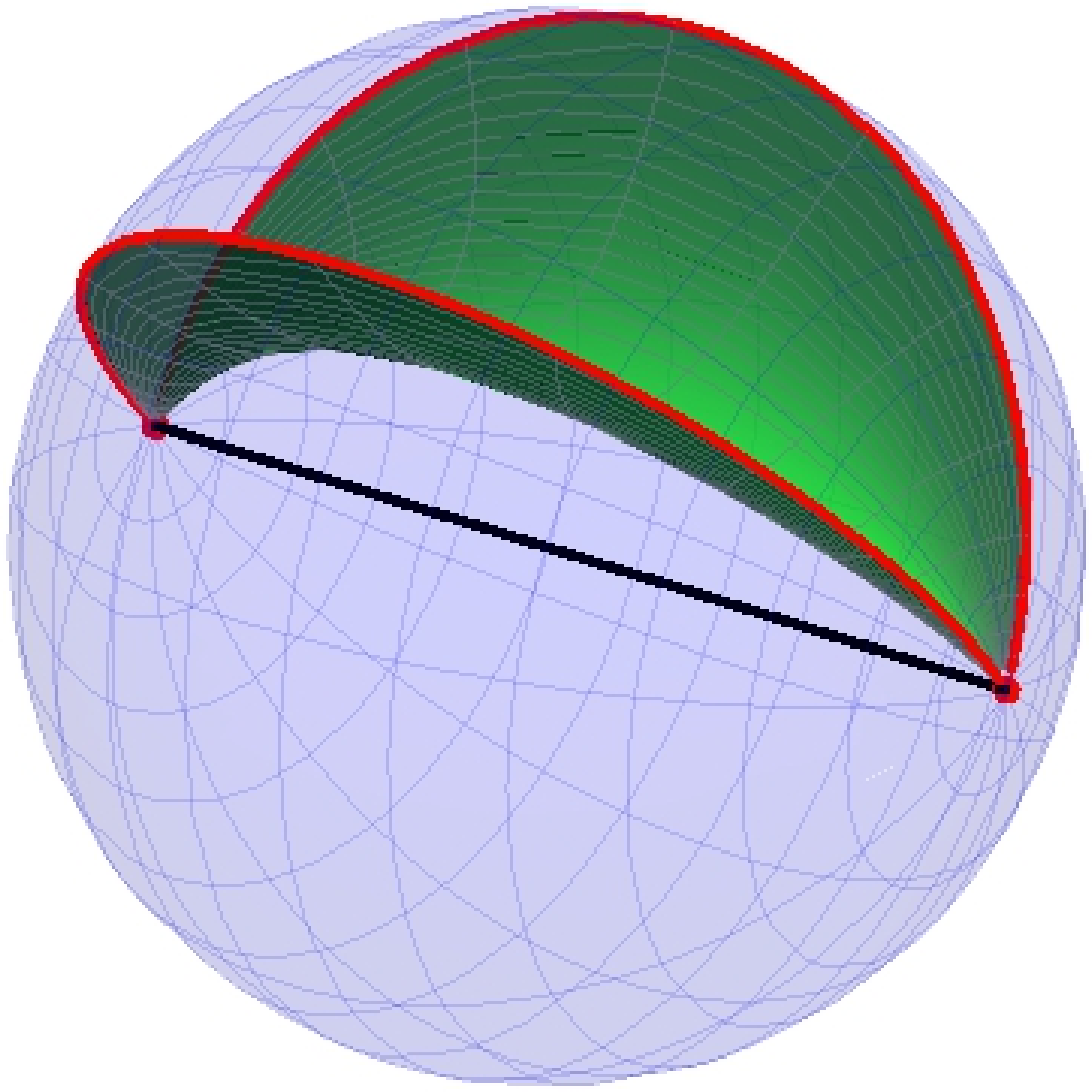}\hspace*{-2pt}%
\includegraphics[width=2in]{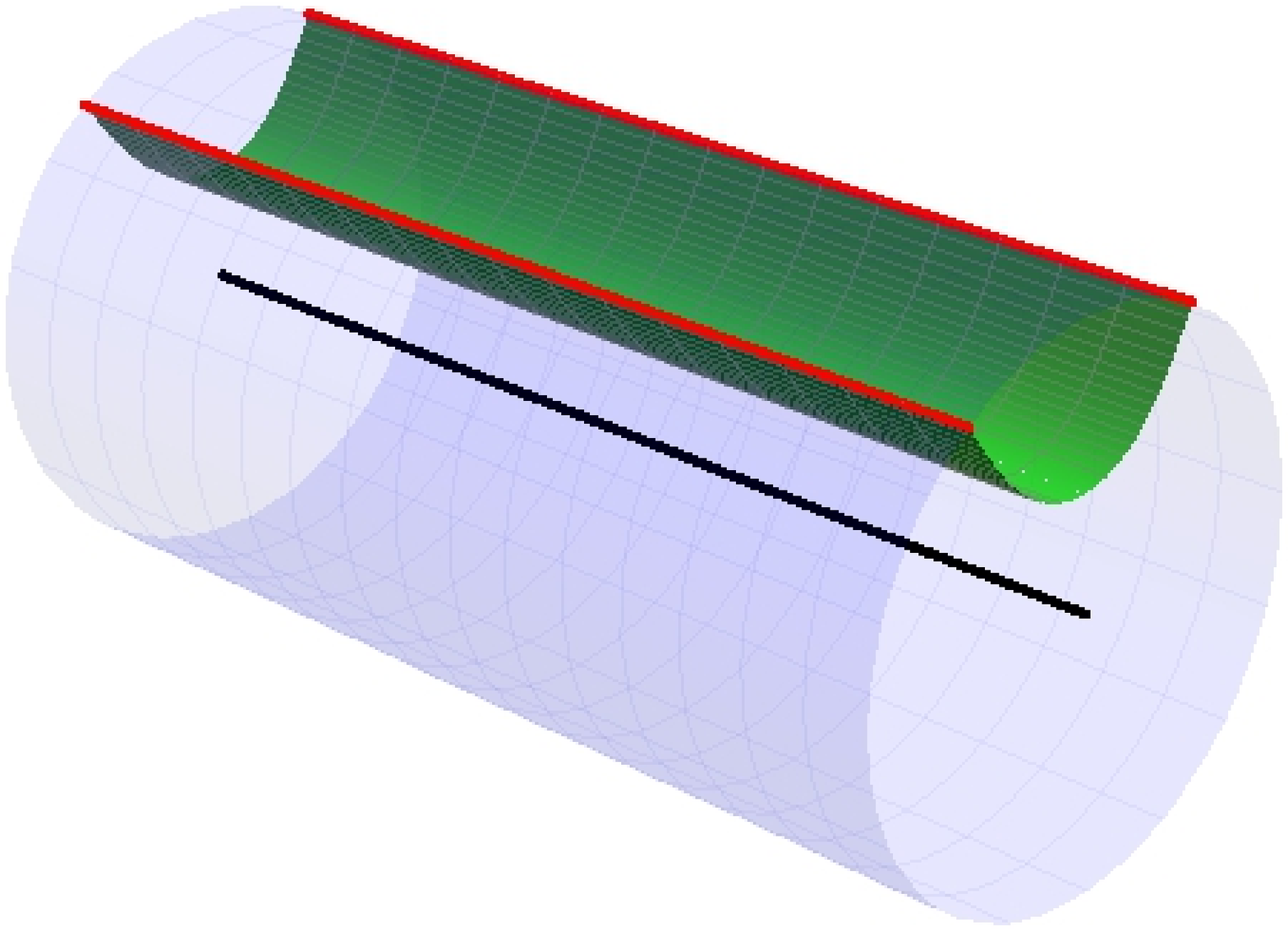}\hspace*{-2pt}%
\includegraphics[width=2in]{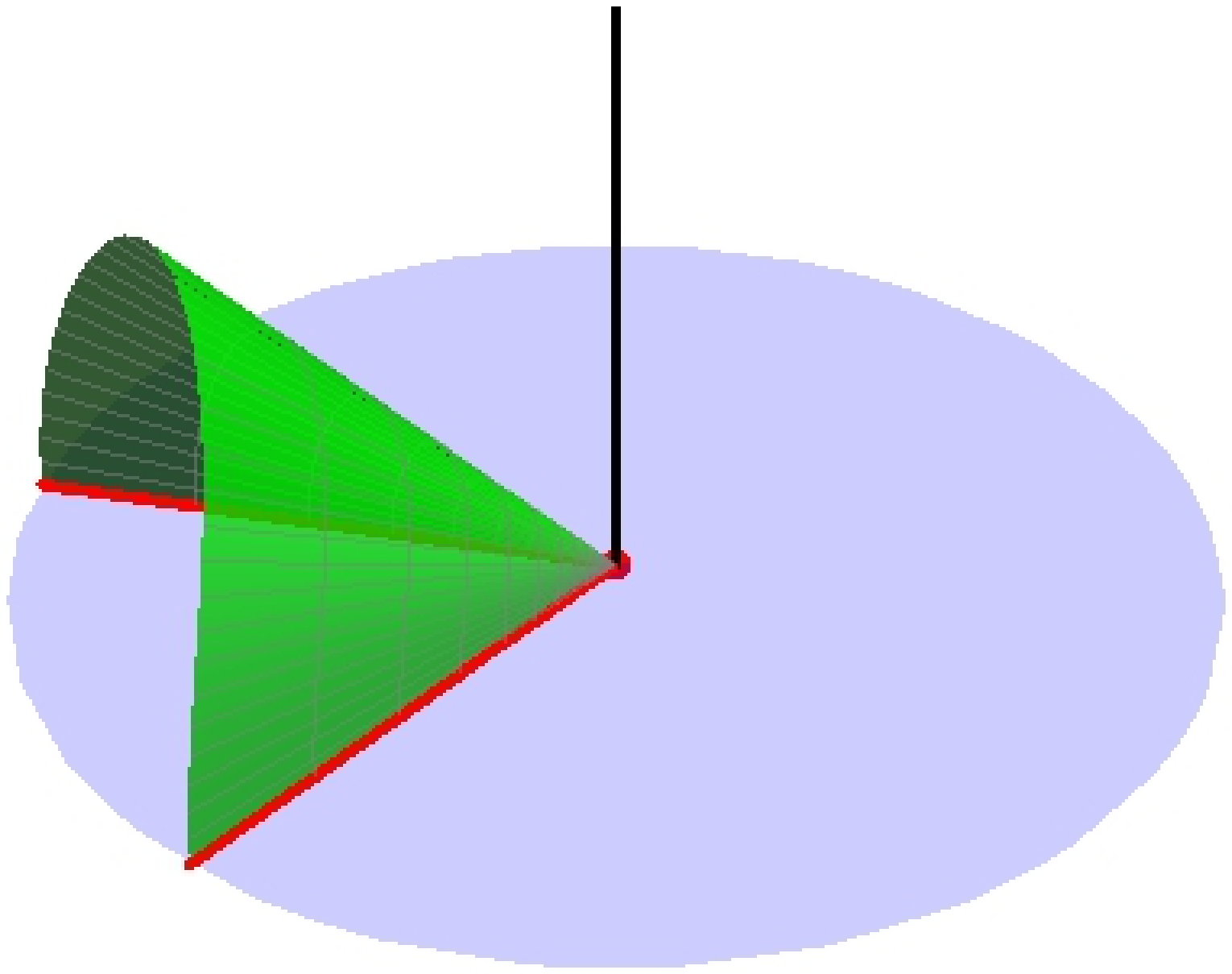}%
\\[1ex]
{\small(a)\hspace{130pt}(b)\hspace{130pt}(c)}%
\end{center}
\caption{\label{fig:mstr}\small
\textbf{Translation-symmetric minimal surface spanned on two crossing boundaries.}\quad
The surface is shown using (a) spherical, (b) cylindric and (c) half-space visualization of the Lobachevsky space (cf.~Fig.~\ref{fig:Lobrepr}). The cylindric visualization is related to the coordinates in which the surface \eqref{trsym-sol} has been found. The section ${\zeta=\text{const}}$ corresponds to Fig.~\ref{fig:prcrtr}. The spherical visualization (a) demonstrates that the `straight' boundaries from diagram (b) actually correspond to two arcs of the crossing circular boundaries at infinity.
}
\end{figure}
\begin{figure}
\begin{center}
\includegraphics[width=3in]{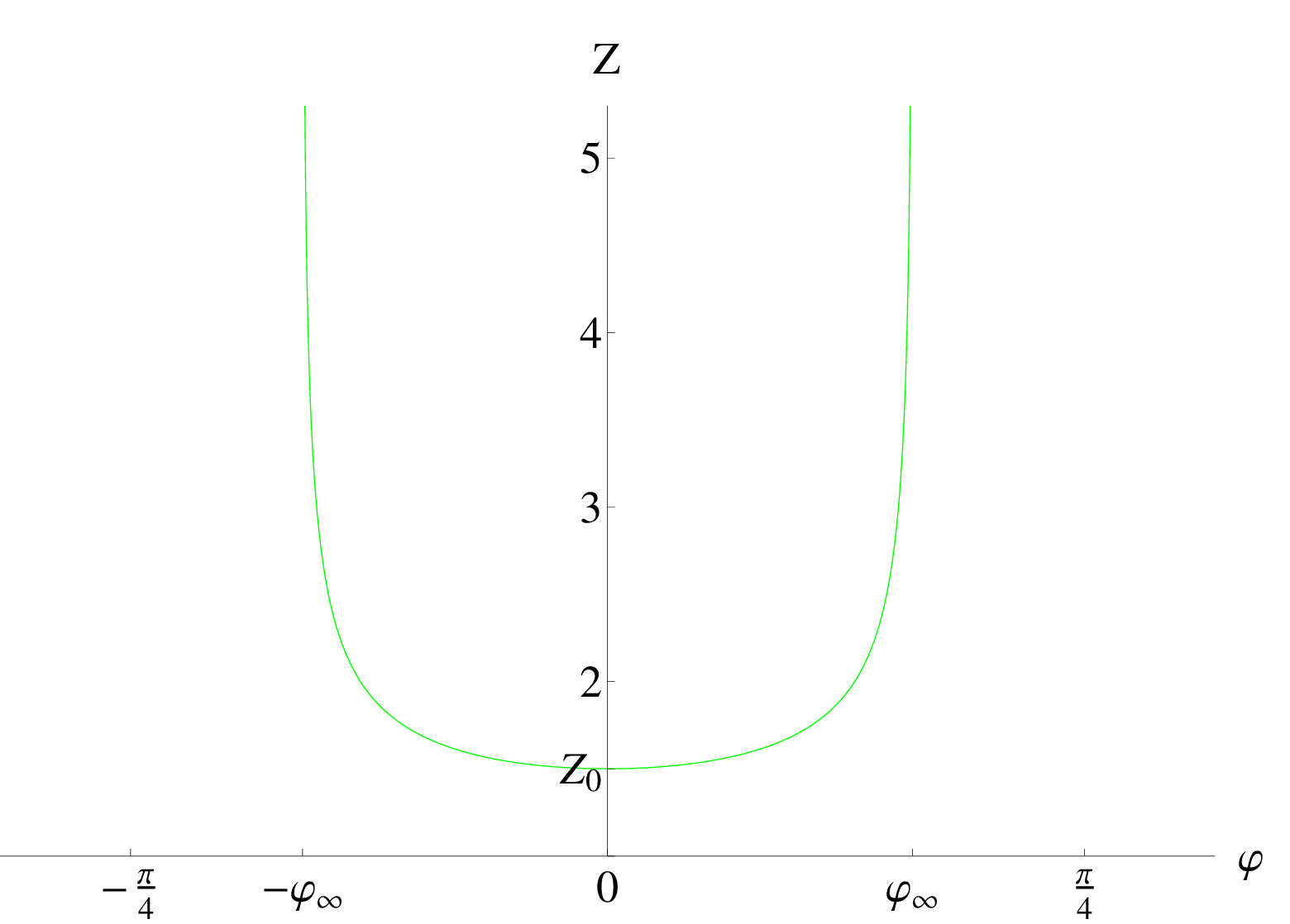}%
\end{center}
\caption{\label{fig:prcrtr}\small
\textbf{Profile curve for translation-symmetric minimal surface.}\quad
The curve is drawn in the ${x}$-plane covered by coordinates ${\ph, Z}$. It is given by solution \eqref{trsym-sol}. The corresponding minimal surface is shown in Fig.~\ref{fig:mstr}.
}
\end{figure}

Similarly to the previous case, two signs correspond to two halves of the profile curve with a turning point at ${Z=Z_0}$, ${\ph=\ph_0}$. The graphs of the corresponding minimal surface embedded into 3-dimensional Lobachevsky space are shown in Fig.~\ref{fig:mstr}. The profile curve in ${x}$-plane is depicted in Fig.~\ref{fig:prcrtr}. Three dimensional graphs demonstrate that the minimal surface is actually spanned on two crossing circular boundaries at infinity; more precisely, spanned on two arcs which intersect in two points.

\begin{figure}
\begin{center}
\includegraphics[width=3in]{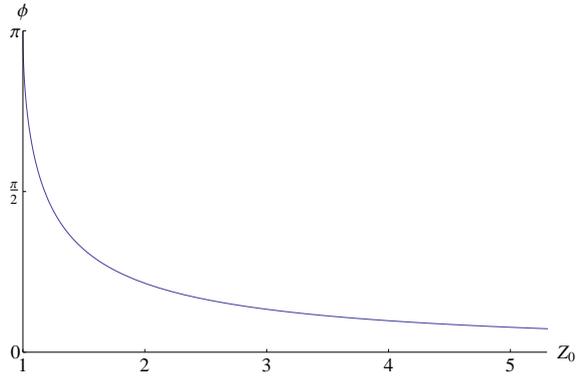}%
\end{center}
\caption{\label{fig:bndangle}\small
\textbf{The angle of two crossing circular boundaries joined by the minimal surface.}\quad
The minimal surface given by \eqref{trsym-sol} reaches infinity in two arcs of the crossing circular boundaries which have angle ${\phi=2\ph_\infty}$. Diagram shows the dependence of ${\phi}$ on the parameter ${Z_0}$ of the solution. The relation is one-to-one.
}
\end{figure}

Values of the angular coordinate in which the profile curve \eqref{trsym-sol} reaches infinity are ${\ph=\ph_0\pm\ph_\infty}$ with
\begin{equation}\label{trsym-bound}
    \ph_\infty = \frac{Z_0}{\sqrt{Z_0^2-1}\sqrt{2Z_0^2-1}}
    \Bigl[
    Z_0^2\, \mathsf{\Pi}\Bigl({\textstyle\frac1{1-Z_0^2},\sqrt{\frac{P_0^2-1}{2Z_0^2-1}}}\Bigr)
    -(Z_0^2-1)\,\mathsf{K}\Bigl({\textstyle\sqrt{\frac{Z_0^2-1}{2Z_0^2-1}}}\Bigr)
    \Bigr]\;.
\end{equation}
The crossing circular boundaries thus form the angle ${\phi=2\ph_\infty}$. The dependence of this angle on parameter ${Z_0}$ is shown in Fig.~\ref{fig:bndangle}. It is monotonous function running ${\phi=0}$ for ${Z_0=\infty}$ to ${\phi=\pi}$ for ${Z_0=1}$. The last value corresponds to a hyperplane spanned on two semi-circles forming the straight angle.

\begin{figure}[t]
\begin{center}
\includegraphics[width=2.5in]{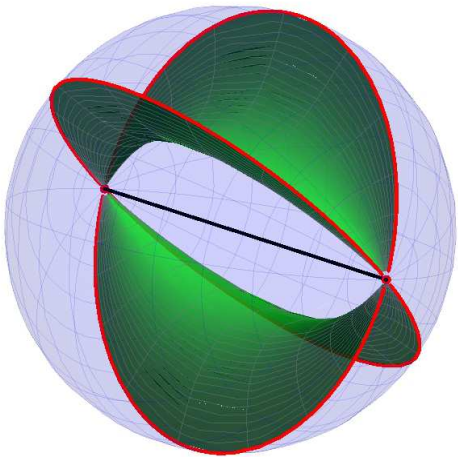}%
\end{center}
\caption{\label{fig:moresheets}\small
\textbf{The minimal surfaces spanned on two crossing circular boundaries}\quad
The minimal surface spanned on two crossing circles (i.e., on four boundary arcs) consists of two sheets. They join the opposite pairs of the boundary arcs. Each of the sheets is given by \eqref{trsym-sol} with appropriately chosen parameters ${\ph_0}$ and ${Z_0}$.\newline
For a given boundary, there exist two minimal surfaces which consist of two non-intersecting sheets. Sheets of the second minimal surface join the complementary opposite pairs of the boundary arcs.
One could consider also the third minimal surface formed by two intersecting hyperplanes spanned on the circular boundaries. This trivial solution is used to renormalize the area of the non-trivial minimal surfaces.}
\end{figure}

As we have already observed, the surface in Fig.~\ref{fig:mstr} is not spanned on whole circular boundaries, but just on two arcs belonging to these boundaries. The complete minimal surface spanned on entire two crossing circles should consist of two sheets spanned on the opposite pairs of arcs joining the intersection points. Each of these sheets is given by surface \eqref{trsym-sol} found above, see Fig.~\ref{fig:moresheets}.

The one-to-one relation \eqref{trsym-bound} between ${\phi}$ on ${Z_0}$ suggests that for a given angle of the crossing circular boundaries there exists only one non-trivial minimal surface. However, it is trivial realization that the second non-trivial surface for angle ${\phi}$ is the surface corresponding to the angle ${\pi-\phi}$. This second minimal surface also consist of two opposite sheets which join the complementary pairs of the boundary arcs.

\begin{figure}
\begin{center}
\includegraphics[width=1.5in]{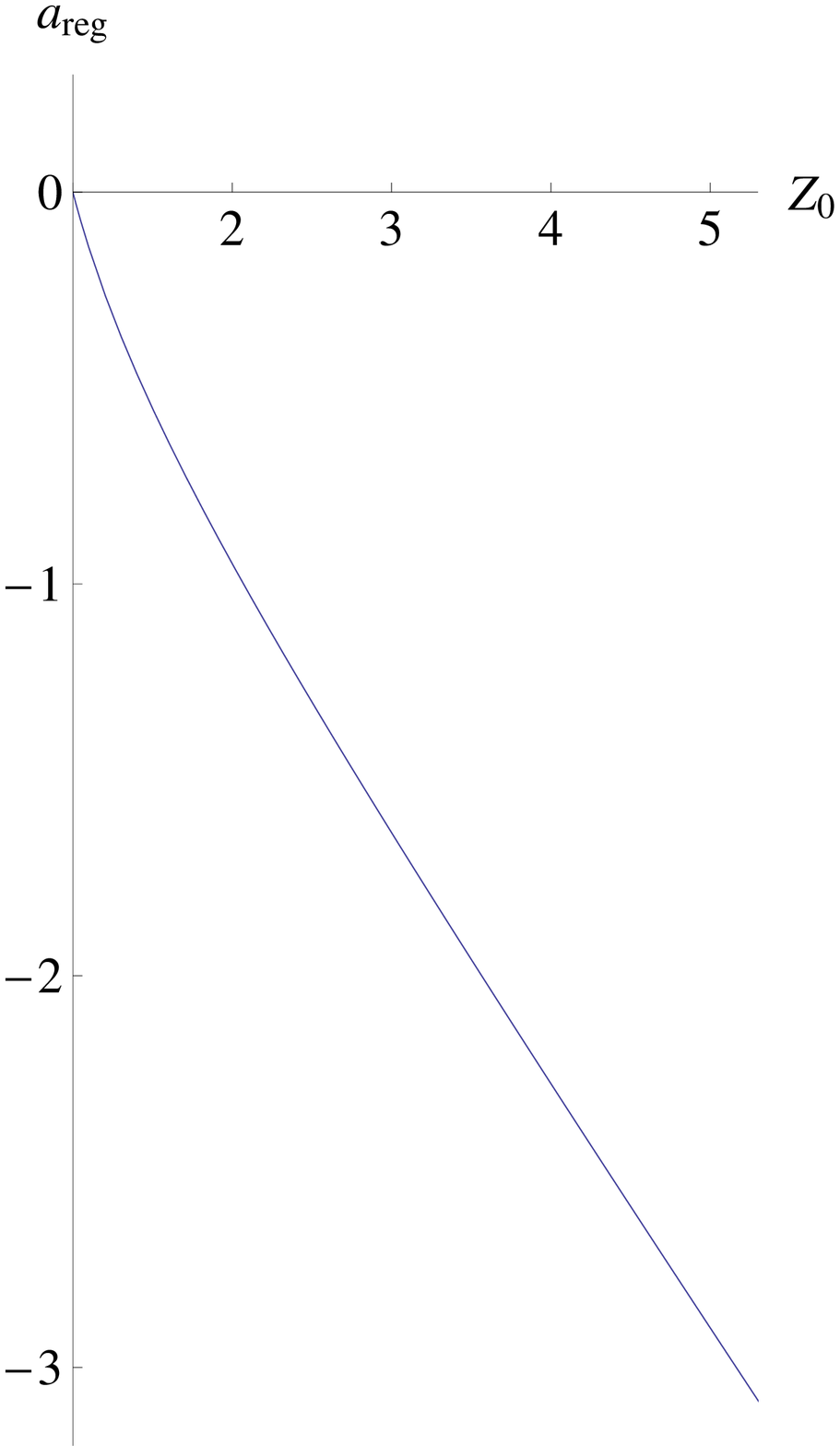}\hspace*{-2pt}%
\includegraphics[width=1.5in]{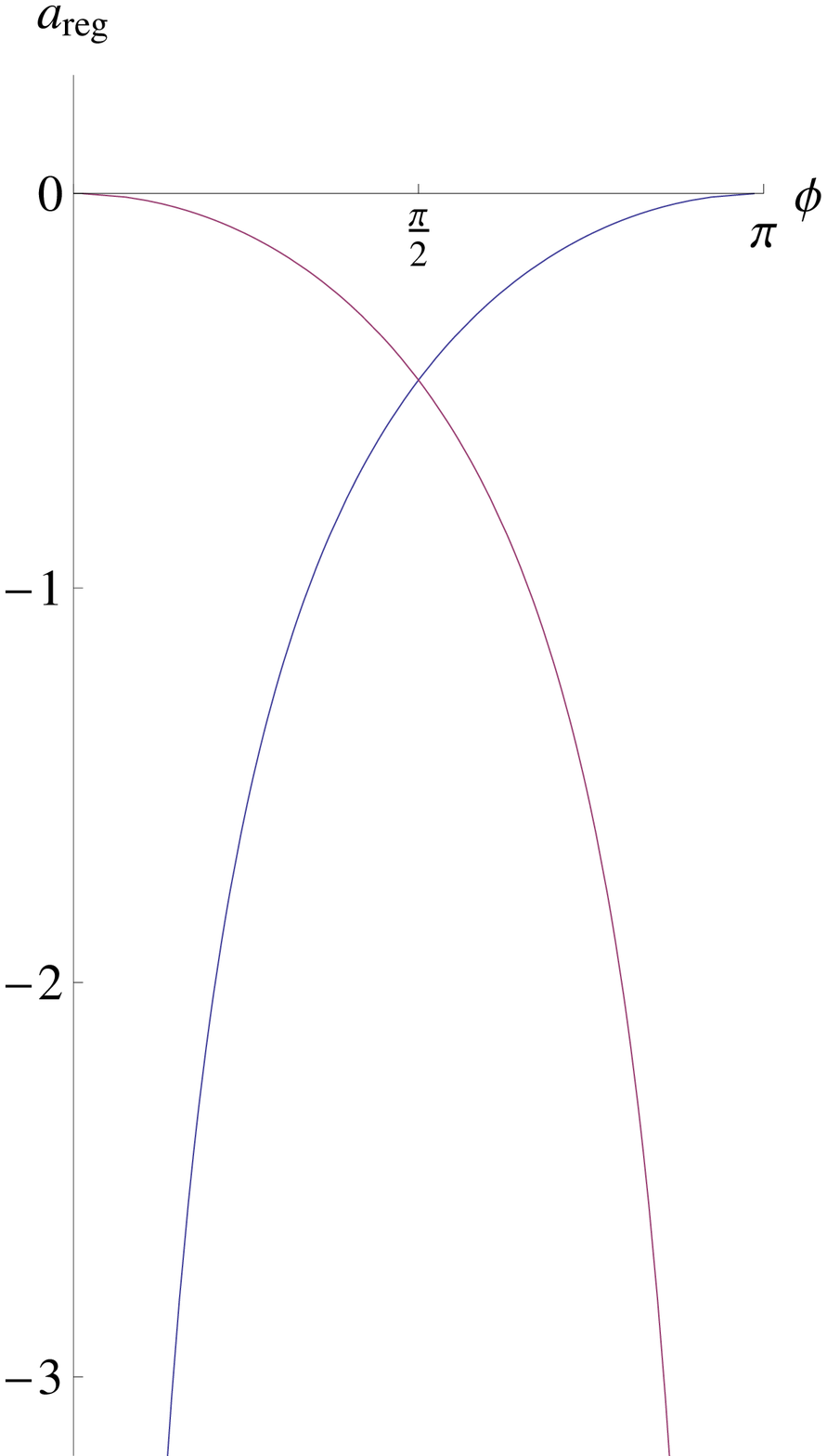}\hspace*{-2pt}%
\includegraphics[width=1.5in]{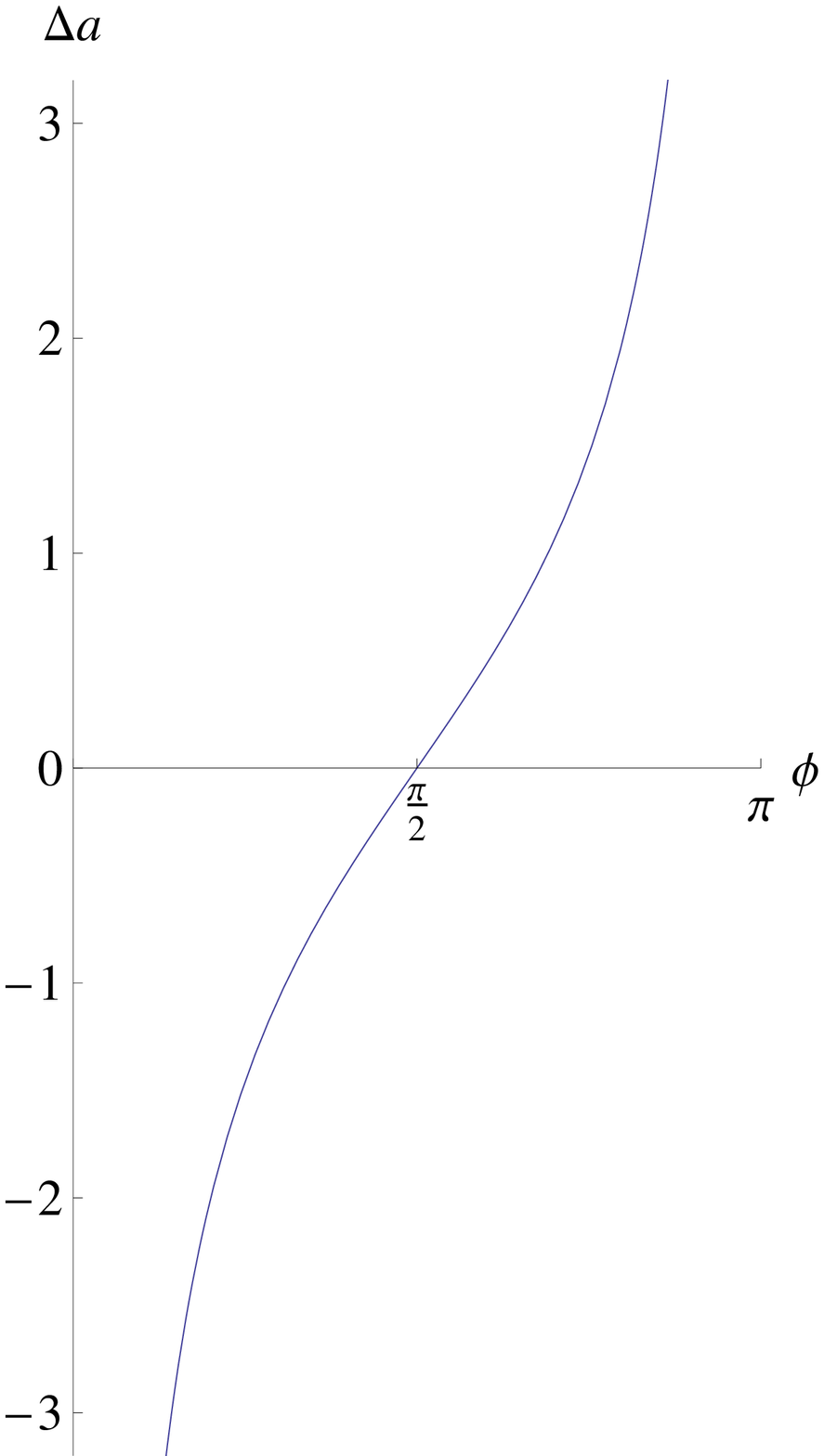}\hspace*{-2pt}%
\includegraphics[width=1.5in]{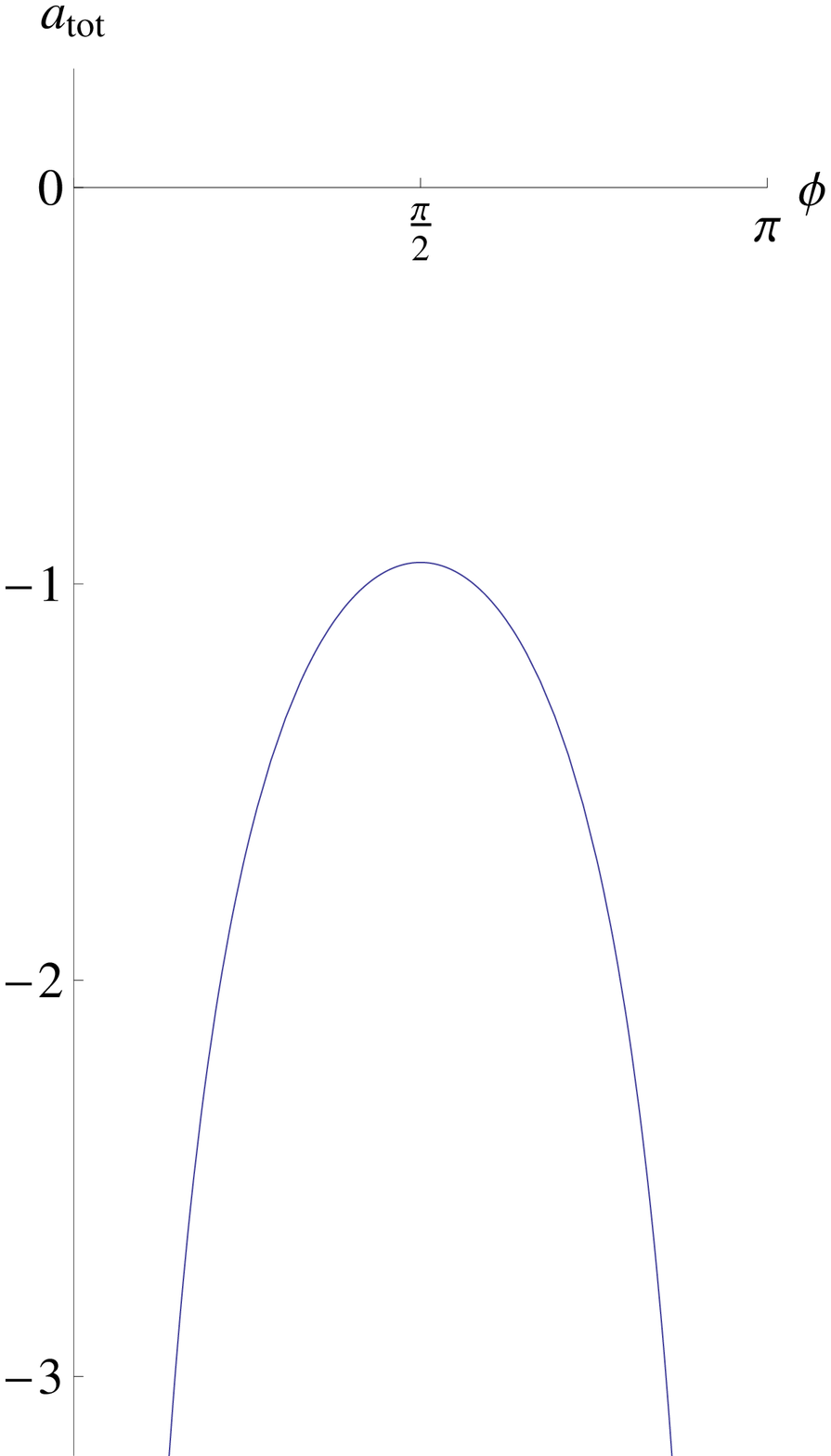}\hspace*{-2pt}%
\\[1ex]
{\small(a)\hspace{1.3in}(b)\hspace{1.3in}(c)\hspace{1.3in}(d)}%
\end{center}
\caption{\label{fig:regareatr}\small
\textbf{Regularized area of the minimal surface spanned on two crossing circles.}\quad
(a)~Regularized area density ${a_\ren}$ as a function of the parameter ${Z_0}$.\quad
(b)~Regularized area density as a function of the angle~${\phi}$ between the circular boundaries. Two branches corresponds to two possible minimal surfaces spanned on the same boundaries, cf.~Fig.~\ref{fig:moresheets}. They are given by complementary angles ${\phi}$ and ${\pi-\phi}$.\quad
(c) Difference of the area densities ${\Delta a}$ of two minimal surfaces spanned on the same boundaries.\quad
(d) Sum ${a_\tot}$ of the regularized area densities of two minimal surfaces.
}
\end{figure}

The area of the minimal surface (composed of two sheets) is given by substituting \eqref{trsym-mtr} into \eqref{areams}. It gives the regularized area evaluated up to radius ${Z}$ in the following form:\\
\begin{equation}\label{trsym-area}
    A(Z) = 4\mathcal{A}\ell^2\int_{Z_0}^Z \frac{Z^2}{\sqrt{Z^4-Z^2-Z_0^4+Z_0^2}}dZ
    =\frac{4\mathcal{A}\ell^2 Z_0^2}{\sqrt{2Z_0^2-1}}\,
      \mathsf{\Pi}\Bigl({\textstyle\arccos\frac{Z_0}{Z},1,\sqrt{\frac{Z_0^2-1}{2Z_0^2-1}}}\Bigr)\;.
\end{equation}
Here, the volume of the symmetry ${y}$-space is given by ${\mathcal{A}=\int d\zeta}$ and it is divergent. Clearly, the surface with a translation symmetry has an infinite length in the symmetric direction and the area suffers the `infrared' divergence. Therefore, we calculate only density ${a=\frac{A}{\mathcal{A}\ell}}$ of the area per unit volume of ${y}$-space. This density ${a(Z)}$ is still diverging for a large ${Z}$ and it must be renormalized by subtracting the area of the trivial solution, i.e., the area of two hyperplanes. The regularized (computed up to radius ${Z}$) density of such an area is
\begin{equation}\label{trsym-planearea}
  a_{\plane}(Z)=\frac{A_\plane(Z)}{\mathcal{A}\ell}
  =2\ell\!\int_0^Z\frac{Z\,dZ}{\sqrt{Z^2-1}}=2\ell\sqrt{Z^2-1}\;.
\end{equation}
Finally, the renormalized area density is
\begin{equation}\label{trsym-regarea}
    a_\ren = (a-2a_{\plane})|_{Z\to\infty}
    =4\ell\Bigl[
    {\textstyle\frac{Z_0^2}{\sqrt{2Z_0^2{-}1}}}\,
    \mathsf{K}\Bigl({\textstyle\sqrt{\frac{Z_0^2{-}1}{2Z_0^2{-}1}}}\Bigr)
    -\sqrt{2Z_0^2{-}1}\,\mathsf{E}\Bigl({\textstyle\sqrt{\frac{Z_0^2{-}1}{2Z_0^2{-}1}}}\Bigr)
    \Bigr]\,.
\end{equation}
The renormalized area density as a function of the parameter ${Z_0}$ and as a function of the angle ${\phi}$ is drawn in Fig.~\ref{fig:regareatr}

We can also evaluate the difference of the area densities of two minimal surfaces spanned on the same crossing circular boundaries, ${\Delta a(\phi)=a_\ren(\phi)-a_\ren(\pi-\phi)}$. The difference is finite and independent of the renormalization of the area densities.

On other side, it can be interesting to look at the total area density $a_\tot(\phi)=a_\ren(\phi)+a_\ren(\pi-\phi)$ of these two minimal surfaces. It corresponds to the renormalized entanglement entropy of the whole space divided into four blocks by two crossing circles. Both quantities ${\Delta a}$ and ${a_\tot}$ are shown in Fig.~\ref{fig:regareatr}.

These diagrams show that, in contrast to the case of two disjoint circular boundaries, the area density (and corresponding entanglement entropy) of the minimal surface spanned on two crossing circles  is always smaller than the area density of the trivial solution.

One could also study inequalities between areas of minimal surfaces and corresponding entanglement entropies (such as strong subadditivity properties \cite{Hirata:2006jx,Headrick:2007km}) spanned on boundaries of various compositions of different domains at infinity. Let us consider domains bounded by two semicircles joining two fixed poles. Such a domain is characterized by the angle $\phi$ between the semicircles. A composition of two such domains with a common semicircle forms again a domain of the same type.\footnote{%
 A similar discussion can be done also in the previous case of domains bounded by two disjoint circles. However, the discussion is more involved since the composition law for the distances between circular boundaries is not so simple: If ${\Omega_{13}=\Omega_{12}\cup\Omega_{23}}$, where $\Omega_{ij}$ is a domain between two circular boundaries $\Sigma_i$ and $\Sigma_j$, 
 the distances ${s_{ij}=s(\Sigma_i,\Sigma_j)}$ between these boundaries are not, in general, in an additive relation. The additivity ${s_{13} = s_{12}+s_{23}}$ holds only if the circular boundaries $\Sigma_i$ are concentric.

 In the case of domains between two arcs is the situation simpler, the angles between arcs satisfy the additivity law.}

The subadditivity property \cite{Ryu:2006bv,Headrick:2007km} translated to language of areas is satisfied in the leading diverging order
\begin{equation}\label{subadd}
   a(\phi_1+\phi_2)\leq a(\phi_1)+a(\phi_1)\;.
\end{equation}
Indeed, the right hand side has more diverging boundary contributions.
It is not a surprise since the subadditivity is a straightforward consequence of the minimality of the area \cite{Ryu:2006bv}. A more subtle situation is the strong subadditivity, where the leading diverging contributions to the area cancel each other and one can compare renormalized values. The strong subaddivity thus reads
\begin{equation}\label{strongsubadd}
   a_\ren(\phi_1+\phi_2+\phi_3)+ a_\ren(\phi_2)\leq a_\ren(\phi_1+\phi_2) + a_\ren(\phi_2+\phi_3)\;,
\end{equation}
where the renormalized area density $a_\ren$ is given by one half of expression \eqref{trsym-area} with the parameter ${Z_0}$ expressed in terms of angle $\phi=2\ph_\infty$ using \eqref{trsym-bound}. For ${\phi>\pi}$ one naturally understands ${a_\ren(\phi)=a_\ren(2\pi-\phi)}$. Evaluating \eqref{strongsubadd} for angles $\phi_1+\phi_2+\phi_3<2\pi$ we have explicitly checked that the strong inequality is satisfied. It is consistent with the general statement of \cite{Headrick:2007km}.

\subsection{Surfaces with horocyclic symmetry}

\begin{figure}[t]
\begin{center}
\includegraphics[width=2in]{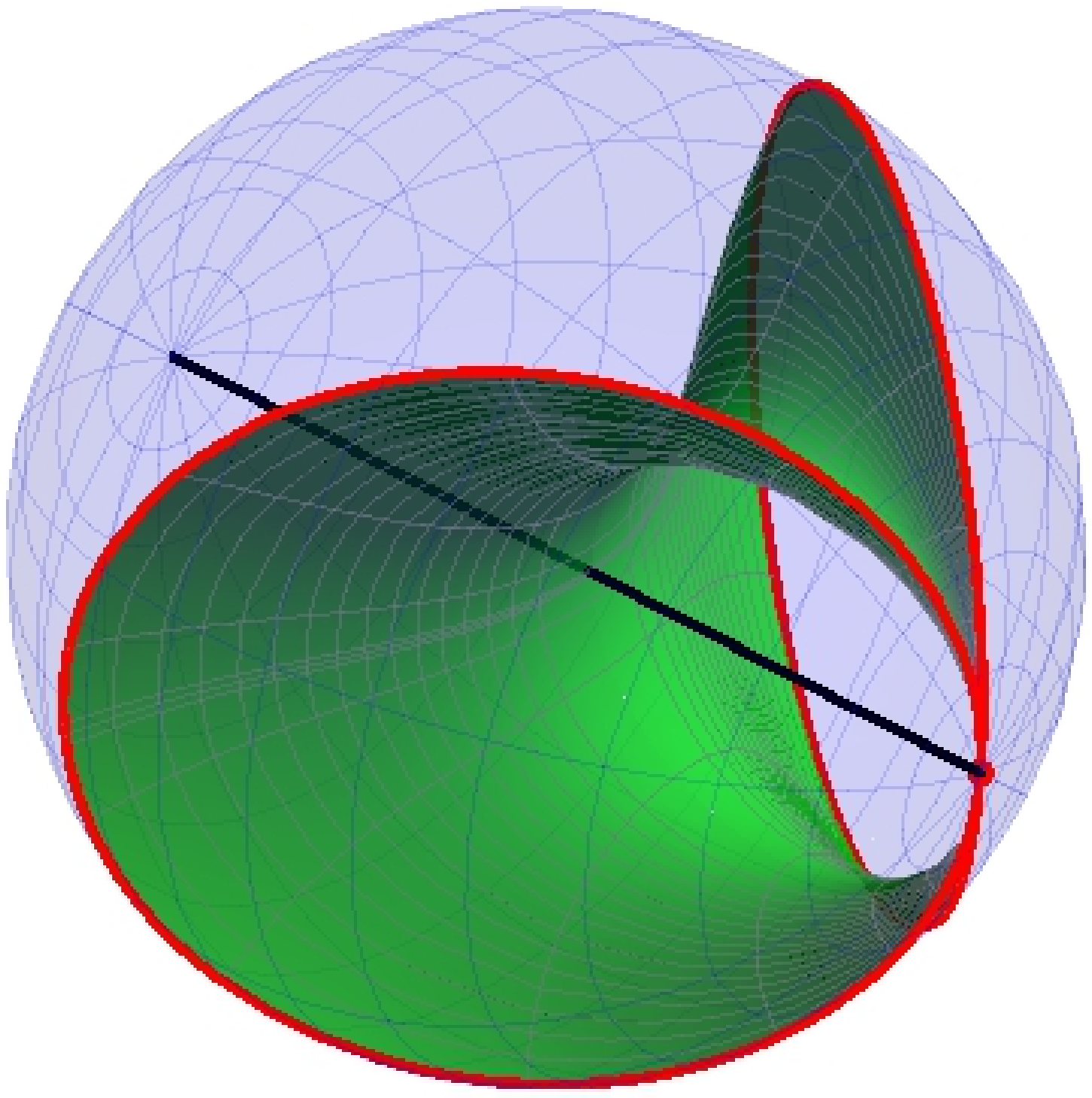}\hspace*{-2pt}%
\includegraphics[width=2in]{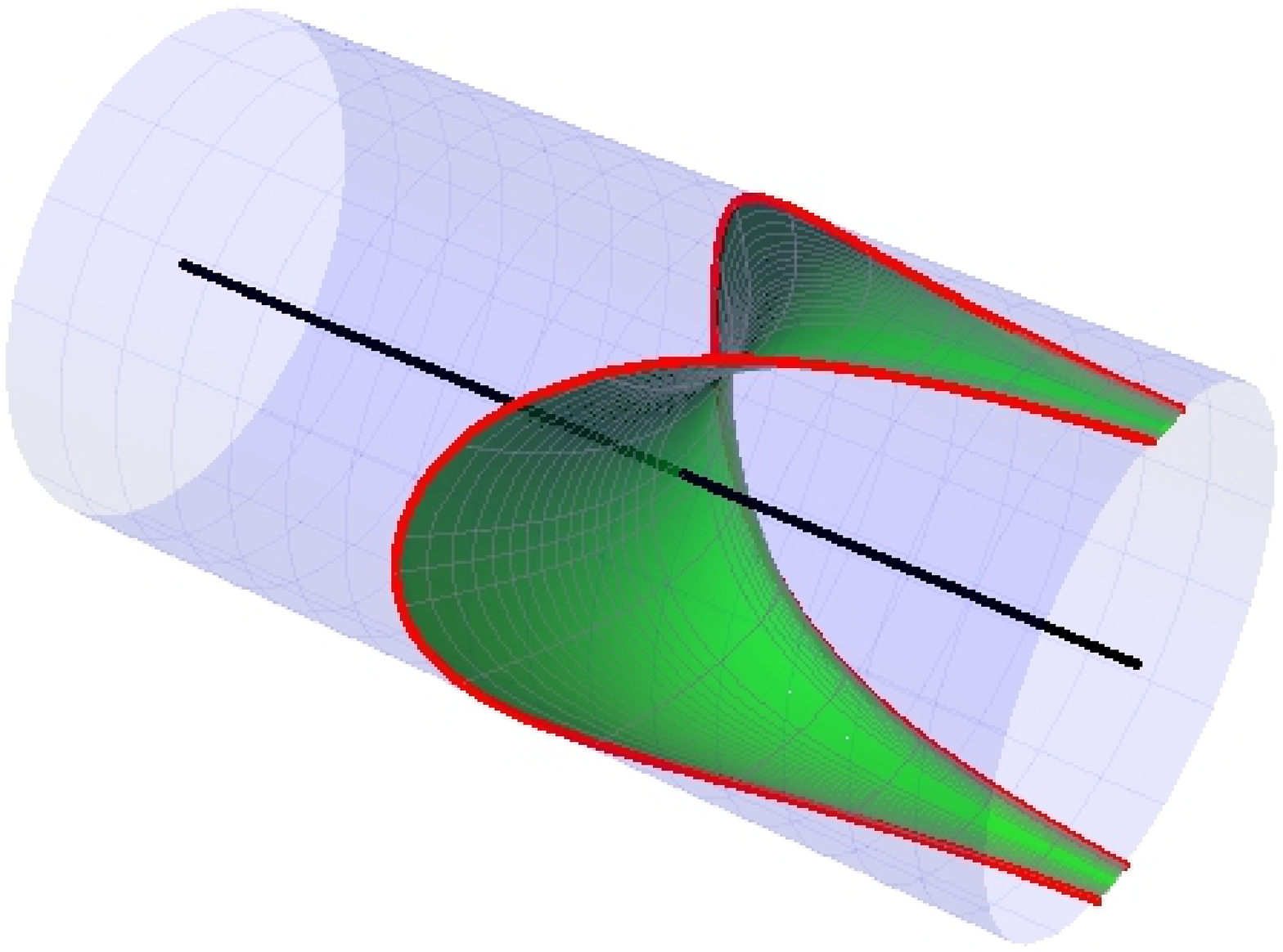}\hspace*{-2pt}%
\includegraphics[width=2in]{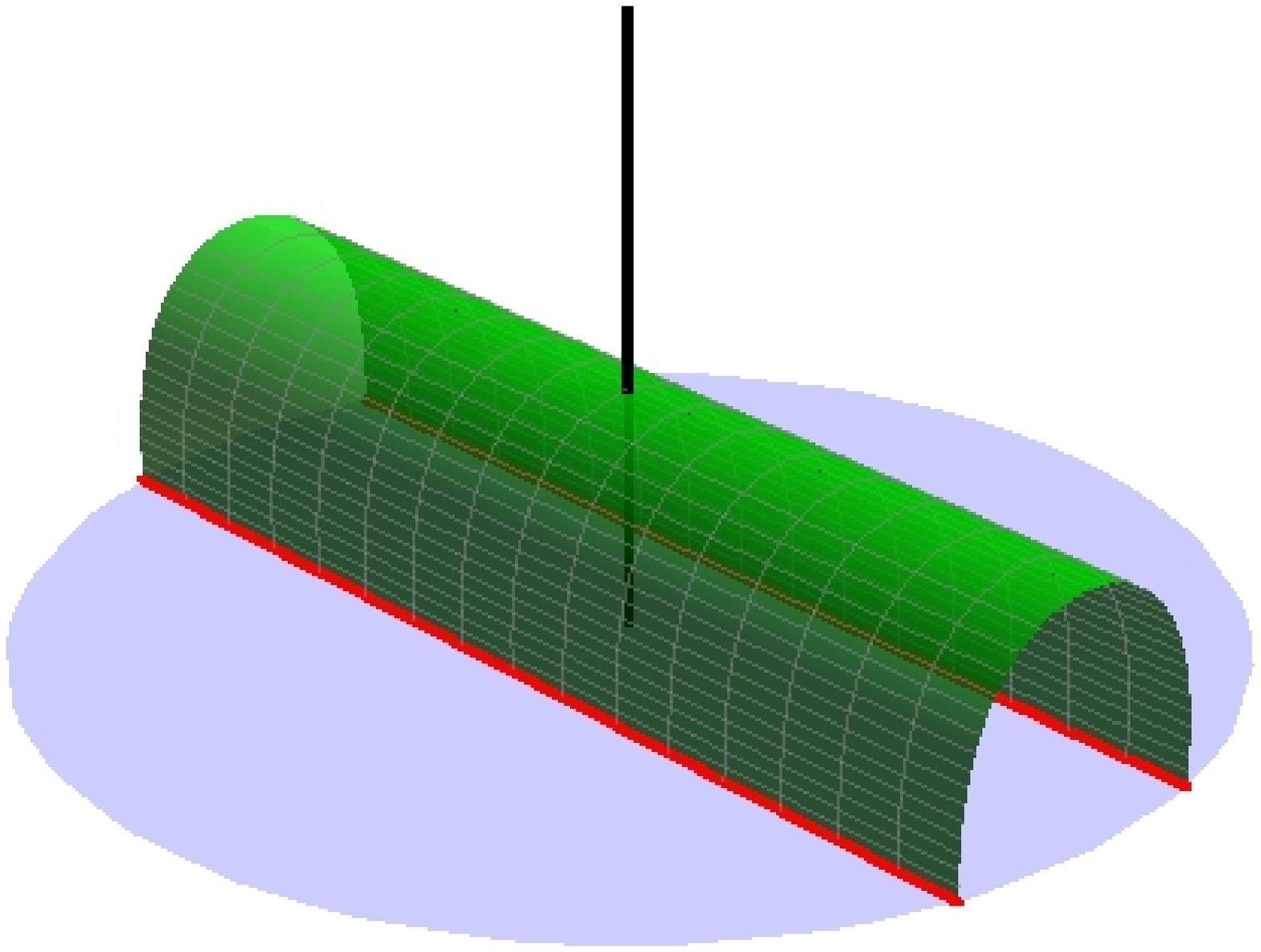}%
\\[1ex]
{\small(a)\hspace{130pt}(b)\hspace{130pt}(c)}%
\end{center}
\caption{\label{fig:mshor}\small
\textbf{Horocyclic-symmetric minimal surface spanned on two touching boundaries.}\quad
The surface is shown using (a) spherical, (b) cylindric and (c) half-space visualization of Lobachevsky space (cf.~Fig.~\ref{fig:Lobrepr}). The half-space visualization is related to the coordinates in which the surface \eqref{horsym-sol} has been found. The section ${\yP=\text{const}}$ corresponds to Fig.~\ref{fig:prcrhor}. The spherical visualization (a) demonstrates that the `straight' boundaries from diagram (c) are actually two circular boundaries touching at one point. All minimal surfaces with this type of boundary conditions are isomorphic.
}
\end{figure}

The last qualitatively different case corresponds to the horocyclic symmetry.
The ${y}$-space can be identified with ${\yP}$ direction in Poincar\'{e}
coordinates \eqref{LobmtrcPoinc}. The additional symmetry of the ${x}$-plane
then corresponds to the horocyclic shift in ${\xP}$ direction:
\begin{equation}\label{horsym-mtr}
\begin{gathered}
    x^1=\xP\;,\qquad x^2=\zP\;,\qquad y^1=\yP\;,\\
    h_{(1)} = \frac\ell\zP\;,\quad
    h_{(2)} = \frac\ell\zP\;,\quad
    R=\frac\ell\zP\;.
\end{gathered}
\end{equation}
The equation of the profile curve \eqref{EoMS} can be again integrated (cf.~3.153.3 of  \cite{GradshteinRyzhik:book})
\begin{equation}\label{horsym-sol}
    \xP(\zP) = \xP_0\pm\int_\zP^{\zP_0}\frac{d\zP}{\sqrt{\frac{\zP_0^4}{\zP^4}-1}}
    =\,\xP_0\pm\zP_0\Bigl[
    \sqrt2\,\mathsf{E}\Bigl({\textstyle\arccos\frac{\zP}{\zP_0},\frac1{\sqrt2}}\Bigr)
    -\frac1{\sqrt2}\mathsf{F}\Bigl({\textstyle\arccos\frac{\zP}{\zP_0},\frac1{\sqrt2}}\Bigr)
    \Bigr]\,.
\end{equation}

\begin{figure}[t]
\begin{center}
\includegraphics[width=3in]{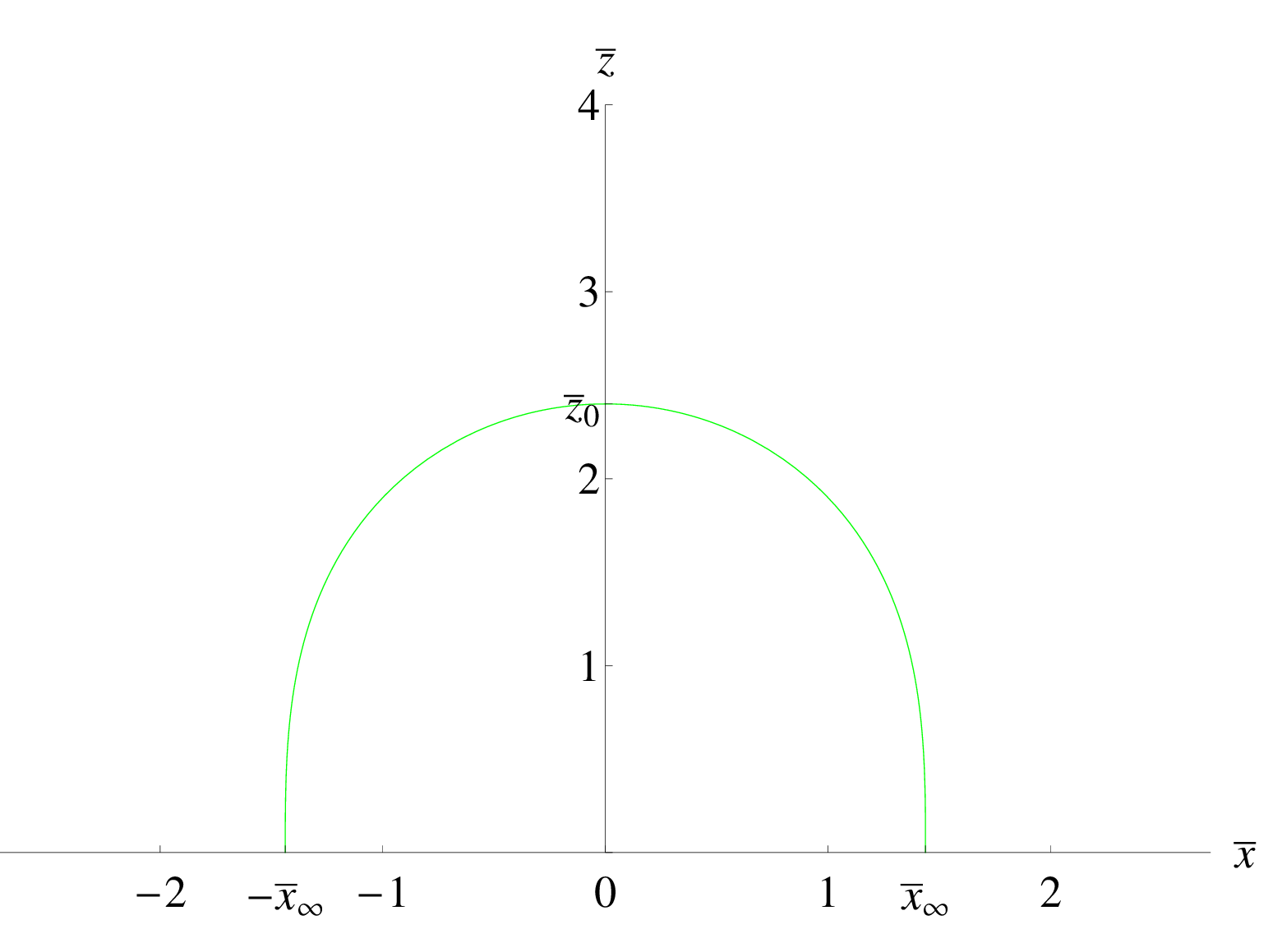}%
\end{center}
\caption{\label{fig:prcrhor}\small
\textbf{Profile curve for horocyclic-symmetric minimal surface.}\quad
The curve is drawn in the ${x}$-plane covered by coordinates ${\zP, \xP}$. It is given by solution \eqref{horsym-sol}. The corresponding minimal surface is shown in Fig.~\ref{fig:mshor}.
}
\end{figure}

A corresponding horocyclic-symmetric minimal surface embedded into the Lobachevsky space is shown in Fig.~\ref{fig:mshor}, the profile curve is depicted in Fig.~\ref{fig:prcrhor}.

The surface is parametrized by the parameter $\zP_0$ which is the maximal value of the coordinate $\zP$ which the surface reaches. It is also the turning point joining two halves of the surface given by two signs in \eqref{horsym-sol}. We call line $\zP=\zP_0$ the top line of the surface, cf.~Fig.~\ref{fig:mshor}c. It is a horocycle in the sense of the hyperbolic geometry, cf.~Fig.~\ref{fig:mshor}a.

The limiting value of coordinate ${\xP}$ at infinity ${\zP=0}$ is
\begin{equation}\label{horsym-inf}
    \xP_\infty = \xP_0\pm X_0\, \zP_0\;,\quad\text{with}\quad
    X_0=\frac{\Gamma(\frac34)^2}{\sqrt{2\pi}}\approx 0.59907\;.
\end{equation}
The minimal surface thus reaches infinity at two straight lines in Poincar\'{e}
coordinates, cf.~Fig.~\ref{fig:mshor}c. However, the spherical representation in
Fig.~\ref{fig:mshor}a shows that these boundaries are actually two circular
boundaries touching at one point (the improper point of planar infinity in the
half-space representation of Fig.~\ref{fig:mshor}c).

From the equation \eqref{horsym-sol} of the profile curve, we can observe that
the combination ${\frac\xP{\zP_0}}$ depends only on ${\frac{\zP}{\zP_0}}$. This
documents that all solutions of this type (i.e., with an arbitrary value of
${\zP_0}$) are isometric. Indeed, the translation ${\zeta\to\zeta+\zeta_0}$
along the Killing vector ${\cv{\zeta}}$ in Poincar\'{e} coordinates acts
\begin{equation}\label{translinPoincare}
    \zP\to\exp\zeta_0\; \zP\;,\quad
    \xP\to\exp\zeta_0\; \xP\;,\quad
    \yP\to\exp\zeta_0\; \yP\;,
\end{equation}
i.e., as a constant rescaling of Poincar\'{e} coordinates. The solutions
\eqref{horsym-sol} for various ${\zP_0}$ are related exactly by this
translation. Parameter ${\zP_0}$ only labels the position of the minimal surface
in the space, not its shape. It is, of course, a consequence of the fact that
all configurations of two touching circular boundaries are equivalent, as we
observed in section~\ref{ssc:boundaries}.

The regularized area function \eqref{areams} of the minimal surface in this case is
\begin{equation}\label{horsym-area}
\begin{split}
    A(\zP) &= 2\mathcal{A}\ell^2\int_{\zP}^{\zP_0} \frac{\zP_0^2}{\zP^2\sqrt{\zP_0^4-\zP^4}}d\zP\\
    &=\frac{2\mathcal{A}\ell^2}{\zP_0}\Biggl[
    \sqrt{\frac{\zP_0^2}{\zP^2}-\frac{\zP^2}{\zP_0^2}}
    -\sqrt2\,\mathsf{E}\Bigl({\arccos\frac{\zP}{\zP_0},\frac1{\sqrt2}}\Bigr)
    +\frac1{\sqrt2}\,\mathsf{F}\Bigl({\arccos\frac{\zP}{\zP_0},\frac1{\sqrt2}}\Bigr)
    \Biggr]\,.
\end{split}\raisetag{15ex}
\end{equation}
The regularized area of the part of the hyperplane ${\xP=\text{const}}$ calculated up to cut-off ${\zP}$ is
\begin{equation}\label{horsym-plane}
    A_\plane(\zP) = \frac{\mathcal{A}\ell^2}{\zP}\;.
\end{equation}
Subtracting two hyperplanes from the minimal surface thus gives the renormalized area
\begin{equation}\label{horsym-regarea}
    A_\ren=(A-2A_\plane)|_{\zP\to0}= -2\mathcal{A} \frac{X_0}{\zP_0}\ell^2\;,
\end{equation}
with the constant ${X_0}$ given in \eqref{horsym-inf}.

However, we have to solve the infrared divergence hidden in the ${y}$-space
volume ${\mathcal{A}=\int d\yP}$. One has to be careful how to treat this
infinity since the choice of the ${\yP}$ coordinate was rather arbitrary.
Indeed, an arbitrary rescaled coordinate ${\yP}$ could have been used since a
constant rescaling correspond to the isometry \eqref{translinPoincare}.

One natural way how to cut-off the ${\yP}$ direction is to calculate the surface area per unit ${\yP}$-length, where this `unit length' is defined by a prescription formulated only in terms of the surface itself, by a prescription which does not employ any additional structure. For example, we can measure ${\yP}$-length ${Y_0}$ at the top line of the surface (i.e., at ${\zP=\zP_0}$). Clearly, ${Y_0=\int\frac{\ell}{\zP_0}d\yP=\ell\frac{\mathcal{A}}{\zP_0}}$. The corresponding area density then reads
\begin{equation}\label{horsym-areadens}
    a_\ren = \frac{A_\ren}{Y_0} = -2 X_0 \ell\;.
\end{equation}
It is independent of the parameter ${\zP_0}$, as could had been expected from the discussion above: ${\zP_0}$ defines only a position of the surface, not its shape, and no additional structure has been introduced which could distinguish among minimal surfaces with different ${\zP_0}$.

Other possibility how to deal with the divergence in the ${\yP}$ direction is to compactify this direction. We can assume ${S^1}$ compactification along the coordinate ${\yP}$ given by a fixed range ${\Delta\yP}$. Then ${\mathcal{A}=\Delta\yP}$ and the regularized area of the compactified minimal surface is
\begin{equation}\label{horsym-areacomp}
    A_\comp = -2\Delta\yP \frac{X_0}{\zP_0} \ell^2
            = -4\frac{\Delta\yP}{\Delta\xP}X_0^2\ell^2\;.
\end{equation}
The dependence on ${\zP_0}$ reflects that the minimal surfaces with various positions ${\zP_0}$ are squeezed into the compactified space in a different way. Since this space is not global Lobachevsky space anymore, the minimal surfaces with various ${\zP_0}$ are not globally isomorphic. In the last equality we expressed $\zP_0$ using the coordinate distance ${\Delta\xP=2X_0\zP_0}$ of the boundaries
of the minimal surface, cf.~\eqref{horsym-inf}.

In both these cases the regularized area is negative, i.e., the area of the minimal surface spanned on two touching circular boundaries is smaller than the area of two corresponding hyperplanes.

By a composition of two or three domains between touching circular boundaries with a common contact point we can check the the subadditivity and the strong subadditivity properties. The subadditivity property is again satisfied in the diverging order.

To check the strong subadditivity we have to consider three domains $\Omega_i$, ${i=1,\,2,\,3}$, located among the circular boundaries separated by the coordinate intervals $\Delta\xP_i$. These domains must be regularized in common way. Therefore we use the compactification of the $\yP$ coordinate to the interval $\Delta\yP$. The strong subadditivity
\begin{equation}\label{strongsubadd-touching}
   A_{\Omega_1\cup\Omega_2\cup\Omega_3}+ A_{\Omega_2}
     \leq A_{\Omega_1\cup\Omega_2} + A_{\Omega_2\cup\Omega_3}\;,
\end{equation}
thus, using \eqref{horsym-areacomp}, translates into
\begin{equation}\label{strongsubadd-touchingreg}
   -\frac{1}{\Delta\xP_1+\Delta\xP_2+\Delta\xP_3}-\frac{1}{\Delta\xP_2}
     \leq -\frac{1}{\Delta\xP_1+\Delta\xP_2} -  \frac{1}{\Delta\xP_2+\Delta\xP_3}\;,
\end{equation}
which is (for positive $\Delta\xP_i$) trivially satisfied.\footnote{%
 One could consider also a composition of domains between touching circular boundaries with different contact points. However, the composed domain would be between two disjoint circles. Both areas \eqref{rotsym-regarea} and \eqref{horsym-area} would enter the subadditivity inequalities. In such a case, the regularization procedure would have to be discussed carefully: all infrared infinities have to be regularized in a consistent way. We leave such a discussion elsewhere.}


\subsection{General position of two circular boundaries}

In the preceding subsections we have found the minimal surfaces for three qualitatively different positions of the circular boundaries. It could seem that we studied only circular boundaries which are specially positioned with respect to the chosen system of coordinates. For example, two disjoint circles are concentric in Fig.~\ref{fig:msrot}. However, it would be a wrong impression. Actually, we have found the minimal surface for a completely arbitrary configuration of two circular boundaries at infinity.

Indeed, any two circles at infinity can be moved by an isometry to the position for which we have already found the solution. Or, in opposite way, we can always construct a coordinate systems which is adjusted to a boundary configuration. Using isometries we can than transform the solution to an arbitrary other frame.

For two disjoint circles spanned on two hyperplanes we can always find a unique line perpendicular to both hyperplanes and use this line as ${\rho=0}$ axis of our cylindrical coordinate system. The circular boundaries become concentric in this frame.

Similarly, for two crossing circles we use the intersection line of the corresponding hyperplanes as the axis of the cylindrical system. For two touching circles we can use any line going through the contact point of both circles as a suitable axis.

\begin{figure*}
\begin{center}
\includegraphics[width=2in]{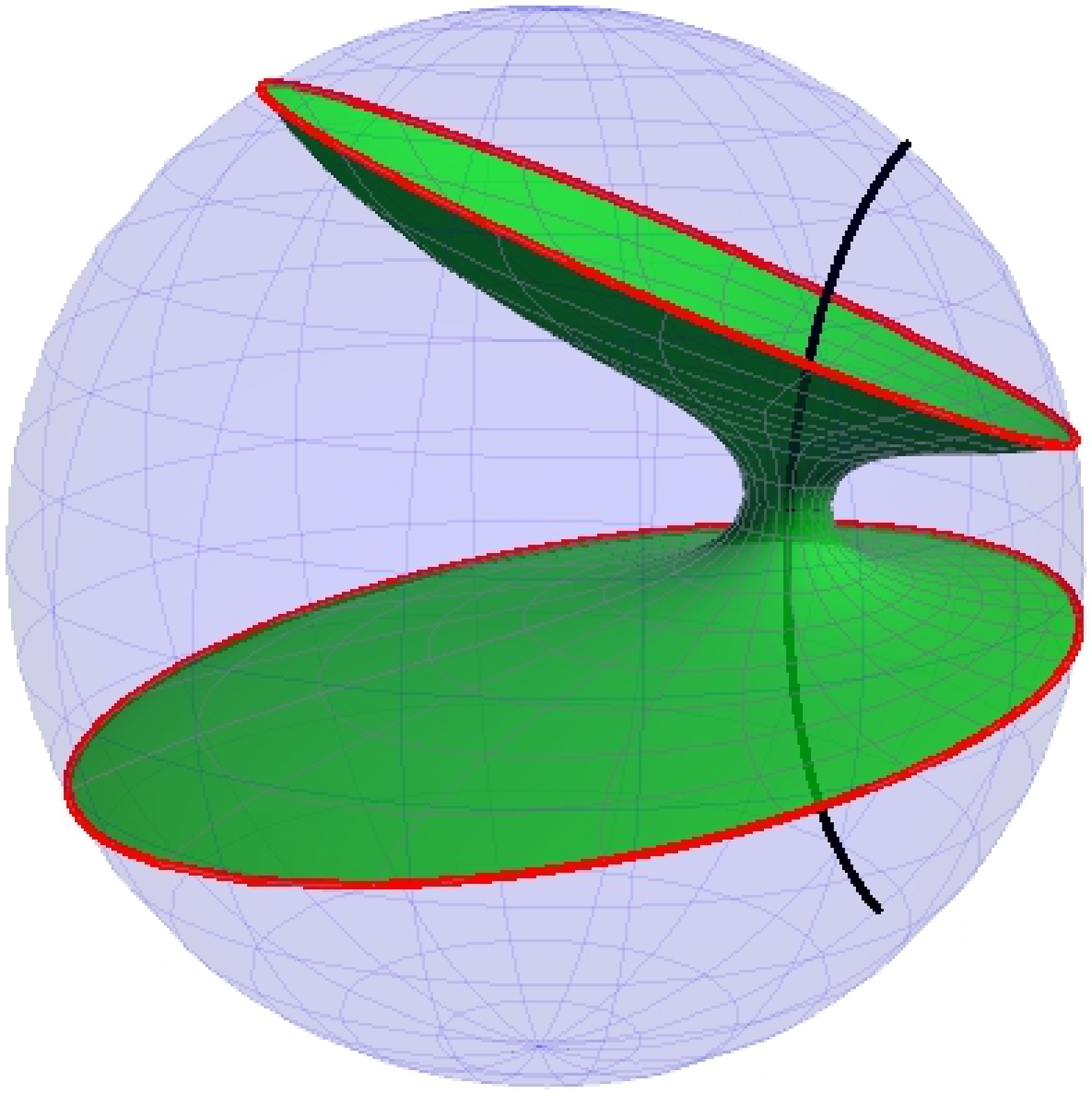}\hspace*{-2pt}%
\includegraphics[width=2in]{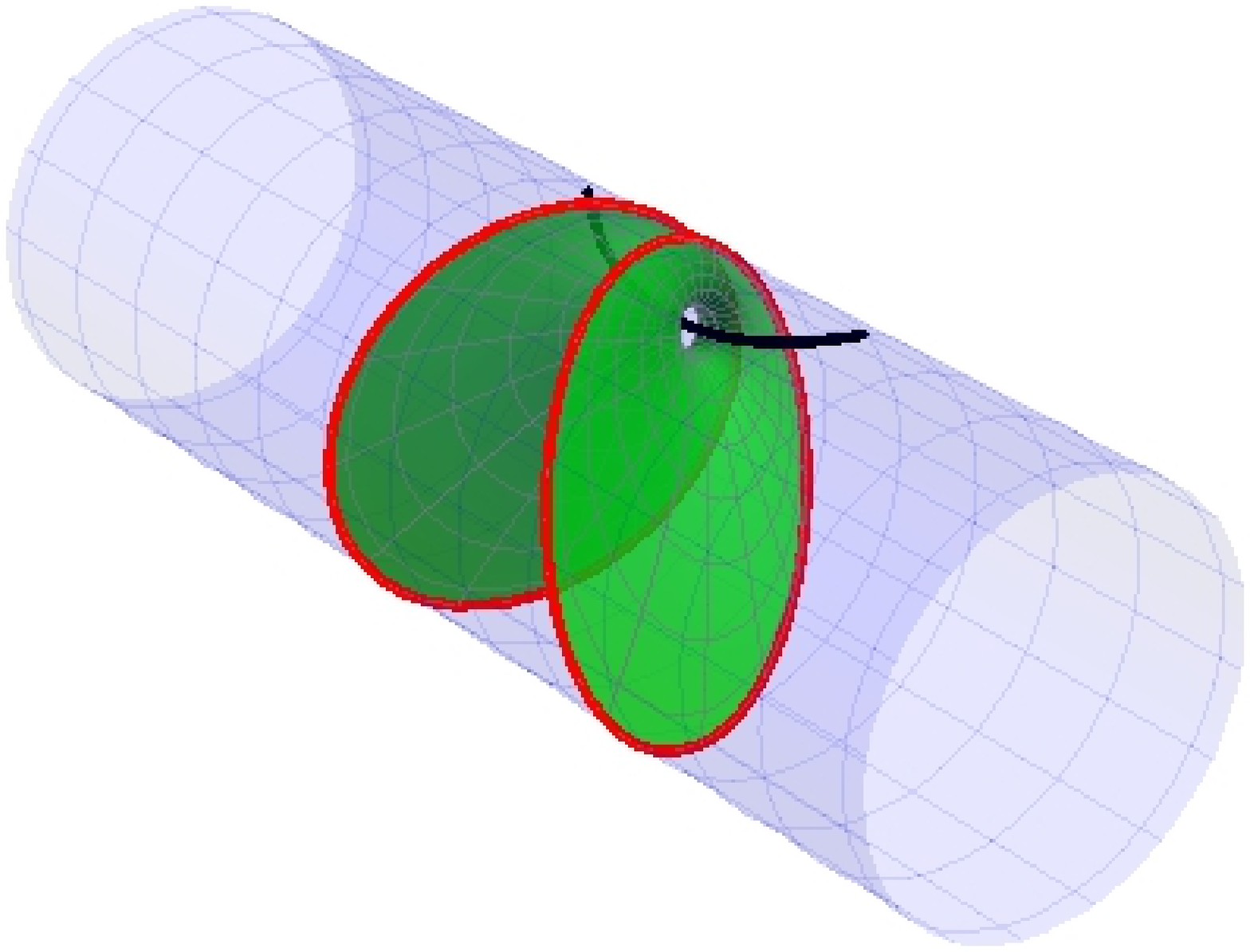}\hspace*{-2pt}%
\includegraphics[width=2in]{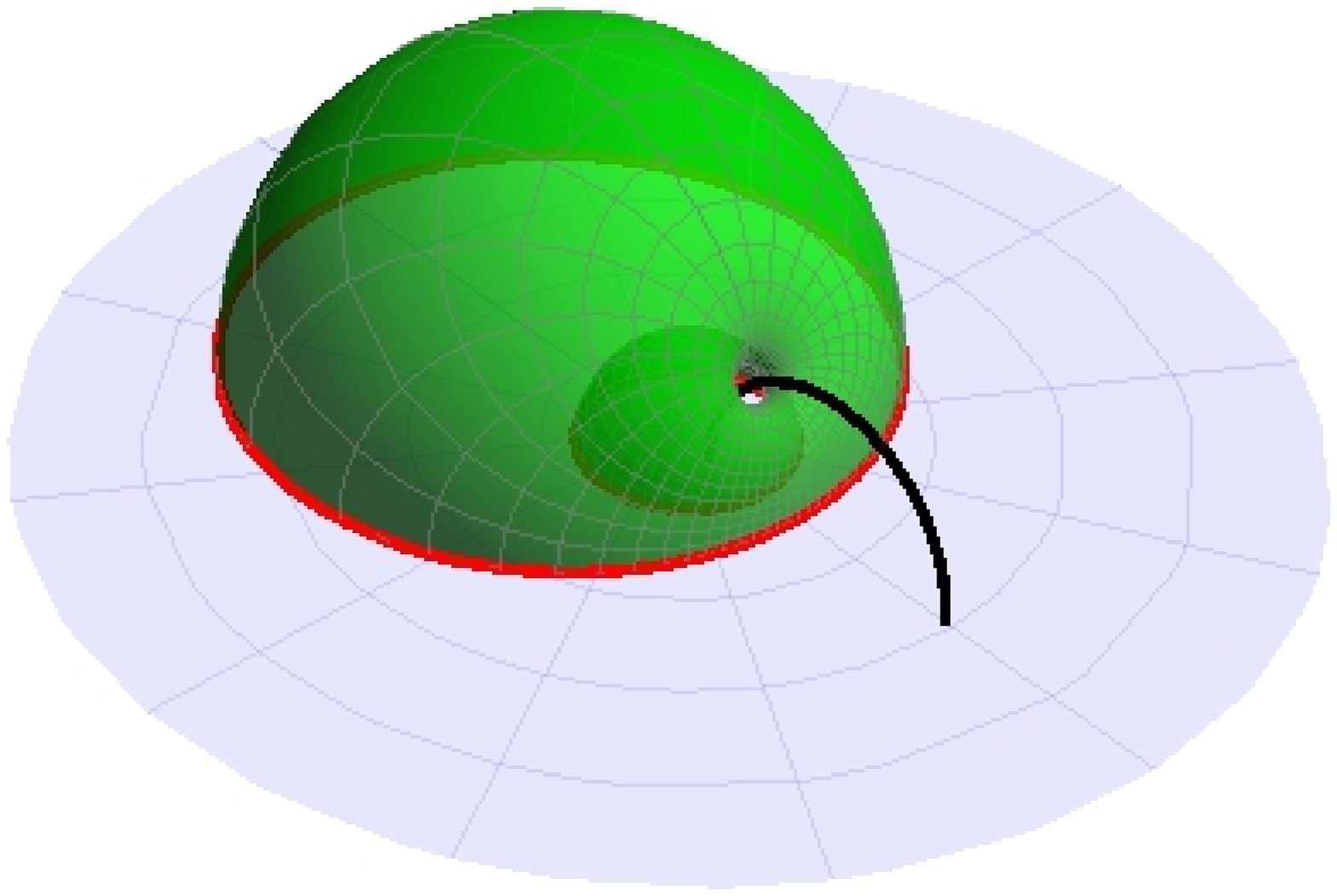}%
\\[-3ex]
\includegraphics[width=2in]{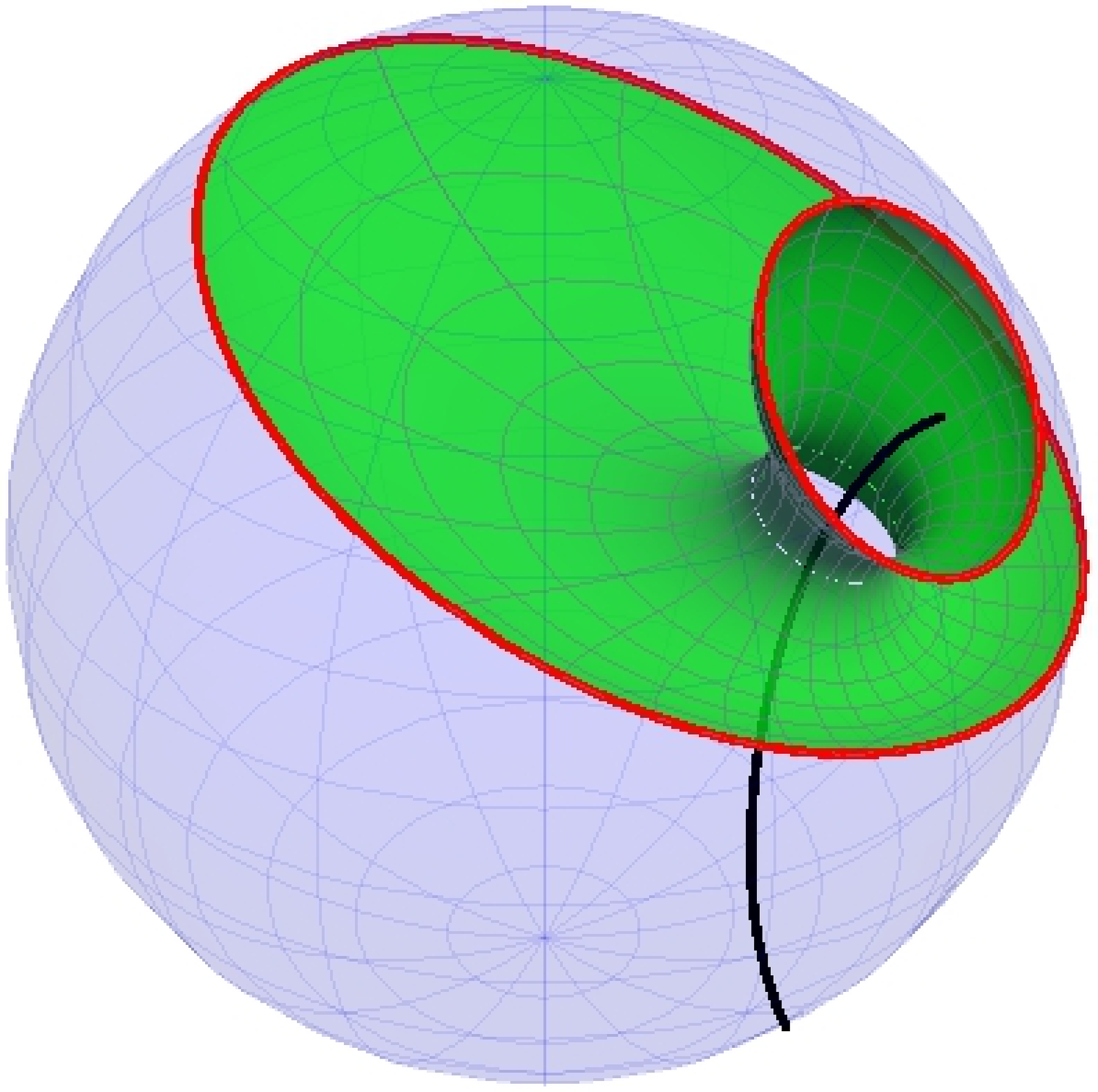}\hspace*{-2pt}%
\includegraphics[width=2in]{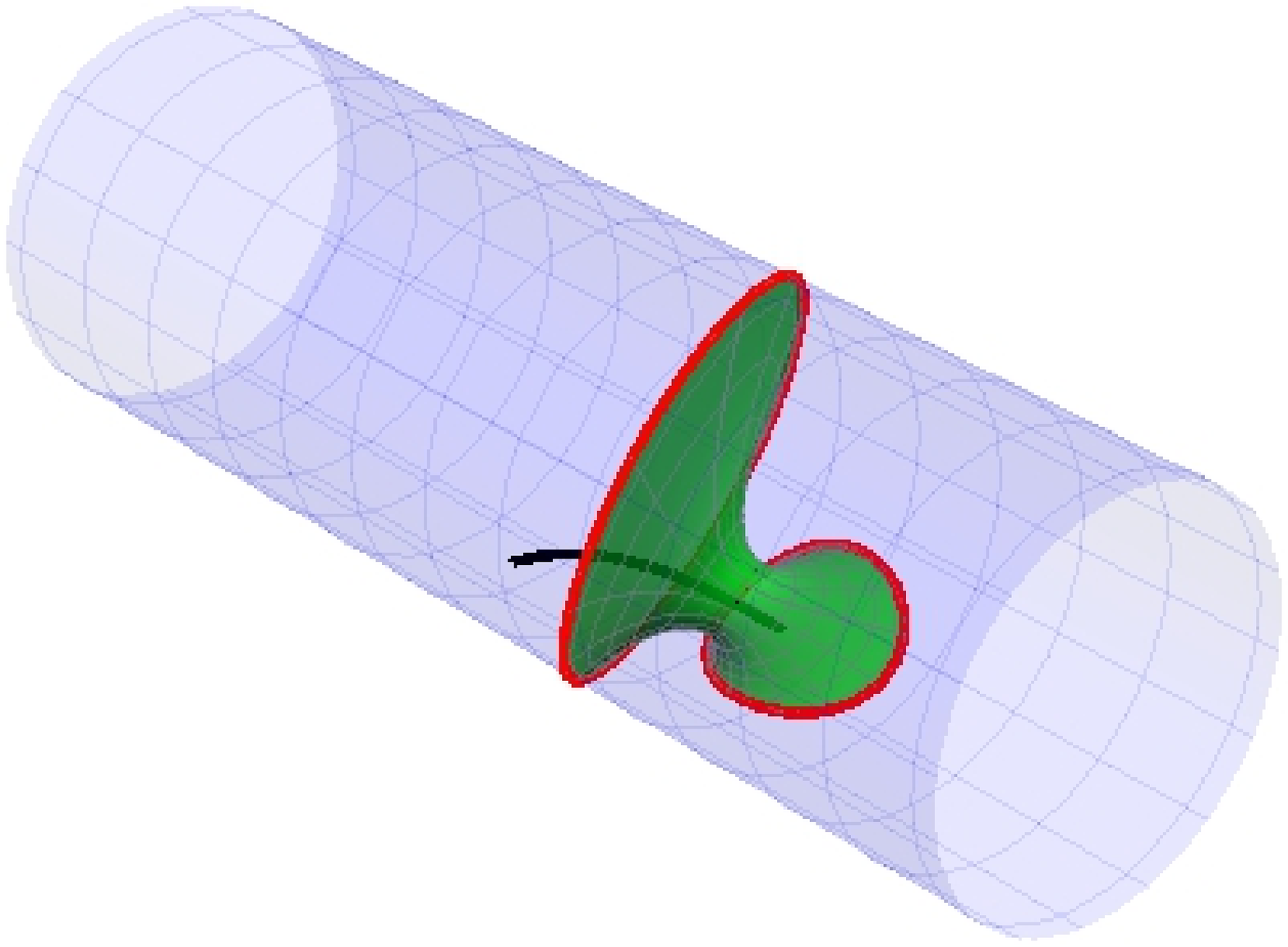}\hspace*{-2pt}%
\includegraphics[width=2in]{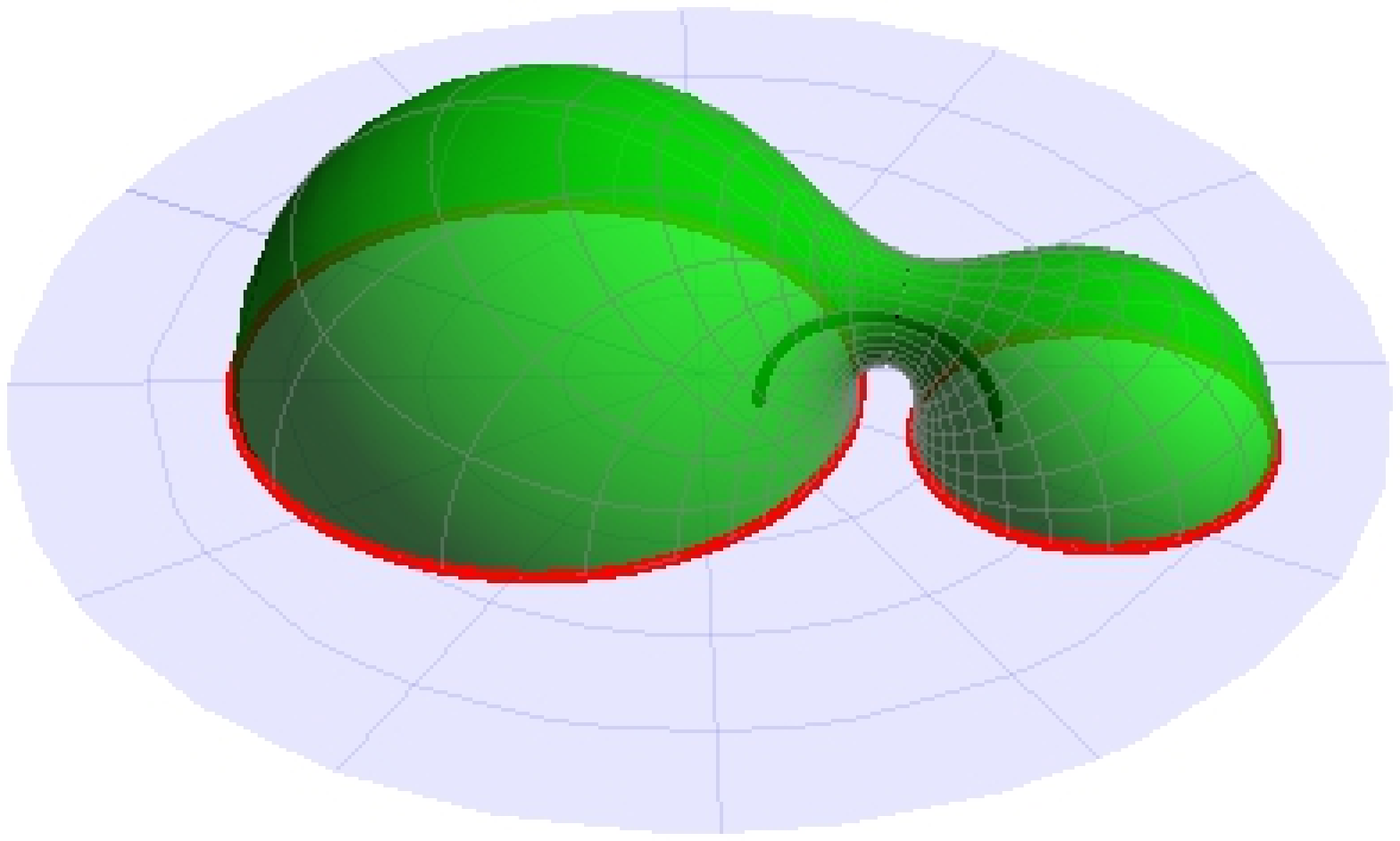}%
\\[-2ex]
{\small(a)\hspace{130pt}(b)\hspace{130pt}(c)}%
\end{center}
\vspace{-3ex}
\caption{\label{fig:msgen}\small
\textbf{Minimal surfaces spanned on two generally located disjoint circles at infinity.}\quad
The minimal surface depicted in Fig.~\ref{fig:msrot} can be shifted using isometries so it reaches any two disjoint circular boundaries at infinity. Each line shows one example of such a configuration of two boundaries and visualizations of the corresponding minimal surface in (a) spherical, (b) cylindrical and (c) half-space model of Lobachevsky space.
}
\begin{center}
\includegraphics[width=2.8in]{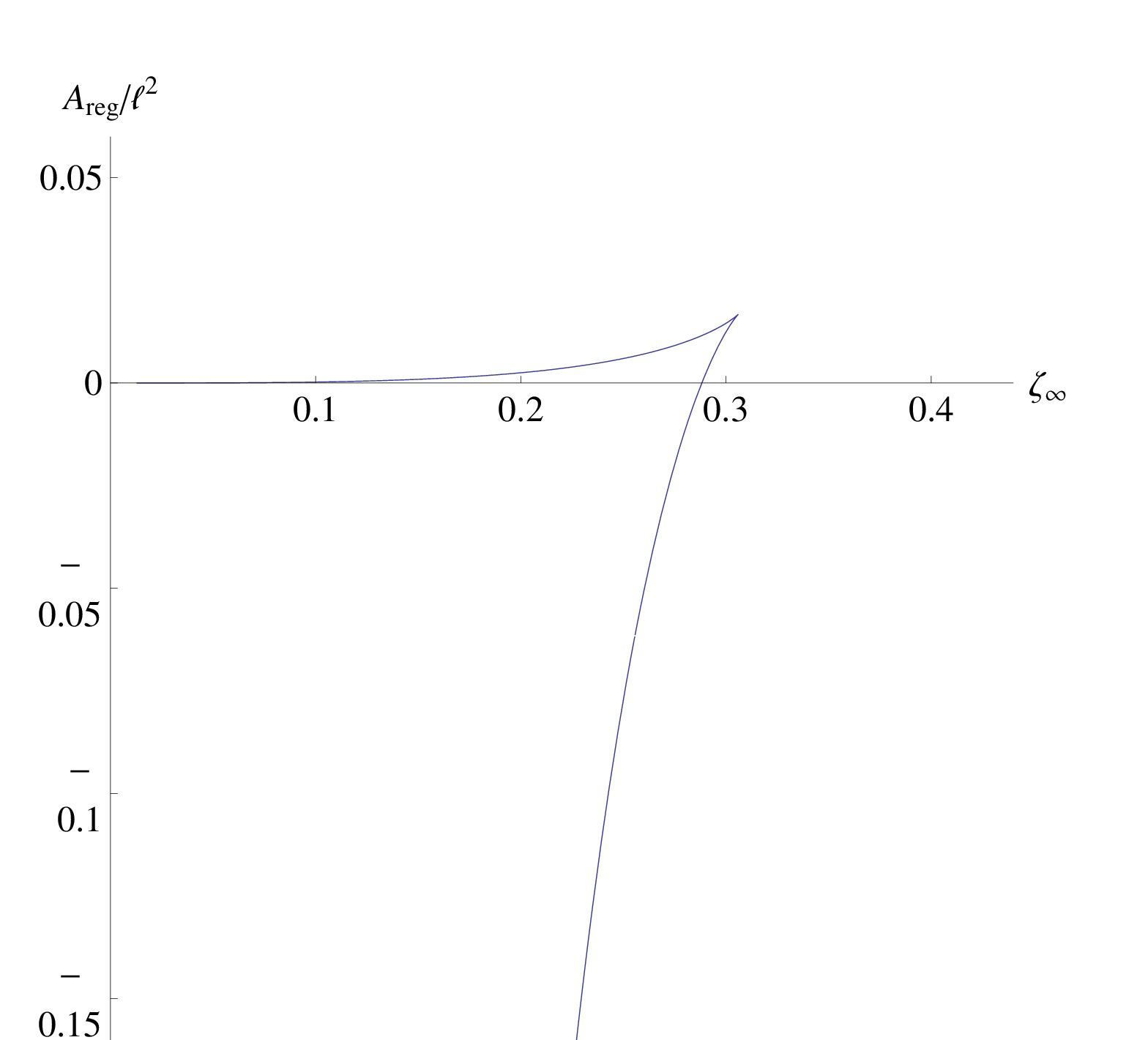}\quad%
\includegraphics[width=2.8in]{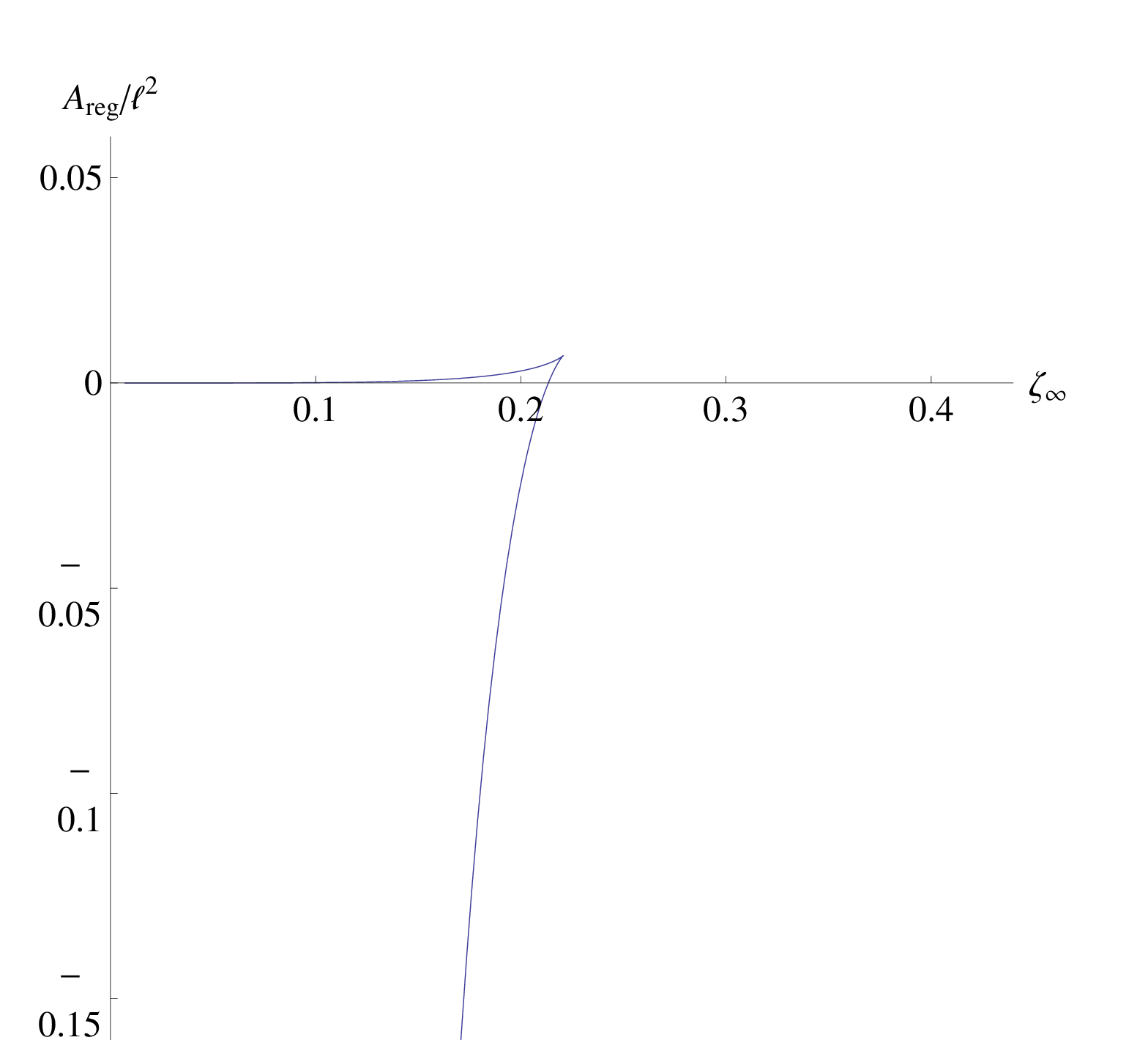}%
\\[0ex]
{\small ${D=4}$\hspace{2.5in}${D=5}$}%
\end{center}
\vspace{-3ex}
\caption{\label{fig:areaHD}\small
\textbf{Regularized area for minimal surface spanned on two disjoint circular
boundaries in higher dimensions}\quad The dependence of the regularized area on
the distance of the circular boundaries is qualitatively the same as in the
dimension ${D=3}$, cf.~Fig.~\ref{fig:arearot}.
}
\end{figure*}

Two examples of the minimal surfaces spanned on two generically positioned disjoint circles at infinity are shown in Fig.~\ref{fig:msgen}.

\subsection{Higher dimensions}

Similar analysis can be done for arbitrary higher dimension ${D}$.
Unfortunately, the integrals for the profile curve and surface area \eqref{EoMS}
and \eqref{areams} become more complicated and cannot be integrated easily in
terms of special functions. However, the numerical integrations show that the
results from the spatial dimension ${D=3}$ remain qualitatively the same in a
higher dimension. For example, Fig.~\ref{fig:areaHD} shows the graphs of the
regularized area versus the distance of boundaries (an analogue of
Fig.~\ref{fig:arearot}b) in the spatial dimensions ${D=4}$ and ${D=5}$.

\section{Minimal surfaces in the anti-de~Sitter spacetime}
\label{sc:msAdS}

\subsection{Two circular boundaries in one static region}

\begin{figure}[b]
\begin{center}
\includegraphics[width=2in]{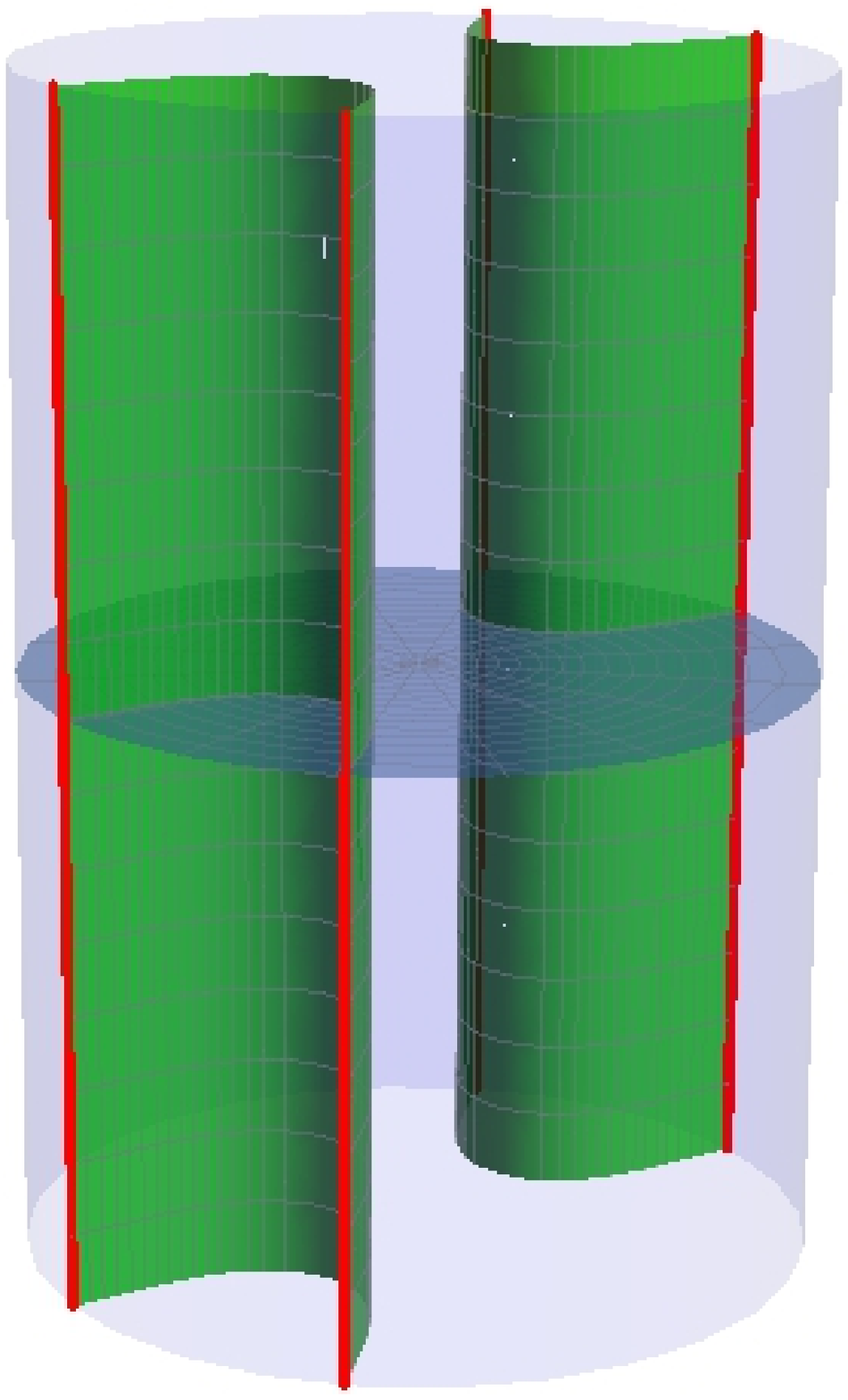}%
\includegraphics[width=2in]{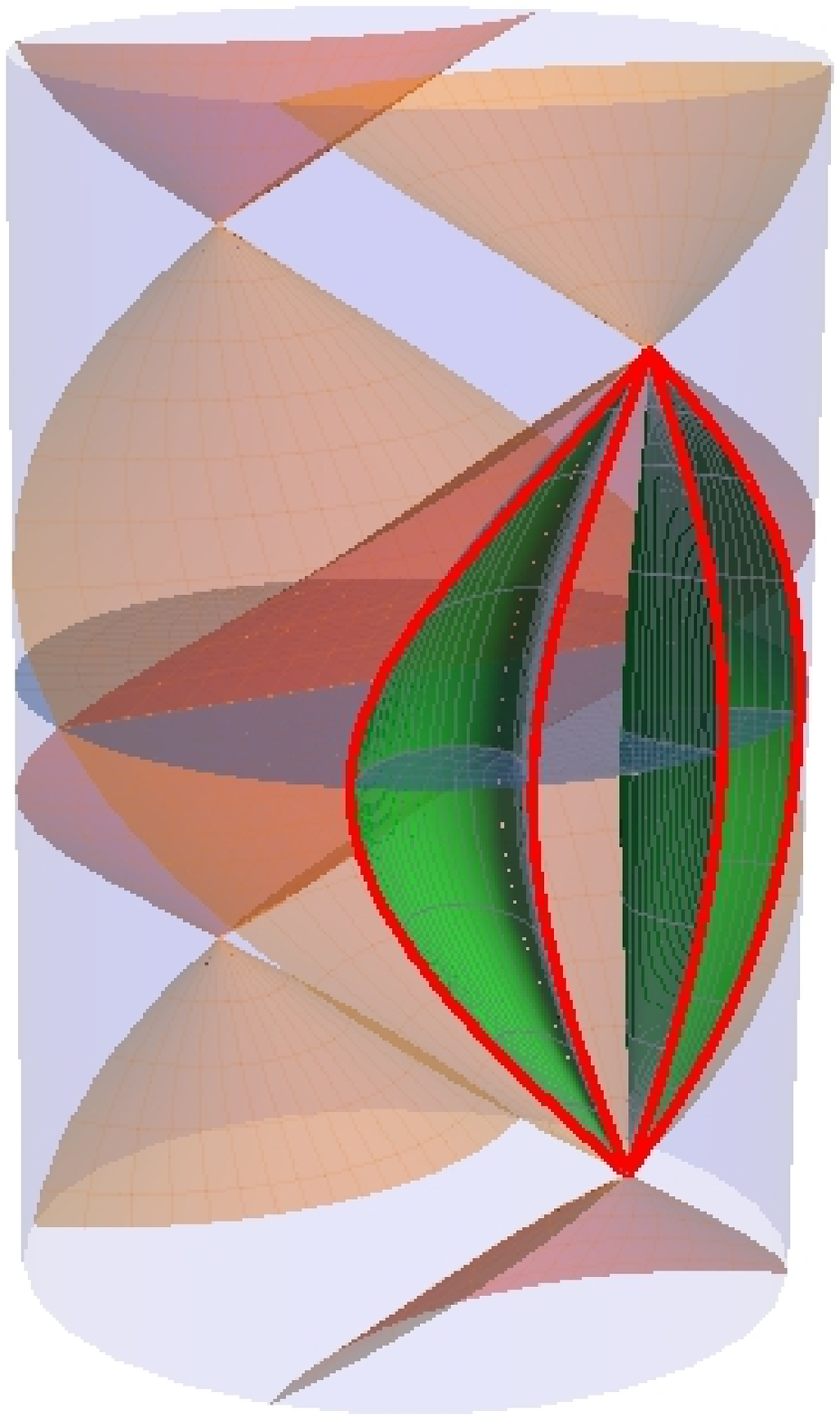}%
\includegraphics[width=2in]{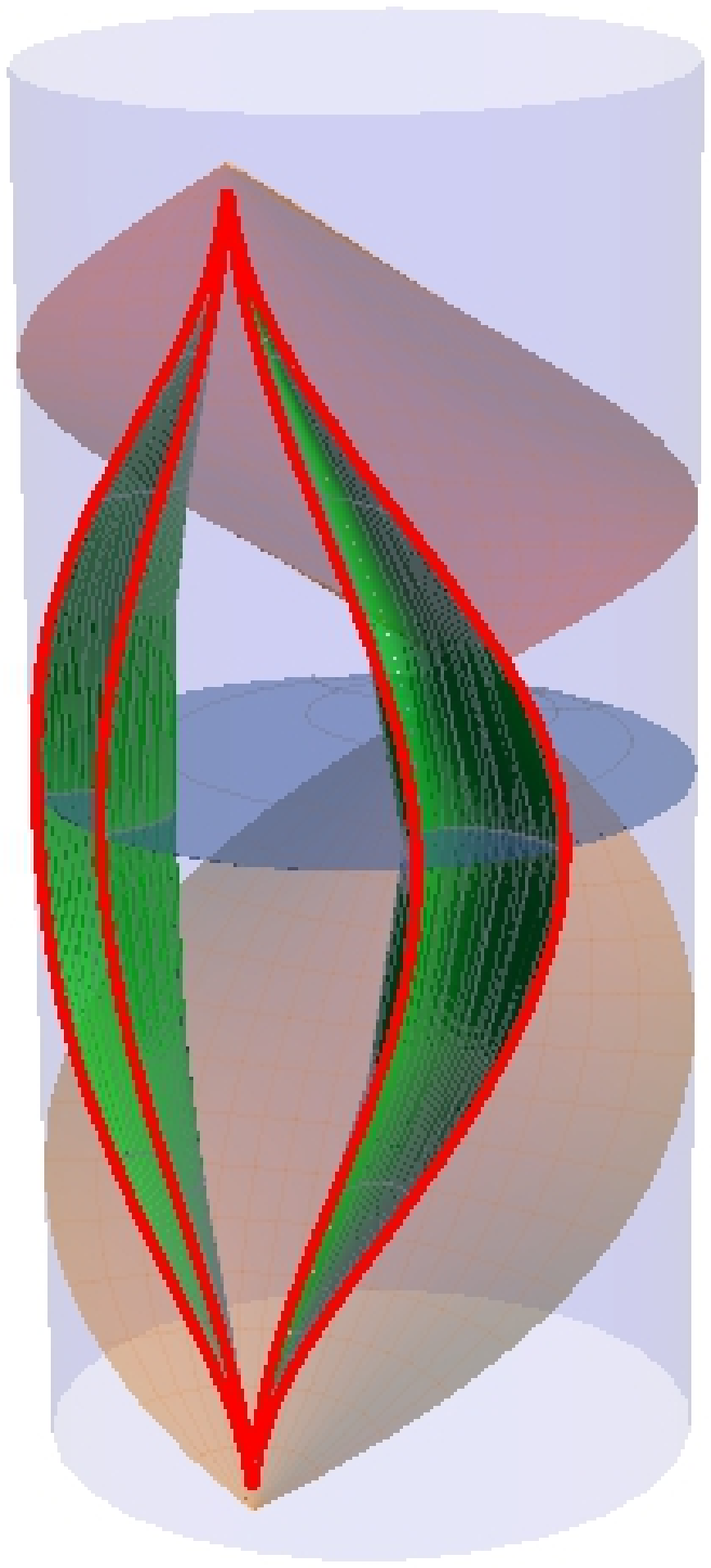}%
\\[1ex]
{\small\hspace*{6pt}(a)\hspace{132pt}(b)\hspace{132pt}(c)}%
\end{center}
\caption{\label{fig:msAdS}\small
\textbf{World-sheets of the minimal surface spanned on two disjoint circles.}\quad
Three diagrams represent three possible extension of the minimal surface into
the temporal direction using three static Killing vectors. In all three cases
both circular boundaries are located in the same static region. The horizons of
the Killing vectors and one slice of a constant static time are indicated. The
rotation-symmetric direction ${\ph}$ i suppressed in these diagrams, the
tube-like minimal surface thus splits into two disconnected pieces. For the same
reason, the world-sheet of each circular boundary is represented just by two
worldlines.  (a) For the static Killing vector of type I there is only one
static region and the minimal surface remains eternally in AdS universe. (b) The
Killing vector of type II possesses Killing horizons which divide the spacetime
into static and non-static regions. Here, both circular boundaries are located
in one static region and the minimal surface remains in this static region. (c)
Horizons of the Poincar\'{e} Killing vector divides AdS spacetime into a
sequence of static regions. Again, both circular boundaries are located in one
static region.
}
\end{figure}

Till now we have discussed minimal surfaces localized in the spatial section of the AdS space. Their area correspond to the entanglement entropy for the holographically associated system at infinity.

However, as we discussed at the beginning of Sec.~\ref{sc:msLobsp}, the spatial section with hyperbolic geometry can be understood as a time slice for three different static Killing vectors in the AdS spacetime. A different choice of the Killing vector should somehow influence the choice of the state of the system at infinity. Surprisingly, this choice does not enter the calculation of the minimal surface and of the entanglement entropy in any way.

In terms of the metric, three qualitatively different static Killing vectors differ by the lapse (red-shift) factor in front of the corresponding time element in the metric. This factor, however, does not enter the characterization of the spatial geometry.

Nevertheless, it could be instructive to visualize the whole history of the minimal surface, even although it is not given by an evolution equation. In the case when the whole boundary of the minimal surface lies at infinity of \emph{one} static region, the procedure is straightforward: the minimal surface is just prolonged along the Killing time coordinate and it spans 3-dimensional world-sheet in one static region of the AdS universe. Of course, by a different choice of the Killing vector one gets different sheets.

The world-sheets of the minimal surface spanned on two disjoint circles which
both lie in the same static region, using static Killing vectors of type I, II,
and Poincar\'{e} one, are depicted in Fig.~\ref{fig:msAdS}.

\begin{figure}[b]
\begin{center}
\includegraphics[width=2in]{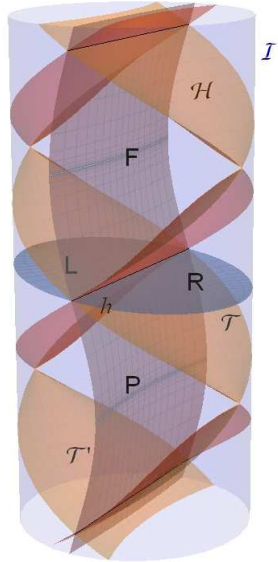}%
\end{center}
\caption{\label{fig:AdSII}\small
\textbf{Hypersurfaces ${T=\text{const}}$ of the static Killing vector of type II.}\quad
The static Killing vector of type II has a bifurcation structure. Its Killing horizons divide AdS spacetime into a sequence of pairs of static regions \textsf{R}, \textsf{L} and non-static regions \textsf{P} and \textsf{F} where the Killing vector is spacelike. The horizons ${\mathcal{H}}$ are null surfaces. Each of them corresponds to a plane of light flying through AdS universe, starting and ending in improper bifurcation points at infinity $\mathcal{I}$. Horizons intersect in bifurcation lines $h$. The Killing vector of type II can play a role of a time translation (in static regions) or of a spatial translation (in non-static regions) or of a boost (near the bifurcation lines). Time slices ${T=\text{const}}$ both in static and non-static regions are indicated. Time slices in two opposite static regions can be joined to form one global Lobachevsky space $\mathcal{T}$. Time slices in non-static regions can be all joined to form 3-dimensional AdS spacetime $\mathcal{T}'$.
}
\end{figure}

\subsection{Two circular boundaries in opposite static regions}

However, for the static Killing vector of type II we can also encounter a more interesting situation. In this case Lobachevsky space corresponds to a time slice ${T=\text{const}}$ in two separate static regions. The Killing vector of type II has a bifurcations structure and its Killing horizons divide the AdS spacetime into separate domains, cf.~Fig.~\ref{fig:AdSII}. There are pairs of static regions \textsf{R} and \textsf{L} which are positioned acausally to each other, but for which their time slices can be joined into one global Lobachevsky space.

We can thus consider circular boundaries localized symmetrically in these opposite static regions \textsf{R} and \textsf{L}. In such a case the world-sheet of the minimal surface spanned on these two circles reaches the horizons of the static regions and it  must be continued into non-static regions \textsf{F} and \textsf{P} above and under the Killing horizons. The equation \eqref{EoMS} for the profile curve of the minimal surface can be solved even in these non-static regions since it does not depend on the character of the Killing vector and signature of the metric. The solution is
\begin{equation}\label{prfcspatial}
    \zeta(P) = {\textstyle \frac{P_0}{\sqrt{1+P_0^2}\sqrt{1+2P_0^2}}
      \mathsf{\Pi}\Bigl(\arccos\frac{P}{P_0},\frac{P_0^2}{1+P_0^2},\sqrt{\frac{P_0^2}{1+2P_0^2}}\Bigr)}\;.
\end{equation}
The solution is closely related to that in static regions \eqref{rotsym-sol}, it is a different branch of the analytic continuation of \eqref{rotsym-sol}.

\begin{figure}
\begin{center}
\includegraphics[width=3in]{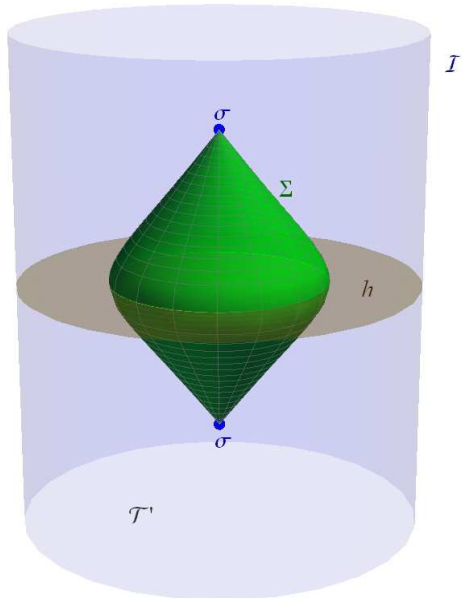}%
\end{center}
\caption{\label{fig:msAdSspT}\small
\textbf{Minimal surface in slice ${\mathcal{T}'}$.}\quad
Minimal surface $\Sigma$ located in time-like slice $\mathcal{T}'$ (joined slices ${T=\text{const}}$ in non-static regions \textsf{P} and \textsf{F}, cf.\ Fig.~\ref{fig:AdSII}) which has 3-dimensional AdS geometry. The minimal surface $\Sigma$ is composed by two symmetric pieces from the regions \textsf{P} and \textsf{F}. It is given by the profile function \eqref{prfcspatial} with parameter ${P_0}$ which characterize the maximum radius of the surface. The surface is non-smooth at two vertexes $\sigma$ on the axis. The surface approaches a null character near these vertices. The spatial intersection $h$ of the Killing horizons $\mathcal{H}$ is shown (cf.\ Fig.~\ref{fig:AdSII}). Another representation of the surface $\Sigma$ (with added ${T}$-direction and suppressed ${\ph}$-direction) is depicted in Fig.~\ref{fig:msAdSII}.
}
\end{figure}

The resulting surface located in the non-static regions \textsf{F} and \textsf{P} can be viewed if restricted into slice ${T=\text{const}}$. Such a slice is actually the 3-dimensional AdS spacetime, cf.~Fig.~\ref{fig:AdSII}. Embedding of the minimal surface into this slice is shown in Fig.~\ref{fig:msAdSspT}. It has a surprising feature that it is not smooth at the vertexes located on the axis.

The world-sheet of the minimal surface could be understood as a collection of trajectories which start from the bottom vertex with the speed of light, slow down, and eventually are spherically collapsing at the top vertex, again with the speed of light. When one add the Killing vector direction, the full surface in the the future non-static region~\textsf{F} has a structure of collapsing cylinder which degenerate along a spatial line (the vertex prolonged for all values of spatial coordinate ${T}$).

\begin{figure}
\begin{center}
\includegraphics[width=3in]{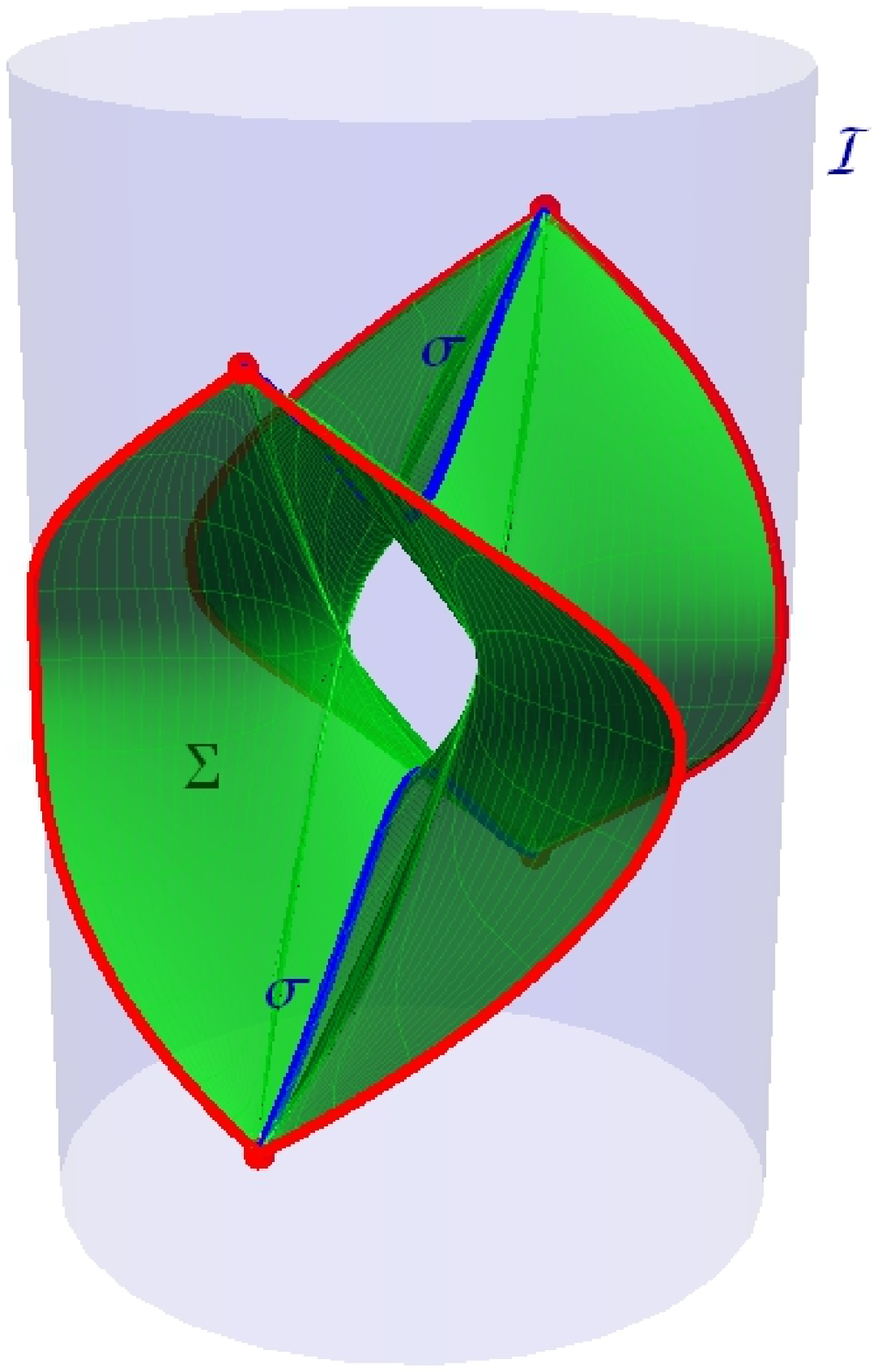}%
\includegraphics[width=3in]{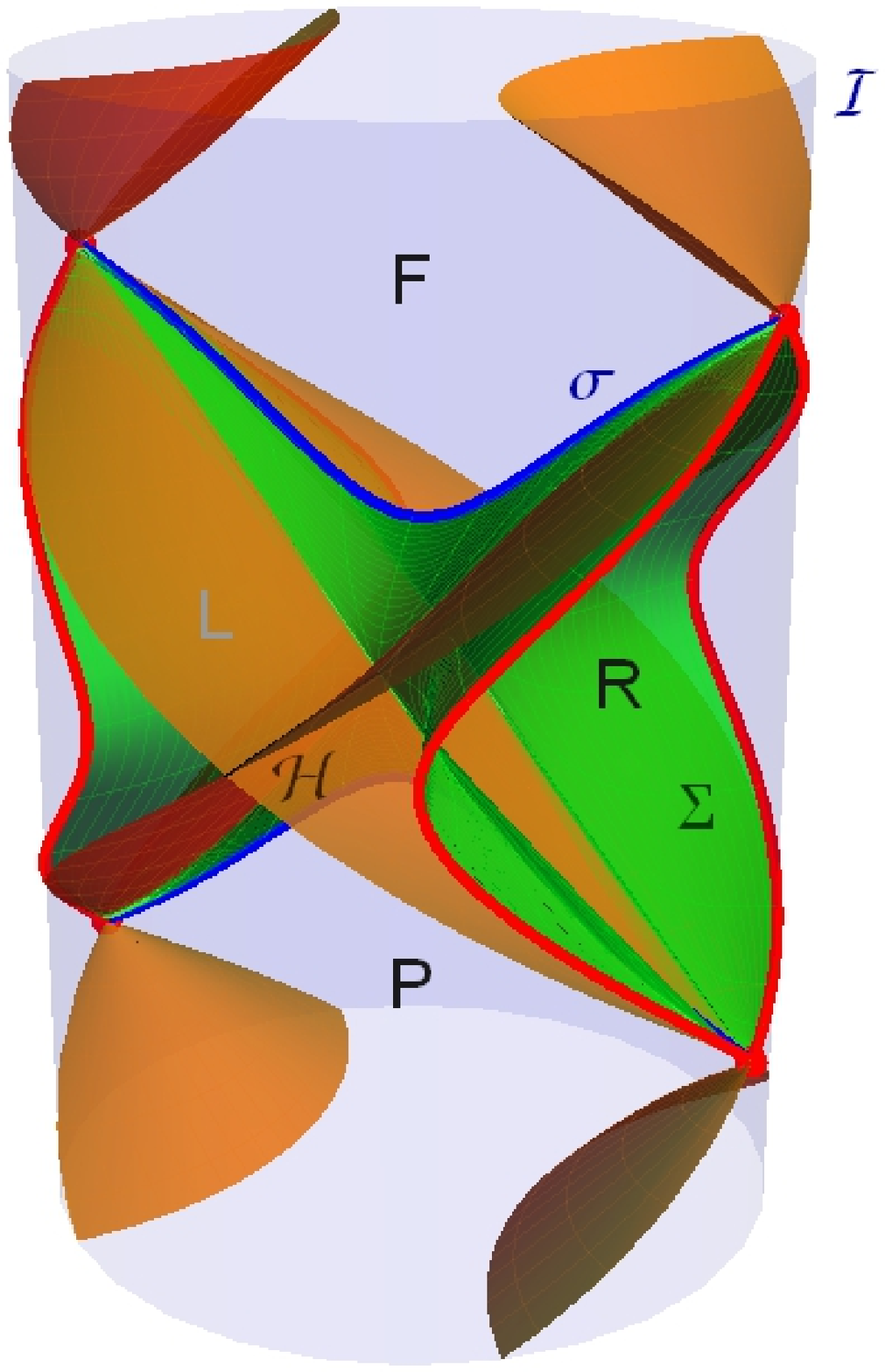}%
\end{center}
\caption{\label{fig:msAdSII}\small
\textbf{World-sheet $\Sigma$ of the minimal surface spanned on two oppositely accelerated circles.}\quad
Diagrams show two views of the same world-sheet. The Killing horizons $\mathcal{H}$ of the static Killing vector of type II are included in the right diagram. The whole world-sheet is obtained by joining pieces located in the static regions \textsf{L}, \textsf{R} (where they are given by the profile function \eqref{rotsym-sol}) and pieces in the non-static regions \textsf{P} and \textsf{F} (where they are given by the profile function \eqref{prfcspatial}). It reaches infinity $\mathcal{I}$ in two circular boundaries located in the opposite static regions \textsf{L} and \textsf{R}. In these diagrams, each boundary is represented by a pair of worldlines.
The world-sheet $\Sigma$ is singular at two spatial lines $\sigma$. The future singular line can be interpreted as a `history' of the rapture of the minimal surface, see discussion in the text.
}
\end{figure}

However the part of the surfaces in the region \textsf{F} also reaches the Killing horizons (for ${T\to\pm\infty}$). Here it has to be joined with the surface in static regions \textsf{R} and \textsf{L}. Similarly the surface located in the past static region \textsf{P} joins the surface in the static regions from below. The complete minimal surface is depicted in Fig.~\ref{fig:msAdSII}. Here, the rotation-symmetric direction ${\ph}$ is suppressed. We see that the surface is indeed singular along two spatial lines in the \textsf{P} and \textsf{F} regions. These singular lines reach infinity of AdS spacetime at points where the Killing vector is bifurcating.

One could try to interpret the world-sheet of the minimal surface as a dynamical process (although we repeat that the surface is not governed by an evolutionary equation, but by the static equation at one time slice). First, we consider hyperplanes in the bulk which correspond to the circular boundaries at infinity. Since they are static in the sense of the Killing vector ${\cv{T}}$, these hyperplanes move with the acceleration larger then ${1/\ell}$. They are coming from infinity towards each other, decelerating from the speed of light to the zero velocity and accelerating back to infinity asymptotically approaching the speed of light. The points at infinity from which the hyperplanes start and where they end are those infinite bifurcation points of the Killing vector ${\cv{T}}$. In this sense we can speak about two accelerating circular boundaries at infinity. Clearly, the world-sheets of these boundaries are not smooth at the infinite bifurcation points.

We can now look at the world-sheet of the minimal surface spanned between these
accelerated circles in terms of the global cosmological time ${\tau}$. Its time
slices correspond to horizontal planes in Fig.~\ref{fig:msAdSII}. Starting in
the middle of the surface (at the closest approach of the circles) the minimal
surface has exactly the shape depicted in Fig.~\ref{fig:msrot}. After that the
circular boundaries are accelerating away from each other and the minimal
surface starts to deform. When the time-slice ${\tau=\text{const}}$ reaches the
top singular line of the world-sheet, the minimal surface tears into two pieces.
At later times, these two pieces are still attached to the circular boundaries
at infinity and they fly from each other. On the other side they are terminated
by the singular vertexes which describe the place where the minimal surface was
torn. These vertexes flies from each other with a superluminal speed (along the
spacelike singular lines ${\sigma}$).
This view corresponds to the earlier observation that the minimal surface can join two disjoint circles at infinity only if they are closer than the critical distance ${s_\crit}$. For the circles accelerating from each other the minimal surface thus cannot exist when they get too far.

However, one should be cautious with such an interpretation since we are mixing here the static picture with respect to one Killing vector with the description in terms of time of another Killing vector. Also, we should remember that the world-sheet of the minimal surface is not a world-sheet of a causally evolving matter.

\section{Summary}
\label{sc:summary}

We found out exact solutions for all types of minimal surfaces spanned on one or two
spherical boundaries at conformal infinity. The relative positions and the sizes of
these spherical boundaries are considered to be arbitrary. The
Ryu-Takayanagi holographic conjecture \eqref{Ryu-Takayanagi} enables us to relate
the areas of minimal surfaces in the bulk of AdS with the entanglement entropy of
any two generally positioned spherical domains at infinity.
There are three qualitatively different cases of mutual
positions
of the spherical domains: (i) two disjoint domains,
(ii) overlapping domains, and (iii) touching domains. In the first case there exist
tube-like minimal surfaces joining the boundaries of these domains.
In this interesting
case we showed that for boundaries
closer than ${s_\mx}$ there are three possible minimal surfaces,
which corresponds to three possibilities (phases) for the holographic
entanglement entropy in CFT.
The transition between these phases occurs
at the critical distance ${s=s_\crit}$, when the area of the tube-like surface
starts to exceed the area of the trivial solution of two hyperplanes.
Thus even in the pure pure AdS background there is a critical
behavior of the entanglement entropy that was
demonstrated \cite{Klebanov:2007ws} for the asymptotically AdS spacetimes with
a black hole in the bulk.

If the entanglement entropy for disjoint subsystems is given
by the area of the absolute minimal surface\footnote{%
See \cite{Hubeny:2007re} for alternative proposals.}
then the renormalized area \eqref{rotsym-regarea} is directly related to the
mutual information
${I(\Omega_1,\Omega_2)=S_{\Omega_1}+S_{\Omega_2}-S_{\Omega_1\cup\Omega_2}}$
which quantifies correlations between the disjoint subsystems. Indeed, since the
entanglement entropy ${S_\Omega}$ of a single spherical domain ${\Omega}$ is
given by the area ${A_\plane}$ of the trivial hyperplane
boundary ${\partial\Omega}$, the renormalized area ${A_\ren}$ of the tube joining
the boundaries of two such domains gives directly the mutual information
${I(\Omega_1,\Omega_2)}$, provided
that the tube does give the minimal area, i.e., for ${s<s_\crit}$.

Although the entanglement entropy changes continuously with the distance between
the
boundaries at ${s=s_\crit}$, the corresponding minimal surface changes
discontinuously.
To see the transition from the trivial phase
to the tube-like phase, one would have to start with two very close
hyperplanes.
At a point, where they almost touch, a very deep tube-like surface can appear.
Thought the topology of the surface changes it does not change the total area
of the surfaces. While we
increase the distance between the boundaries, the tube grows wider.
It follows
the upper branch of the curve in Fig.~\ref{fig:arearot}b and
Fig.~\ref{fig:areaHD} up to the maximal possible distance ${s_\mx}$ of the
boundaries. This branch corresponds to the locally minimal surface, but it's
not an absolute minimum because there is another solution for a tube-like
minimal surface with the same boundaries but lesser area.
When one reaches the ${s_\mx}$ and starts to decrease the distance between the
boundaries the tube grows even wider (following the lower branch in
Fig.~\ref{fig:arearot}b and Fig.~\ref{fig:areaHD}). After decreasing the
distance under
${s_\crit}$ one obtains the physical tube-like phase.

In addition to the case of two spherical domains one can investigate even more
complicated situations, for example, a set of spherical domains
${\Omega_i}$,
each of them being a subdomain of all the subsequent ones:
${\Omega_i\subset\Omega_j}$ for ${i<j}$. They may not be all
simultaneously concentric. The circular boundaries of these domains correspond
to ultraparallel hyperplanes in the bulk.
For such a configuration we know the minimal surfaces for any pair of the
boundaries. Employing \eqref{Ryu-Takayanagi} we find that the renormalized
entropy
depends only on the distance between the boundaries, cf.~\eqref{rotsym-bound},
\eqref{rotsym-area}. We can thus test properties of the entropy for domains
obtained by a combination of several subdomains. Namely, one can check
the strong subadditivity inequalities to find that they are satisfied, as
expected from general considerations \cite{Hirata:2006jx}.
Similarly, one can study systems of strips between several semicircles joined
at the same poles.

To summarize, the obtained exact analytical solutions for minimal
surfaces in AdS provide us with a classical geometric tool of probing quantum
properties of CFT.

The holographic entanglement entropy can be applied to testing phase
transitions in QFT, similar to the confinement/deconfinement phase
transition at a finite temperature \cite{Klebanov:2007ws,Lewkowycz:2012mw}.
It can useful in generalizations of c-theorems in higher dimensions
\cite{Casini:2012ei,Myers:2010tj}. One can use the properties of
the entanglement entropy the other way around and even to `derive'
gravitational dynamics from
entanglement \cite{Lashkari:2013koa,Faulkner:2013ica}.

\acknowledgments
P.~K. was supported by Grant GA\v{C}R 14-37086G
and appreciates the hospitality of the Theoretical
Physics Institute of the University of Alberta.
A.~Z. thanks the Natural Sciences and Engineering
Research Council of Canada and the Killam Trust for
the financial support and appreciates the hospitality
and support of the Institute of Theoretical Physics
of the Charles University in Prague.


\appendix

\section{Coordinates in Lobachevsky space}
\label{apx:Lobsp}

The geometry of the hyperbolic space in spherical coordinates is given by the metric \eqref{Lobmtrcsph}, i.e.,
\begin{equation}\label{Lobmtrcsph-apx}
    \frac1{\ell^2}\,
    \tens{g}_{\Lob} = \grad r^2 + \sh^2\!r\, \bigl( \grad\tht^2+\sin^2\!\tht\,\grad\ph^2\bigr)\;.
\end{equation}
Here, ${r}$ is the radial distance from the origin. It can be redefined
\begin{equation}\label{rchirel}
   \sh r = \tan\chi
\end{equation}
to obtain the metric conformally related to the homogeneous metric on the hemisphere,
\begin{equation}\label{LobMconfsph}
    \frac1{\ell^2}\,
    \tens{g}_{\Lob} = \frac1{\cos^2\!\chi}
      \Bigl(\grad\chi^2 + \sin^2\!\chi\, \bigl( \grad\tht^2+\sin^2\!\tht\,\grad\ph^2\bigr)\Bigr)\;.
\end{equation}
The boundary ${\chi=\frac\pi2}$ of the hemisphere corresponds to the conformal infinity of the hyperbolic space.

The spherical coordinates ${\chi,\tht,\ph}$ on the hemisphere can be replaced by other spherical coordinates ${\bar\chi,\bar\tht,\ph}$ around a new pole on the equator of the hemisphere,
\begin{equation}\label{chithtbarrel}
\begin{aligned}
    \cos\chi&=\sin\bar\chi\cos\bar\tht\;,&    \cos\bar\chi&=-\sin\chi\tht\;,\\
    \tan\tht&=-\tan\bar\tht\sin\bar\tht\;,&   \tan\bar\tht&=\tan\chi\sin\tht\;.
\end{aligned}
\end{equation}
The Lobachevsky metric becomes:
\begin{equation}\label{LobMconfsphbar}
    \frac1{\ell^2}\,
    \tens{g}_{\Lob} = \frac1{\sin^2\!\bar\chi\cos^2\bar\tht}
      \Bigl(\grad\bar\chi^2 + \sin^2\!\bar\chi\, \bigl( \grad\bar\tht^2+\sin^2\!\bar\tht\,\grad\ph^2\bigr)\Bigr)\;.
\end{equation}
These coordinates are related to the cylindric coordinates by a redefinition of the coordinate~${\bar\chi}$,
\begin{equation}\label{barchizetarel}
    \tanh\zeta=-\cos\bar\chi\;.
\end{equation}
This new coordinate ${\zeta}$ is the Killing coordinate corresponding to the translation in the hyperbolic space along the axis ${\bar\tht=0}$. One can introduce several variants of the axial coordinate which are related as
\begin{equation}\label{axialcoorrel}
\begin{gathered}
    \tan\bar\tht=\sh\rho=P=\sqrt{Z^2-1}\;,\\
    \cos^{\!-1}\bar\tht=\ch\rho=\sqrt{1+P^2}=Z\;.  
\end{gathered}
\end{equation}
The metric takes forms \eqref{Lobmtrccyl}, \eqref{LobmtrccylP}, and \eqref{LobmtrccylZ}, respectively.

Introducing yet another ``radial'' coordinate ${\bar r}$ in the metric \eqref{LobMconfsphbar},
\begin{equation}\label{zetachirbar}
    \rP = \tan\frac{\bar\chi}{2} = \exp\zeta\;,
\end{equation}
one obtains a conformally flat form of the Lobachevsky metric
\begin{equation}\label{LobMconfflat}
    \frac1{\ell^2}\,
    \tens{g}_{\Lob} = \frac1{\rP^2\cos^2\bar\tht}
      \Bigl(\grad\rP^2 + \rP^2\, \bigl( \grad\bar\tht^2+\sin^2\!\bar\tht\,\grad\ph^2\bigr)\Bigr)\;.
\end{equation}
Introducing the ``Cartesian'' coordinates ${\xP,\yP,\zP}$
\begin{equation}\label{xyzrPrel}
    \xP = \rP \sin\bar\tht\,\cos\ph\;,\quad
    \zP = \rP \sin\bar\tht\,\sin\ph\;,\quad
    \zP = \rP \cos\bar\tht\;.
\end{equation}
for the conformally related flat metric, one obtains the Poincar\'{e}
coordinates on the hyperbolic space with the metric \eqref{LobmtrcPoinc}.

Finally, if we introduce the coordinate ${\bar\zeta}$ measuring the distance along the ${\zP}$-direction,
\begin{equation}\label{zPzetaPrel}
    \zP = \exp\bar\zeta\;,
\end{equation}
the metric reads
\begin{equation}\label{LobMPoinczeta}
    \frac1{\ell^2}\,\tens{g}_{\Lob} = \grad{\bar\zeta}^2 + e^{-2\bar\zeta} \bigl( \grad\xP^2+\grad\yP^2\bigr)\;.
\end{equation}

The coordinates ${\ph}$, ${\zeta}$, and ${\xP}$, ${\yP}$ are Killing coordinates
corresponding to rotational, translational and horocyclic symmetries,
respectively.

\section{Coordinates in anti-de~Sitter spacetime}
\label{apx:AdS}

The anti-de~Sitter spacetime is described in global cosmological coordinates ${\tau,\,r,\,\tht,\,\ph}$ by the metric \eqref{AdSmtrc},
\begin{equation}\label{AdSmtrc-apx}
    \tens{g}_{\AdS} = \ell^2\,\bigl(-\ch^2\!r\,\grad\tau^2+\grad r^2+
    \sh^2\!r\,(\grad\tht^2+\sin^2\!\tht\,\grad\ph^2)\bigr)\;.
\end{equation}
The Killing vector ${\cv{\tau}}$ represents the global translation symmetry in temporal direction. Orbits of this vectors represent uniformly accelerated static observers with the acceleration smaller than the cosmological acceleration ${1/\ell}$, cf.\ Fig.~\ref{fig:KVAdS}a.

One can introduce another coordinates ${T,R,\bar\tht,\ph}$---the static coordinates of type II---associated with the uniformly accelerated static observers with acceleration larger than ${1/\ell}$. In these coordinates the AdS metric reads
\begin{equation}\label{AdSmtrcII}
    \tens{g}_{\AdS} = \frac{\ell^2}{R^2\cos^2\bar\tht}\,\biggl(
    -\bigl({\textstyle 1-\frac{R^2}{\ell^2}}\bigr)\grad T^2+ \bigl({\textstyle1-\frac{R^2}{\ell^2}}\bigr)^{\!\!-1}\grad R^2
    + R^2\bigl(\grad\bar\tht^2 +\sin^2\bar\tht\,\grad\ph^2\bigl)\biggr)\;.
\end{equation}
The Killing vector ${\cv{T}}$ is timelike in domains ${R^2<\ell^2}$. It has a bifurcation character and its orbits are visualized in Fig.~\ref{fig:KVAdS}b. The spatial section ${T=\text{const}}$, ${R^2<\ell^2}$ possesses the spatial metric
\begin{equation}\label{LobmtrcII}
    \tens{g}_{\Lob} = \frac{\ell^2}{\cos^2\bar\tht}\,\biggl(
    \frac{1}{R^2\bigl(1-\frac{R^2}{\ell^2}\bigr)}\grad R^2
    + \grad\bar\tht^2 +\sin^2\bar\tht\,\grad\ph^2\biggr)\;,
\end{equation}
which describes the geometry of the hyperbolic space. It can related to the
Lobachevsky metric \eqref{LobMconfsphbar} by
\begin{equation}\label{Rchizetarel}
    R= \sin\bar\chi=\ch^{\!-1}\zeta\;.
\end{equation}

Relations between the global cosmological coordinates ${\tau, r, \tht, \ph}$ and the static coordinates of type II ${T,R,\bar\tht,\ph}$ can be split into two steps. First, at the spatial section ${\tau=\text{const}}$ one introduces the conformally spherical coordinates ${\chi,\tht,\ph}$ and the rotated coordinates ${\bar\chi,\bar\tht,\ph}$ by the relations \eqref{rchirel} and \eqref{chithtbarrel}. In the second step, one mixes ${\tau}$--${\bar\chi}$ sector introducing the coordinate ${T}$ and ${R}$,
\begin{equation}\label{tauchiTRrel}
    \frac{R}{\ell} = \frac{\sin\bar\chi}{\sin\tau}\;,\quad
    \frac{T}{\ell} = \frac12\log\biggl|\frac{\cos\tau-\cos\bar\chi}{\cos\tau+\cos\bar\chi}\biggr|\;.
\end{equation}

Another well-known coordinates on the AdS spacetime are the Poincar\'{e}
coordinates in which the metric takes the conformally flat form
\begin{equation}\label{AdSmtrcPoinc}
    \tens{g}_{\AdS} = \frac{\ell^2}{\zP^2}\Bigl(-\grad\tP^2+\grad\xP^2+\grad\yP^2+\grad\zP^2\Bigr)\;.
\end{equation}
If we introduce the spherical Poincar\'{e} coordinates ${\tP,\rP,\bar\tht,\ph}$
by relations~\eqref{xyzrPrel}, one can relate the Poincar\'{e} coordinates to
the coordinates ${\tau,\bar\chi,\bar\tht,\ph}$ as
\begin{equation}\label{tauchitrrel}
\begin{aligned}
    \tP &= \frac{\ell\cos\tau}{\cos\tau+\cos\bar\chi} \;,\quad&
    \tan\tau &= \frac{2\ell\tP}{\ell^2-\tP^2+\rP^2}\;,\\
    \rP &= \frac{\ell\sin\bar\chi}{\cos\tau+\cos\bar\chi} \;,\quad&
    \tan\bar\chi &= \frac{2\ell\rP}{\ell^2+\tP^2-\rP^2}\;.
\end{aligned}
\end{equation}
The Killing vector ${\cv{\tP}}$ represents static observers with the uniform acceleration ${1/\ell}$. Its orbits are shown in Fig.~\ref{fig:KVAdS}c.

\bigskip
\vfill


\begin{thebibliography}{10}

\bibitem{Bekenstein:1972tm}
J.~Bekenstein, {\it {Black holes and the second law}},  {\em Lett. Nuovo Cim.}
  {\bf 4} (1972) 737--740.

\bibitem{Bekenstein:1973ur}
J.~D. Bekenstein, {\it {Black holes and entropy}},  {\em Phys. Rev. D} {\bf 7}
  (1973) 2333--2346.

\bibitem{Bekenstein:1974ax}
J.~D. Bekenstein, {\it {Generalized second law of thermodynamics in black hole
  physics}},  {\em Phys. Rev. D} {\bf 9} (1974) 3292--3300.

\bibitem{Hawking:1971vc}
S.~Hawking, {\it {Black holes in general relativity}},  {\em Commun. Math.
  Phys.} {\bf 25} (1972) 152--166.

\bibitem{Hawking:1974rv}
S.~Hawking, {\it {Black hole explosions}},  {\em Nature} {\bf 248} (1974)
  30--31.

\bibitem{Hawking:1974sw}
S.~Hawking, {\it {Particle Creation by Black Holes}},  {\em Commun. Math.
  Phys.} {\bf 43} (1975) 199--220.

\bibitem{Bombelli:1986rw}
L.~Bombelli, R.~K. Koul, J.~Lee, and R.~D. Sorkin, {\it {A Quantum Source of
  Entropy for Black Holes}},  {\em Phys. Rev. D} {\bf 34} (1986) 373--383.

\bibitem{Frolov:1993ym}
V.~P. Frolov and I.~Novikov, {\it {Dynamical origin of the entropy of a black
  hole}},  {\em Phys.Rev.} {\bf D48} (1993) 4545--4551,
  [\href{http://xxx.lanl.gov/abs/gr-qc/9309001}{{\tt gr-qc/9309001}}].

\bibitem{Srednicki:1993im}
M.~Srednicki, {\it {Entropy and area}},  {\em Phys. Rev. Lett.} {\bf 71} (1993)
  666--669, [\href{http://xxx.lanl.gov/abs/hep-th/9303048}{{\tt
  hep-th/9303048}}].

\bibitem{Barvinsky:1994jca}
A.~Barvinsky, V.~P. Frolov, and A.~Zelnikov, {\it {Wavefunction of a Black Hole
  and the Dynamical Origin of Entropy}},  {\em Phys. Rev. D} {\bf 51} (1995)
  1741--1763, [\href{http://xxx.lanl.gov/abs/gr-qc/9404036}{{\tt
  gr-qc/9404036}}].

\bibitem{Susskind:1994sm}
L.~Susskind and J.~Uglum, {\it {Black hole entropy in canonical quantum gravity
  and superstring theory}},  {\em Phys. Rev. D} {\bf 50} (1994) 2700--2711,
  [\href{http://xxx.lanl.gov/abs/hep-th/9401070}{{\tt hep-th/9401070}}].

\bibitem{Sakharov:1967pk}
A.~Sakharov, {\it {Vacuum quantum fluctuations in curved space and the theory
  of gravitation}},  {\em Sov. Phys. Dokl.} {\bf 12} (1968) 1040--1041.

\bibitem{Jacobson:1994iw}
T.~Jacobson, {\it {Black hole entropy and induced gravity}},
  \href{http://xxx.lanl.gov/abs/gr-qc/9404039}{{\tt gr-qc/9404039}}. preprint.

\bibitem{Frolov:1996aj}
V.~P. Frolov, D.~Fursaev, and A.~Zelnikov, {\it {Statistical origin of black
  hole entropy in induced gravity}},  {\em Nucl. Phys.} {\bf B486} (1997)
  339--352, [\href{http://xxx.lanl.gov/abs/hep-th/9607104}{{\tt
  hep-th/9607104}}].

\bibitem{Frolov:1996qh}
V.~P. Frolov, D.~Fursaev, and A.~Zelnikov, {\it {Black hole statistical
  mechanics and induced gravity}},  {\em Nucl. Phys. Proc. Suppl.} {\bf 57}
  (1997) 192--196.

\bibitem{Frolov:2003ed}
V.~P. Frolov, D.~Fursaev, and A.~Zelnikov, {\it {CFT and black hole entropy in
  induced gravity}},  {\em JHEP} {\bf 0303} (2003) 038,
  [\href{http://xxx.lanl.gov/abs/hep-th/0302207}{{\tt hep-th/0302207}}].

\bibitem{Ryu:2006bv}
S.~Ryu and T.~Takayanagi, {\it {Holographic derivation of entanglement entropy
  from AdS/CFT}},  {\em Phys. Rev. Lett.} {\bf 96} (2006) 181602,
  [\href{http://xxx.lanl.gov/abs/hep-th/0603001}{{\tt hep-th/0603001}}].

\bibitem{Ryu:2006ef}
S.~Ryu and T.~Takayanagi, {\it {Aspects of Holographic Entanglement Entropy}},
  {\em JHEP} {\bf 0608} (2006) 045,
  [\href{http://xxx.lanl.gov/abs/hep-th/0605073}{{\tt hep-th/0605073}}].

\bibitem{Nishioka:2009un}
T.~Nishioka, S.~Ryu, and T.~Takayanagi, {\it {Holographic Entanglement Entropy:
  An Overview}},  {\em J. Phys.} {\bf A42} (2009) 504008,
  [\href{http://xxx.lanl.gov/abs/0905.0932}{{\tt arXiv:0905.0932}}].

\bibitem{Fursaev:2006ih}
D.~V. Fursaev, {\it {Proof of the holographic formula for entanglement
  entropy}},  {\em JHEP} {\bf 0609} (2006) 018,
  [\href{http://xxx.lanl.gov/abs/hep-th/0606184}{{\tt hep-th/0606184}}].

\bibitem{Headrick:2010zt}
M.~Headrick, {\it {Entanglement Renyi entropies in holographic theories}},
  {\em Phys. Rev. D} {\bf 82} (2010) 126010,
  [\href{http://xxx.lanl.gov/abs/1006.0047}{{\tt arXiv:1006.0047}}].

\bibitem{Lewkowycz:2013nqa}
A.~Lewkowycz and J.~Maldacena, {\it {Generalized gravitational entropy}},  {\em
  JHEP} {\bf 1308} (2013) 090, [\href{http://xxx.lanl.gov/abs/1304.4926}{{\tt
  arXiv:1304.4926}}].

\bibitem{Myers:2013lva}
R.~C. Myers, R.~Pourhasan, and M.~Smolkin, {\it {On Spacetime Entanglement}},
  {\em JHEP} {\bf 1306} (2013) 013,
  [\href{http://xxx.lanl.gov/abs/1304.2030}{{\tt arXiv:1304.2030}}].

\bibitem{Hartman:2013mia}
T.~Hartman, {\it {Entanglement Entropy at Large Central Charge}},
  \href{http://xxx.lanl.gov/abs/1303.6955}{{\tt arXiv:1303.6955}}. preprint.

\bibitem{Faulkner:2013yia}
T.~Faulkner, {\it {The Entanglement Renyi Entropies of Disjoint Intervals in
  AdS/CFT}},  \href{http://xxx.lanl.gov/abs/1303.7221}{{\tt arXiv:1303.7221}}.
  preprint.

\bibitem{Headrick:2007km}
M.~Headrick and T.~Takayanagi, {\it {A Holographic proof of the strong
  subadditivity of entanglement entropy}},  {\em Phys. Rev. D} {\bf 76} (2007)
  106013, [\href{http://xxx.lanl.gov/abs/0704.3719}{{\tt arXiv:0704.3719}}].

\bibitem{Hubeny:2007re}
V.~E. Hubeny and M.~Rangamani, {\it {Holographic entanglement entropy for
  disconnected regions}},  {\em JHEP} {\bf 0803} (2008) 006,
  [\href{http://xxx.lanl.gov/abs/0711.4118}{{\tt arXiv:0711.4118}}].

\bibitem{Tonni:2010pv}
E.~Tonni, {\it {Holographic entanglement entropy: near horizon geometry and
  disconnected regions}},  {\em JHEP} {\bf 1105} (2011) 004,
  [\href{http://xxx.lanl.gov/abs/1011.0166}{{\tt arXiv:1011.0166}}].

\bibitem{Balasubramanian:2013lsa}
V.~Balasubramanian, B.~D. Chowdhury, B.~Czech, J.~de~Boer, and M.~P. Heller,
  {\it {A hole-ographic spacetime}},  {\em Phys.Rev.} {\bf D89} (2014) 086004,
  [\href{http://xxx.lanl.gov/abs/1310.4204}{{\tt arXiv:1310.4204}}].

\bibitem{Myers:2014jia}
R.~C. Myers, J.~Rao, and S.~Sugishita, {\it {Holographic Holes in Higher
  Dimensions}},  \href{http://xxx.lanl.gov/abs/1403.3416}{{\tt
  arXiv:1403.3416}}.

\bibitem{Klebanov:2007ws}
I.~R. Klebanov, D.~Kutasov, and A.~Murugan, {\it {Entanglement as a probe of
  confinement}},  {\em Nucl. Phys.} {\bf B796} (2008) 274--293,
  [\href{http://xxx.lanl.gov/abs/0709.2140}{{\tt arXiv:0709.2140}}].

\bibitem{Lewkowycz:2012mw}
A.~Lewkowycz, {\it {Holographic Entanglement Entropy and Confinement}},  {\em
  JHEP} {\bf 1205} (2012) 032, [\href{http://xxx.lanl.gov/abs/1204.0588}{{\tt
  arXiv:1204.0588}}].

\bibitem{Hirata:2006jx}
T.~Hirata and T.~Takayanagi, {\it {AdS/CFT and strong subadditivity of
  entanglement entropy}},  {\em JHEP} {\bf 0702} (2007) 042,
  [\href{http://xxx.lanl.gov/abs/hep-th/0608213}{{\tt hep-th/0608213}}].

\bibitem{Krtous:2013vha}
P.~Krtous and A.~Zelnikov, {\it {Entanglement entropy of spherical domains in
  anti-de Sitter space}},  {\em Phys.Rev.} {\bf D89} (2014) 104058,
  [\href{http://xxx.lanl.gov/abs/1311.1685}{{\tt arXiv:1311.1685}}].

\bibitem{Hawking:1973uf}
S.~Hawking and G.~Ellis, {\it {The Large scale structure of space-time}}, .

\bibitem{GradshteinRyzhik:book}
I.~Gradshteyn and I.~Ryzhik, {\em {Table of integrals, series, and products.
  Transl. from the Russian by Scripta Technica, Inc. 5th ed.}}
\newblock Boston, MA: Academic Press, Inc., 5th ed.~ed., 1994.

\bibitem{ByrdFriedman:book}
P.~Byrd and M.~Friedman, {\em Handbook of elliptic integrals for engineers and
  scientists}.
\newblock Grundlehren der mathematischen Wissenschaften. Springer-Verlag, 1971.

\bibitem{Casini:2012ei}
H.~Casini and M.~Huerta, {\it {On the RG running of the entanglement entropy of
  a circle}},  {\em Phys.Rev.} {\bf D85} (2012) 125016,
  [\href{http://xxx.lanl.gov/abs/1202.5650}{{\tt arXiv:1202.5650}}].

\bibitem{Myers:2010tj}
R.~C. Myers and A.~Sinha, {\it {Holographic c-theorems in arbitrary
  dimensions}},  {\em JHEP} {\bf 1101} (2011) 125,
  [\href{http://xxx.lanl.gov/abs/1011.5819}{{\tt arXiv:1011.5819}}].

\bibitem{Lashkari:2013koa}
N.~Lashkari, M.~B. McDermott, and M.~Van~Raamsdonk, {\it {Gravitational
  dynamics from entanglement 'thermodynamics'}},  {\em JHEP} {\bf 1404} (2014)
  195, [\href{http://xxx.lanl.gov/abs/1308.3716}{{\tt arXiv:1308.3716}}].

\bibitem{Faulkner:2013ica}
T.~Faulkner, M.~Guica, T.~Hartman, R.~C. Myers, and M.~Van~Raamsdonk, {\it
  {Gravitation from Entanglement in Holographic CFTs}},  {\em JHEP} {\bf 1403}
  (2014) 051, [\href{http://xxx.lanl.gov/abs/1312.7856}{{\tt
  arXiv:1312.7856}}].

\end{thebibliography}
\end{document}